\documentclass[11pt,a4paper]{article} 

\usepackage{natbib,hyperref,caption,geometry,lineno,graphicx,amsmath,booktabs,amsthm}
\usepackage{amssymb}
\usepackage{xcolor}
\usepackage{multicol}
\usepackage{color}
\usepackage{lscape}
\usepackage{eurosym}
\usepackage{siunitx}
\usepackage{subcaption}
\usepackage{empheq}
\usepackage{xr}
\newcommand{\footremember}[2]{%
    \footnote{#2}
    \newcounter{#1}
    \setcounter{#1}{\value{footnote}}%
}
\newcommand{\footrecall}[1]{%
    \footnotemark[\value{#1}]%
} 

\begin{document}

\title{\emph{e4clim 1.0}: The Energy for CLimate Integrated Model: Description and Application to Italy}

\author{
  Alexis Tantet\footremember{lmd}{LMD/IPSL, \'Ecole polytechnique, IP Paris, Sorbonne Universit\'e, ENS, PSL University, CNRS, Palaiseau, France}
  \and Marc St\'efanon\footrecall{lmd}
  \and Philippe Drobinski \footrecall{lmd}
  \and Jordi Badosa\footrecall{lmd}
  \and Silvia Concettini\footremember{iji}{IRJI, Universit\'e de Tours, Tours, France}\footremember{depeco}{D\'epartement d'Economie, \'Ecole polytechnique, IP Paris, Palaiseau, France}
  \and Anna Cret\`i\footrecall{depeco}\footremember{dauphine}{Universit\'e Paris Dauphine, PSL, Leda-CGEMP, Paris, France}
  \and Claudia D'Ambrosio\footremember{lix}{LIX, \'Ecole polytechnique, IP Paris, CNRS, Palaiseau, France}
  \and Dimitri Thomopulos\footremember{crest}{CREST, ENSAE, \'Ecole polytechnique, IP Paris, Palaiseau, France}
  \and Peter Tankov$^{6}$
}

\maketitle

\begin{abstract}
  We develop an open-source Python software integrating flexibility needs from Variable Renewable Energies (VREs) in the development of regional energy mixes. It provides a flexible and extensible tool to researchers/engineers, and for education/outreach. It aims at evaluating and optimizing energy deployment strategies with high shares of VRE; assessing the impact of new technologies and of climate variability; conducting sensitivity studies. Specifically, to limit the algorithm's complexity, we avoid solving a full-mix cost-minimization problem by taking the mean and variance of the renewable production-demand ratio as proxies to balance services. Second, observations of VRE technologies being typically too short or nonexistent, the hourly demand and production are estimated from climate time-series and fitted to available observations. We illustrate \emph{e4clim}'s potential with an optimal recommissioning-study of the 2015 Italian PV-wind mix testing different climate-data sources and strategies and assessing the impact of climate variability and the robustness of the results.
\end{abstract}

\section{Introduction}\label{sec:introduction}

The world net electricity generation is expected to increase by 45\% between 2015 and 2040 \citep{iea_world_2017}.
In view of climate change and energy security concerns, the renewable energies will inevitably play a major role in satisfying this growing demand.
Variable Renewable Energy (VRE) such as solar photovoltaics (PV) and wind are the fastest-growing energy sources for new generation capacity and their share is expected to grow from 7\% of total world generation in 2015 to 15\% in 2040, with more than half of this growth coming from the wind power~\citep{iea_world_2017}.

However, while the size of VRE projects varies greatly, the harvesting of wind and solar energy is necessarily spread in space.
As they develop, VRE systems interact in a non-trivial way with a number of actors, such as citizens, ecosystems, markets and electricity networks~\citep{labussiere_energy_2018}.
Leaving aside critical social and political aspects of energy transitions, we focus on the integration of VRE to an existing interconnected system.
A key issue is the variable nature of the VRE production and the need for a constant supply-demand balance.
For systems historically dimensioned to face the variability of the demand only, VRE variability may lead to local power shortages or increased transmission congestion.
Today, this must be compensated at all times by an increased flexibility of the conventional generation systems such as coal plants or combined cycle gas turbines~\citep{huber_integration_2014}.
On the other hand, it brings higher price instability along with a reduction of the wholesale prices.
In the long run falling prices may `erode' the returns of both renewable and conventional producers, pushing the latter out of the market.
Yet, in the absence of non-fossil flexibility solutions, the latter are essential to smooth out the fluctuations of renewable-power output and ensure system stability.
Thereby, the possibilities for a future large-scale renewable capacity are still controversial~\citep{hirth_market_2013, spiecker_future_2014}.

Technological and spatial diversification are possible strategies to circumvent the problem of intermittency.
In Europe, wind and solar-generated electricity roughly have negatively correlated seasonal cycles, solar generation being maximal in summer and wind generation in winter~\citep{heide_seasonal_2010}.
Spatial diversification is only applicable at large scale, whenever the VRE variability is sufficient (see~\citet{widen_correlations_2011} for a study focusing on Sweden and~\citet{tsuchiya_electricity_2012} analyzing Japan).

Technological and geographical optimization of renewable energy systems within a multi-objective framework has been discussed by several authors at continent and country scales.
Complete electrical systems have been designed to quantify the requirements in installed power, transmission grid and storage capacity for a 100\% of renewable energy scenario over Europe.
For example, at the European scale,~\citet{heide_reduced_2011} optimize the wind-solar mix in a fully renewable future European power system to reduce the storage and balancing needs;
\citet{rodriguez_transmission_2014} do the same for the cross-border transmission capacities in the future;
and~\citet{becker_transmission_2014} investigate the change in the optimal wind-solar mix in Europe as the transmission grid is enhanced.
\citet{becker_features_2014} optimize the wind-solar mix in the US to reduce storage needs and \citet{nelson_high-resolution_2012} simulate how a range of generation technologies, storage and transmission may meet the projected energy demand in the US at the least societal cost.
~\citet{elliston_simulations_2012} analyze how the Australian renewable mix should change in order to reduce the need for backup generation;
and~\citet{lund_energy_2009} discuss feasible energy mix scenarios for a fully renewable electricity supply in Denmark.
Finally,~\citet{francois_complementarity_2016} analyse the complementary of run-of-the river hydro-power (RoR) with PV in northern Italy, while~\citet{raynaud_energy_2018} evaluate optimal RoR-PV-wind mixes in 12 independent Euro-Mediterranean regions.

Other conceptual frameworks with less ambitious energy targets have been explored at continental and regional scale by repowering the current installed renewable energy capacity.
Repowering consists in fully decommissioning current renewable energy capacity
and in re-allocating this capacity according to specific objectives~\citep{del_rio_policies_2011}.
For example,~\citet{beltran_modern_2009} applies the mean-variance optimization techniques to infer the optimal energy mix;
\citet{roques_optimal_2010} use similar methods to determine optimal wind power deployment among 5 European countries;
\citet{thomaidis_optimal_2016} and~\citet{santos-alamillos_exploring_2017} use mean-variance optimization to assess the optimal wind and solar deployment and repowering actions in Spain.
These studies use the Markovitz mean-variance portfolio theory or analogous methodologies to define the optimal full re-allocation of existing power plants among regions.
It relies on a trade-off between maximizing the mean renewable productivity while minimizing the aggregate renewable energy supply risk (i.e.~variability).
Note that these studies evaluate the risk of a given renewable energy mix using power production data only, whereas this risk clearly depends not only on power production, but also on the power consumption, which is also sensitive to
climate.

Many assessments of the optimal renewable energy mix are based on the statistical properties of the historical production and demand.
Due to the only recent deployment and monitoring of wind and solar energy systems, the length of regional production and demand time series is often limited to a few years at best.
This is not sufficient to properly take into account the effect of interannual climate fluctuations~\citep{jourdier_wind_2015} on the covariance used in mean-variance analyses.
In addition, relying on energy observations does not allow for including new technologies (e.g.~offshore wind-energy), for which no observations exist, in analyses.

To alleviate these two issues, other studies~\citep[e.g.][]{bremen_large-scale_2010}, rely on weather observations or climate simulations (such as reanalyses or projections) to estimate the renewable production and the electricity demand.
Times series from observations are often still too short, however, to include contributions from all significant time-scales to the variance.
On the other hand, strong biases exist between different climate simulations, so that multi-model approaches are essential to estimate the robustness of energy estimates on the climate data.

Last, most studies do not provide an open-source software that would allow one to reproduce published results, perform sensitivity analyses, or answer new research questions based on existing methodologies.
\citep{wiese_open_2014} reviews existing energy models and proposes a novel open-source software for public participation in the development of strategies adapted to the German and European electricity system.
Their study shows the need for tools that are both accessible (open-source, open-data, documented, etc.) and with a level of complexity amenable to sensitivity studies, as opposed to operational electricity-system models used by system operators.

We share with the aforementioned studies the overarching goal of assessing the techno-economic feasibility of existing energy transition strategies and exploring different energy transition solutions integrating high shares of VRE.
This work differs, however, both by the methodology that is developed and by the design of the associated \emph{e4clim} software.
The latter aim at:
\begin{itemize}
\item Providing an integrated energy-system modeling-framework of intermediate complexity amenable to sensitivity analyses.
\item Evaluating different optimal strategies taking into account, or not, the potential for geographic and technological diversification.
\item Integrating new technologies which have not been monitored in real conditions.
\item Taking into account flexibility needs associated both with a variable energy production and a variable electricity demand.
\item Assessing the impact of interannual to intraday climate variability and, potentially, climate change on electricity mixes.
\item Tracking uncertainties stemming from the climate data into account through a multi-model approach.
\item Providing a flexible tool open to the research, engineering and education communities that is open-source, uses open-data, and is user-friendly.
\end{itemize}
How these objectives are achieved and the decisions behind the design of the
software are explained in the following sections.
Note that, at the moment, we focus on electricity mixes only, but that interactions between different vectors may be considered in the future.

To illustrate the present capabilities and the potential of this software and its methodology, we show an application to the recommissioning of the 2015 Italian PV-wind mix.
Furthermore, we provide new insights regarding the distribution of PV-wind capacities in Italy that minimize flexibility needs.


The remainder of the paper is structured as follows.
Section~\ref{sec:software_design} outlines the methodology and the general software design.
A concrete implementation of the complete chain of modules allowing to perform a climate-aware mean-variance analysis is presented in Section~\ref{sec:method}.
In Section~\ref{sec:application} we illustrate the capabilities of \emph{e4clim} with an optimization of geographical distribution of wind and PV generation in Italy for historical climate conditions, with a comparison with the actual PV-wind mix.
We also show how \emph{e4clim} can be used to assess the impact of climate variability on the optimal mixes and the sensitivity of the results to the climate data.
Conclusions are drawn in Section~\ref{sec:conclusion}.

Furthermore, an extensive supplementary material is provided. There, Appendix~\ref{sec:data_model_description} details the energy and climate datasets, the production and demand models.
The mean-variance optimization problem, its mathematical formulation and the algorithm are presented in Appendix~\ref{sec:meanVariance}.
The robustness of the numerical results is tested against observations in Appendix~\ref{sec:evaluation}.
Finally, the \emph{e4clim} source-code is available at \url{https://gitlab.in2p3.fr/alexis.tantet/e4clim} and its documentation --- including all cases from this study --- at \url{https://alexis.tantet.pages.in2p3.fr/e4clim/}.

\section{Methodology and software design}\label{sec:software_design}

The purpose of this section is to outline the design of \emph{e4clim}.
To show its full potential, this description remains voluntarily abstract,
while the next Section~\ref{sec:method} gives the concrete implementation of the climate-aware mean-variance analysis.

Summarizing some of the methodology's main objectives (Sect.~\ref{sec:introduction}), the \emph{e4clim} software is a flexible tool allowing for the evaluation of energy mixes, taking into account the variability of the demand and production, from both mature and emerging technologies, on a broad range of time scales.
An energy mix is defined here as a set of georeferenced capacities (e.g.~at the scale of bidding zones or states) per energy source and may be prescribed,  e.g.~taking the actual mix of a given area, or optimized, e.g.~with the mean-variance analysis detailed in the next Section~\ref{sec:method}.
We are interested in properties of the energy mixes such as the mean penetration, the risk, the frequency of occurrence of critical situations, etc.
These properties may provide the objectives of an optimization problem or may be computed ex post.
They are computed from georeferenced energy data, such as demand and capacity factors per source and electrical region.

In \emph{e4clim}, energy time-series may directly be taken from observations.
In order to consider new technologies and to resolve the impact of low-frequency variability on the production, however, it is also possible to estimate energy data by applying statistical models to climate time-series, e.g.~of temperature, wind speed or irradiance.
To summarize, an \emph{e4clim} project is thus divided in three phases:
\begin{enumerate}
\item Computing georeferenced energy-time-series from historic or climate data,
\item Distributing capacities spatially and technologically,
\item Post-processing and analyzing the resulting mixes.
\end{enumerate}
        
Energy mixes are based on several \emph{components}, i.e.~loads or sources (wind, PV, etc.) for which georeferenced time-series of relevant \emph{variables} (demand, capacity factors, etc.) must be estimated or parsed, for a given area (e.g.~a country or a macro region).
The algorithms used to compute these variables are composed of statistical models made of sequences of blocks, and of data sources required by the models.
Statistical models and data sources are, however, independent from each other and connected through a standard interface.
New algorithms may thus be composed by assembling different sources and models.
In particular, it is possible to either use energy observations provided by utilities directly, or to rely on statistical learning to fit demand/production models to observations and make predictions over a longer/future period from climate data.

These time series are then used in the optimization step and for the mix analysis, together with installed capacities.
In the future, controllable solutions (production, storage) could be dispatched at this post-processing stage to compute economic/carbon costs associated with the satisfaction of the mismatch between the demand and the VRE production.

\section{A concrete implementation for mean-variance analyses}\label{sec:method}

\begin{figure}
  \centering
  \includegraphics[width=\linewidth]{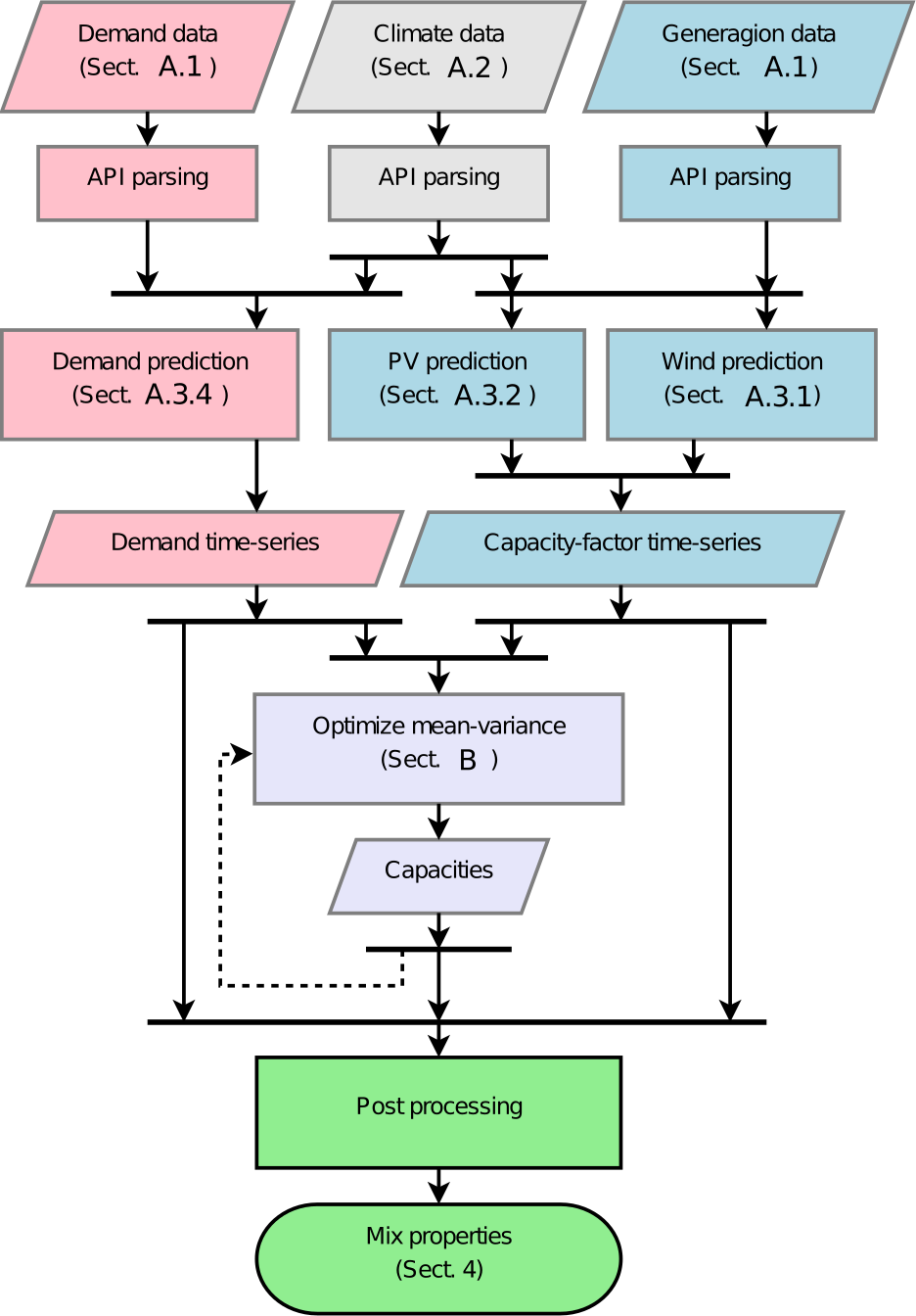}
  \caption{
    Flow chart of the concrete implementation of the mean-variance analysis for the Italian PV-wind mix.
  }\label{fig:flow_chart}
\end{figure}

We now describe the implementation in \emph{e4clim} of the mean-variance analysis applied in the next Section~\ref{sec:application}.
The corresponding flow chart is given in Figure~\ref{fig:flow_chart}.
We proceed backward from the end of the chart to describe this program.

\paragraph{Mix analysis}
The ``post-processing'' step translates capacities into mix properties such as the PV fraction and shortage and saturation occurrence frequencies (see plots for Italy in Section~\ref{sec:application}).
This step may be further developed to compute economic costs and GHG emissions associated with a mix.

\paragraph{Mix optimization}
In \emph{e4clim}, a mix is either prescribed or obtained as the solution of an optimization problem.
In order to isolate the optimization of the VRE capacity from the rest of the energy systems, we use a ``mean-variance'' analysis of the VRE production with respect to the demand.
The latter is based on two measures:
the \emph{mean penetration},

\begin{align}
  \mu := \frac{\mathbb{E}\left[\sum_\mathbf{k} w_\mathbf{k} \eta_\mathbf{k}(\cdot) \right]}{\mathbb{E}\left[\sum_i D_i(\cdot)\right]}
  = \frac{\sum_\mathbf{k} w_\mathbf{k} \mathbb{E}\left[\eta_\mathbf{k}(\cdot)\right]}{\mathbb{E}\left[\sum_i D_i(\cdot)\right]},
  \label{eq:meanPenetration}
\end{align}
and the variance, or squared \emph{risk},

\begin{align}
  \sigma_\mathrm{global}^2(\mathbf{w})
  := \mathbb{V}\left[\frac{\sum_\mathbf{k} w_\mathbf{k} \eta_\mathbf{k}(\cdot)}{\sum_i D_i(\cdot)}\right]
  .\label{eq:globalRisk}
\end{align}
In other words, the mean penetration is given by the ratio of the mean total PV and wind production over the mean total demand,
and the risk is given by the standard deviation of the total PV and wind production over the total demand.
Here, $\mathbf{k} = (i, j)$ is the multi-index composed of an index $i$ running over zones, or electricity regions, and a technological index $j$ running over technologies, the $w_\mathbf{k}$ are the installed capacities for each region and technology, the $\eta_\mathbf{k}(t)$ are the corresponding predicted time-dependent zonal capacity factors and the $D_i(t)$ are the predicted zonal demands (Sect.~\ref{sec:model_description}).
Note that, in the following numerical applications, statistics such as the expectation or the variance are replaced by sample estimates from the climate record.

Taking the mean penetration and risk as two objectives, the mean-variance analysis translates into an optimization problem distributing PV and wind capacities:

\begin{subequations}\label{eq:optimization}
\begin{empheq}[box=\fbox]{align}
    \min_\mathbf{w} \quad
    & \sigma^2_{\mbox{global}|\mbox{technology}|\mbox{base}}(\mathbf{w})\label{eq-biobj:of1}\\
    \max_\mathbf{w} \quad
    & \sum_\mathbf{k} w_\mathbf{k}~\mathbb{E}[\eta_\mathbf{k}]\label{eq-biobj:of2}\\
    \mbox{subject to} \quad
    &{\color{gray} \sum_\mathbf{k} w_\mathbf{k}
      = w_{\mbox{total}}}\label{eq-biobj:constr_sum_w}\\
    &\quad w_\mathbf{k} \geq~0 \quad
    \forall~\mathbf{k},\label{eq-biobj:positive_w}
\end{empheq}
\end{subequations}
where $\sigma^2_{\mbox{technology}}$ and $\sigma^2_{\mbox{base}}$ are variants (see below) of the global squared-risk $\sigma^2_{\mbox{global}}$ defined in~\eqref{eq:globalRisk}, and $w_{\mbox{total}}$ is a total VRE capacity used to constrain the sum of the capacities (see below).

This problem is equivalent to minimizing both the mean and the variance of the mismatch between the demand and the VRE production.
Together, the latter give an proxy of the implied costs, or GHG emissions, necessary to meet the electricity demand\footnote{
  Note, however, that costs are usually higher for shortage than for surplus situations, so that an asymmetric statistic may be more appropriate than the variance.}.
In Section~\ref{sec:application}, results with and without the total VRE capacity constraint~\eqref{eq-biobj:constr_sum_w} are discussed.
Moreover, the production and the demand are summed ignoring both network constraints and exchanges with other countries.

By minimizing the variance~\eqref{eq:globalRisk}, weaker covariances between regions and technologies are leveraged.
We refer to this case as the \emph{global} strategy.
Two other strategies are considered, one in which only covariances between technologies within the same region are processed, the \emph{technology} strategy, and one in which no covariances are taken into account, the \emph{base} strategy.
Comparing mixes from these strategies allows us to assess the benefits from technological and geographical diversification.

\begin{figure}
  \centering
  \includegraphics[width=\linewidth]{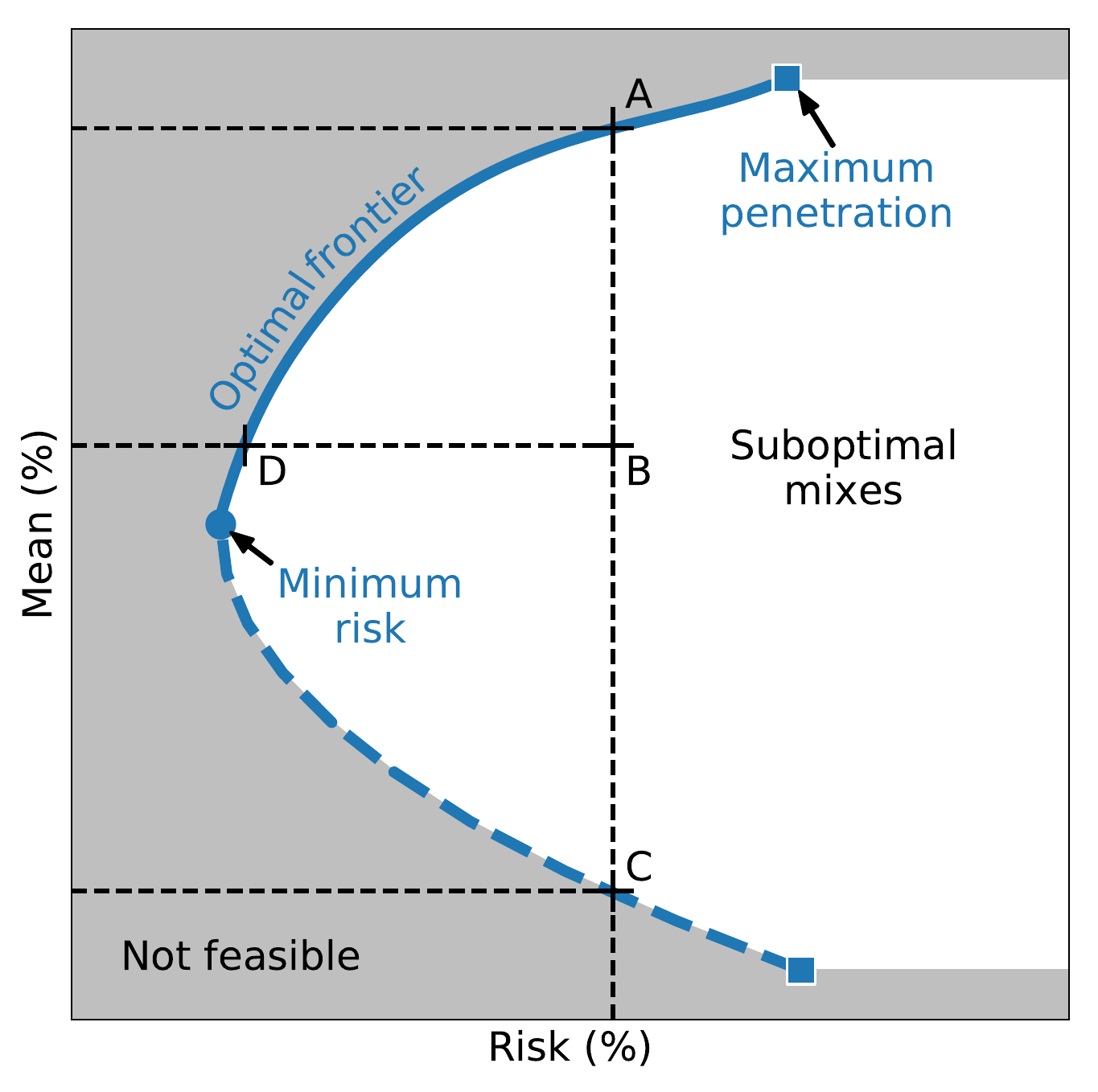}
  \caption{
    Example of the optimal frontier of a mean-variance bi-objective optimization problem.
    The optimal frontier is one-dimensional and represented by a plain blue line.
    Mixes in the white region to the right of the frontier are suboptimal.
    Points in the gray region to the left of the frontier are not feasible.
    In this example, the optimal frontier is bounded below by a minimum-risk optimal-mix (blue dot) below which the risk may only increase.
    The optimal frontier is bounded above by a maximum-penetration optimal-mix above which higher penetration mixes are not feasible due to the constraints of the problem.
    The point B is an example of suboptimal mix, since a higher mean penetration is achievable for the same risk (point A) and a lower risk is achievable for the same mean penetration (point D).
    The dashed blue line is obtained by minimizing the risk for a range of target mean penetration values.
    These solutions are, however, not Pareto optimal.
    For instance, point C yields the same risk as point A but achieves a lower mean penetration.
    Thus, A ``dominates'' C.
  }\label{fig:pareto}
\end{figure}

As a bi-objective optimization problem~\citep{miettinen_nonlinear_1999}, there exists a set of Pareto-optimal mixes, the optimal frontier.
Each optimal mix may be represented in a mean-variance chart, as illustrated in Figure~\ref{fig:pareto}.
A solution is said to be Pareto optimal if there exists no feasible solution with a better or equal value for each of the objective functions.
The points under or to the right of the frontier are by definition suboptimal and will be discarded by a rational investor.
The area above or to the left of the frontier cannot be reached.
A detailed description of the mean-variance analysis procedure is given in Appendix~\ref{sec:meanVariance}.

\paragraph{Energy models}
A proper estimation of the mean penetration and the risk is key to the mean-variance analysis.
``Demand'' and ``capacity-factor time-series'' are thus needed.
While computing the mean does not require data at a particular sampling frequency, the variance should be computed from long time-series at a high sampling-frequency to measure variability on a sufficiently wide range of time scales.
Indeed, the variance of the renewable production, and, to some extent, of the demand, stems from climate variability and is distributed over a broad range of spatial and temporal scales.
In order to take into account time scales ranging from hours to decades and to be able to integrate new technologies for which no or little data is available, time series of the demands and capacity factors per zone are computed from climate data.
To do so:
\begin{itemize}
\item the ``wind'' production is ``predicted'' from wind data fed to a power curve at each grid point (of the climate data), summed over each zone, and bias corrected against wind production observations (Sect.~\ref{sec:wind_model}),
\item the ``PV'' production is computed from surface irradiance and temperature data fed to an electric model, summed over each zone, and bias corrected against PV production observations (Sect.~\ref{sec:pv_model}),
\item the ``demand'' is estimated via a linear Bayesian regression model taking as input warming and cooling degree days averaged over each zone and fitted to demand observations (Sect.~\ref{sec:demandModel}).
\end{itemize}

Importantly, when used with daily-mean climate data, these models include a parameterization of intraday fluctuations.
These models are evaluated against observations in Appendix~\ref{sec:evaluation}.

\paragraph{Energy and climate data}

The energy models rely on ``demand'', ``generation'' and ``climate data''.
Application Programming Interfaces (APIs) are developed to download and format the required data (``API parsing'').
The data sources used for the application of the next Section~\ref{sec:application} are described in Section~\ref{sec:GMEGSE}-\ref{sec:climate_data}.
Downstream \emph{e4clim} data follow a standard format allowing to use different data sources for the same purpose.
In particular, several climate data sources may be used to assess biases stemming from the latter.

\section{Application: Italian PV-wind optimal recommissioning}\label{sec:application}

\begin{figure}
  \centering
  \begin{subfigure}{0.53\linewidth}
    \includegraphics[width=\linewidth]{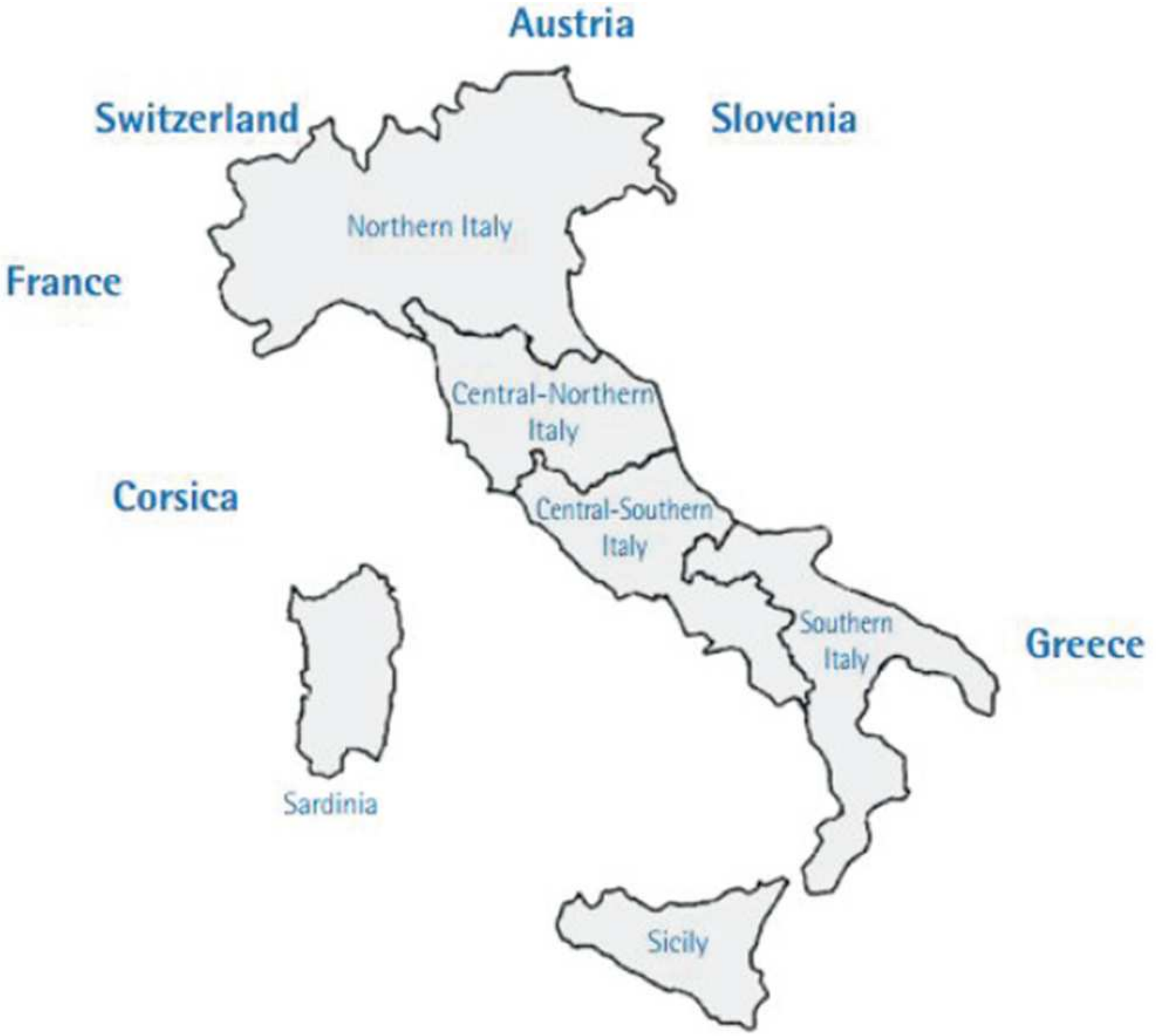}
    \caption{Italian bidding zones.\label{ElecRegions}}
  \end{subfigure}
  \begin{subfigure}{0.45\linewidth}
    \includegraphics[width=\textwidth]{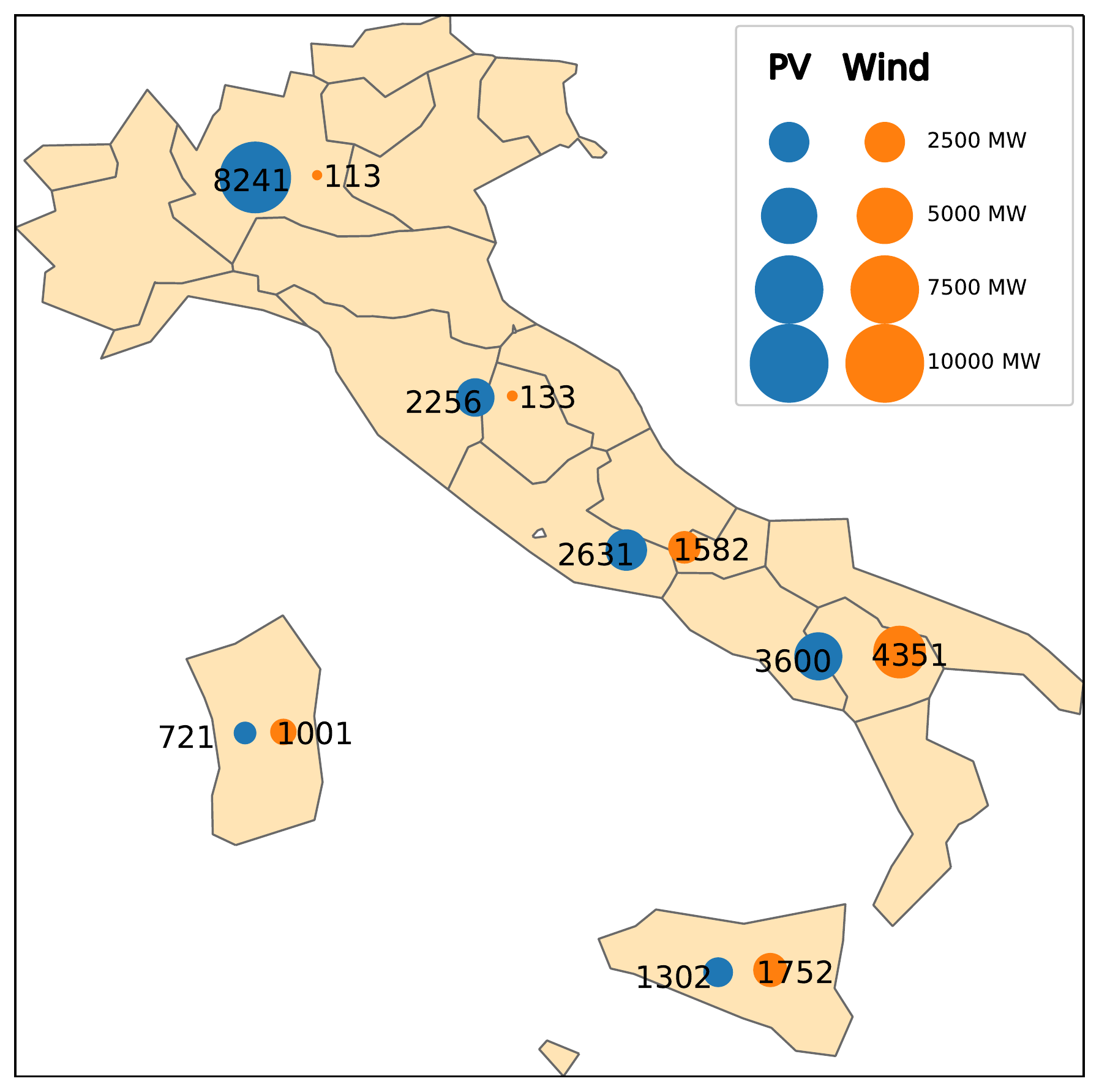}
    \caption{
      PV-wind capacities installed by the end of 2015 in Italy (Sect.~\ref{sec:GMEGSE}).
    }\label{fig:capacityMapGSE}
  \end{subfigure}
\end{figure}

We now present the application to the Italian PV-wind mix illustrating the potential of the methodology and the software.
We focus on Italy and its 6 bidding zones, or electrical regions, as shown in Figure~\ref{ElecRegions}.
Italy offers an interesting case study of a market with high renewables penetration as it has reached its quota of 17\% renewables in final energy consumption in 2014,
therefore implementing the 2009 Climate Package six years ahead of the 2020 horizon~\citep{gse_rapporto_2015}.
Figure~\ref{fig:capacityMapGSE} represents the geographical
distribution of the installed PV-wind capacities.
The PV (wind) installed capacity is \SI{18.8}{GW} (\SI{8.9}{GW}).
The share of the renewable energy production in the electricity demand
over the six zones is 19.4\% in 2015.

\subsection{General results}\label{sec:optimization}

\begin{figure}
  \centering
  \begin{subfigure}{0.49\linewidth}
    \includegraphics[width=\textwidth]{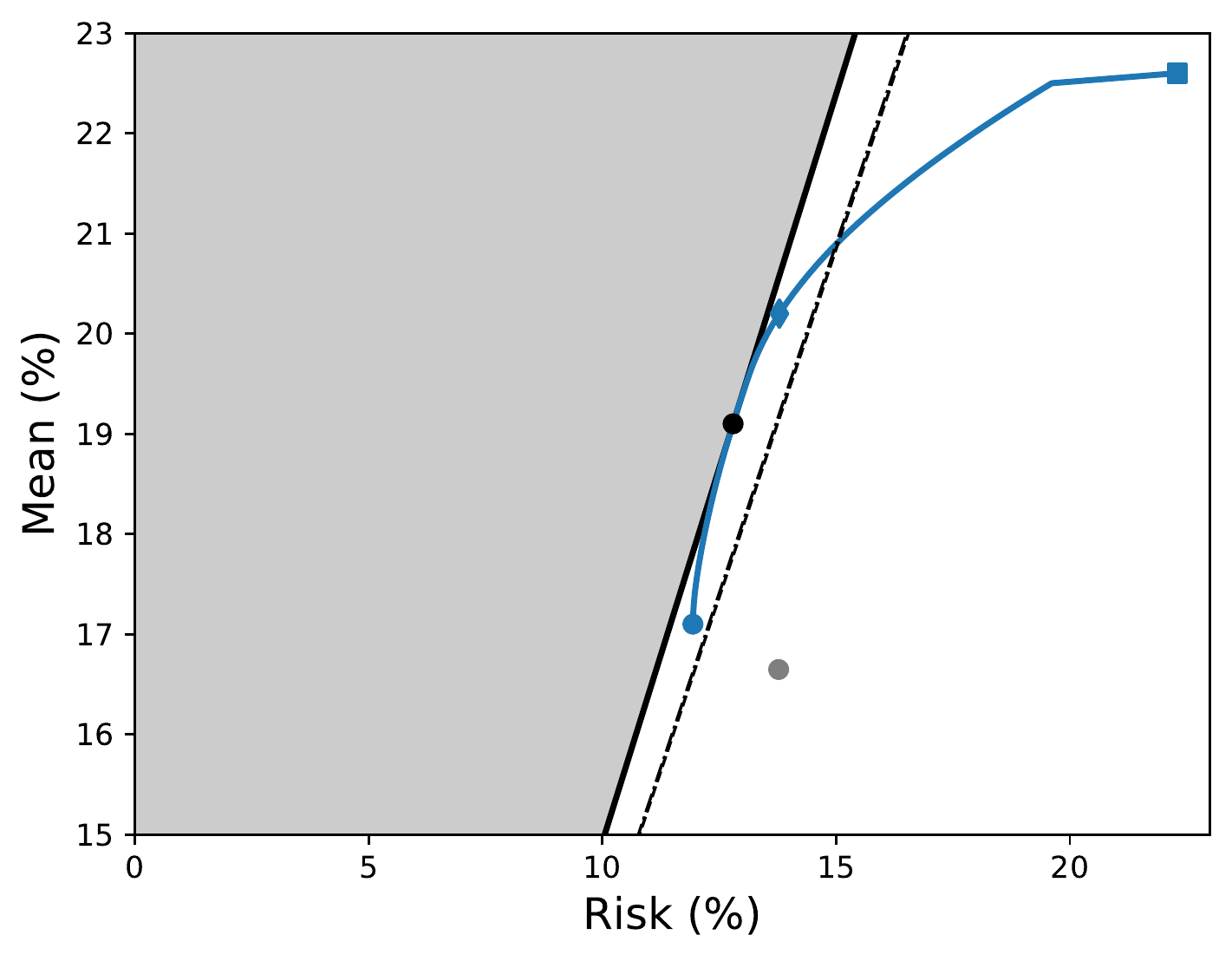}
    \caption{}\label{fig:meanRiskGlobal}
  \end{subfigure}
  \begin{subfigure}{0.49\linewidth}
    \includegraphics[width=\textwidth]{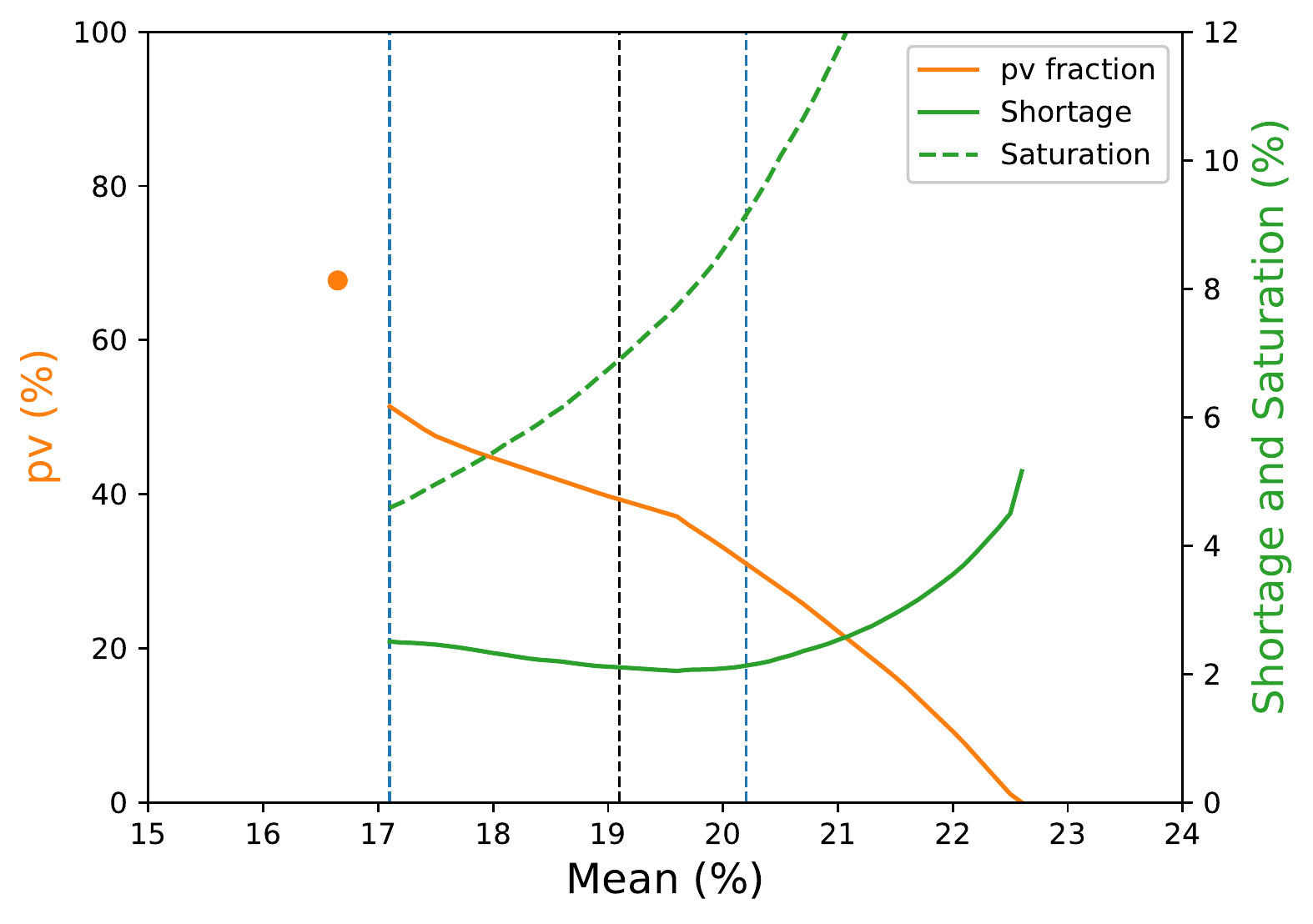}
    \caption{}\label{fig:ratioGlobal}
  \end{subfigure}
  \caption{
    Left:
    Approximations of the optimal frontiers from the CORDEX hourly data.
    The thick plain curves represent numerical approximations of the frontiers for the global strategy with (plain blue line) and without (plain black curve) total capacity constraint.
    The dashed and point-dashed black lines represent the optimal frontier
    without the total capacity constraint but for the technology and the base
    strategies.
    The approximations were obtained using a discretization step of $0.1\%$.
    The black dot where the thick black curve is tangent to the blue one corresponds
    to the optimal electricity mix for which the total capacity constraint is inactive.
    The blue dot in panel (a) corresponds to the optimal energy mix for which the risk is minimized while satisfying the total capacity constraint.
    The gray dot in panel (a) is obtained from the same capacity factor and demand data but applying the actual capacities installed in Italy in 2015.
    The blue diamond in panel (a) corresponds to the optimal mix achieving
    the same level of risk as the actual mix in gray while 
    maximizing the mean penetration.\\
    Right:
    Fraction of PV capacity in the mix (plain orange line);
    shortage frequency (plain green line);
    saturation frequency (dashed green line);
    versus the mean penetration and for the global strategy with total capacity constraint.
    The blue and black dashed vertical lines mark the mean penetration values corresponding to the blue and black dots and the blue diamond on the left panels.
    The orange dot represents the PV ratio for the actual capacities installed in Italy in 2015.
  }\label{fig:mix}
\end{figure}

We represent the optimal frontiers obtained from the CORDEX data~\citep{ruti_med-cordex_2016} with the intraday parameterizations over the 1989--2012 period (Appendix~\ref{sec:data_model_description}) in Figure~\ref{fig:mix}.
Each point of the frontiers represents an optimal distribution of the PV and wind capacities.
Representing frontiers rather than single optimal mixes leaves more space for arbitrages between mixes with high shares of VREs and mixes requiring less flexibility (low risk).
The latter could in term be guided by associated costs, GHG emissions, expert knowledge, values, etc.

Two variants are represented: one in which the total installed capacity is constrained to its observed 2015-value of $\SI{27.7}{GW}$ (plain blue curve), and one without such constraint (plain black curve).
We can see that the optimal frontier without total capacity constraint (thick black line) is a straight line passing through the origin.
Its slope, the \emph{mean-risk ratio}, is of $1.43$.
In other words, letting the risk increase by $1.00\%$ results in an increase of the  mean penetration by $1.43\%$, at best.

With the addition of the total capacity constraint (plain blue curve), the frontier bends away from the unconstrained one.
The point at which both curves intersect represents the mix for which the total capacity constraint is satisfied without the need to force it.
It is thus the optimal mix satisfying the total capacity constraint
that has the maximum mean-risk ratio.
If no preference is put on maximizing the mean penetration or on minimizing
the risk, this optimal mix is the most attractive mix satisfying the total capacity constraint, the \emph{maximum mean-risk ratio mix}.
One may be interested in allowing for the deterioration of the mean-risk ratio in order to either decrease the risk or increase the mean penetration.
The blue dot in Figure~\ref{fig:meanRiskGlobal} corresponds to the optimal mix minimizing the risk, the \emph{minimum-risk mix}.
For comparison with the actual mix the blue diamond in Figure~\ref{fig:meanRiskGlobal} represents the optimal mix that satisfies the same level of risk as the actual mix (gray dot) while maximizing the penetration rate,
the \emph{high-penetration mix}.

Benefits from interconnections between zones and synergies between technologies can be assessed by comparing the frontiers obtained for the global, technology and base strategies.
The technology and base frontiers without total capacity constraints (dashed and point-dashed thin black lines, respectively) almost coincide with each other.
Thus, taking local correlations between the PV-demand ratio and the wind-demand ratios into account do not significantly reduce the risk.
However, with a mean-risk ratio of $1.39$, these frontiers lie to the right of the global frontier.
Thus, for a given level of mean penetration, taking correlations between zones into account allows one to reduce the risk by about 5\%.

Key properties of the optimal mixes may then be derived.
This is shown in Figure~\ref{fig:ratioGlobal}, for the global strategy with total capacity constraint.
The fraction of PV capacity in the mix, is plotted in orange.
The plain and dashed green curves represent the frequency of occurrence of
shortage and saturation, respectively.
Here, it is assumed that conventional generation units are able to meet up to $80\%$ of the maximum demand modeled.
Shortage then occurs if the PV and wind generation is not able to meet the rest of the demand.
The second critical situation corresponds to network saturation, when PV and wind production exceeds technical limits of renewable energy fraction in the mix.
In this study, saturation is defined to occur if more than $40\%$ of the demand is met by PV and wind sources.

\begin{figure}
  \centering
  \begin{subfigure}{0.32\linewidth}
    \includegraphics[width=\textwidth]{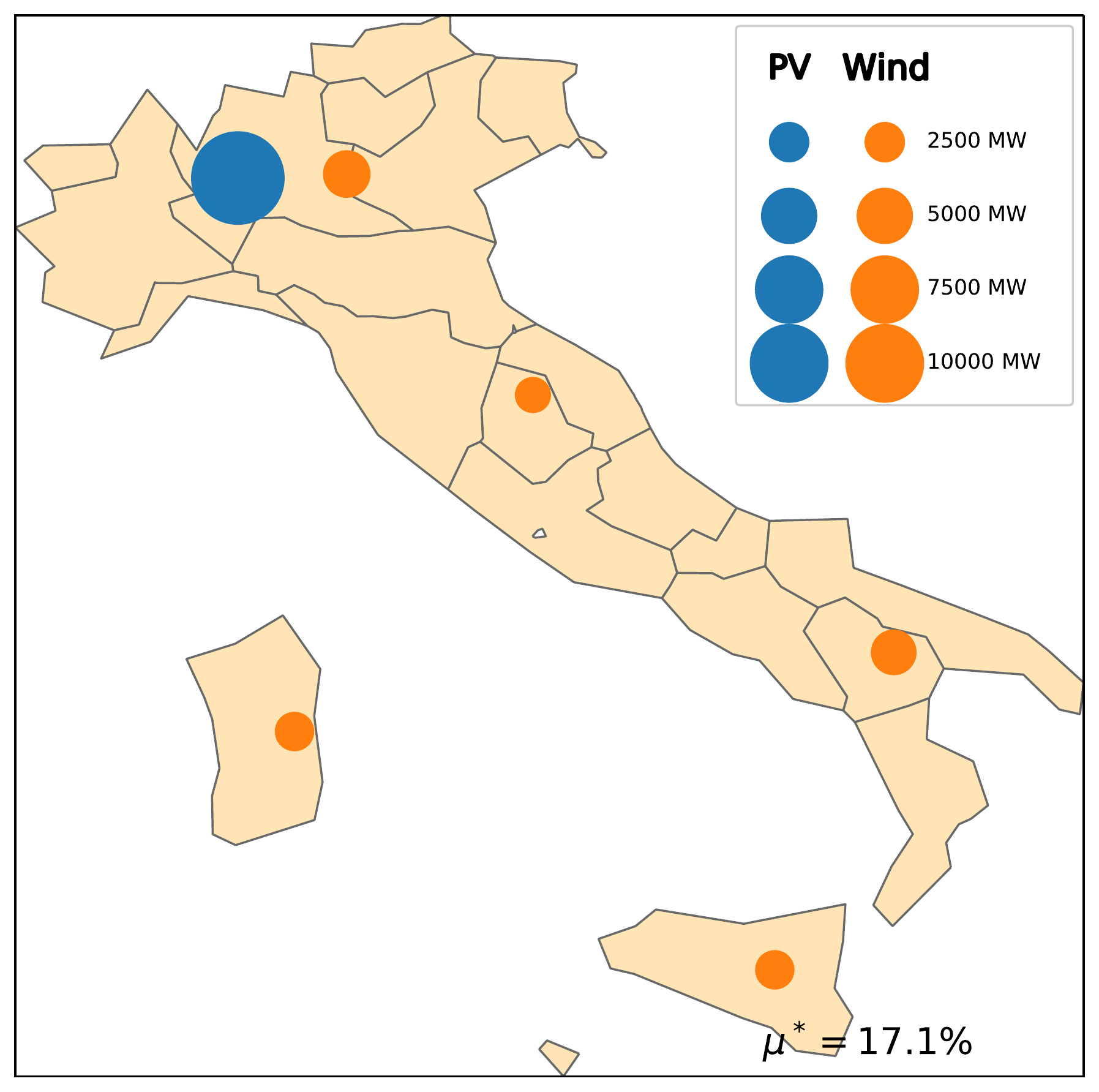}
    \caption{}\label{fig:mapGlobalMinRisk}
  \end{subfigure}
  \begin{subfigure}{0.32\linewidth}
    \includegraphics[width=\textwidth]{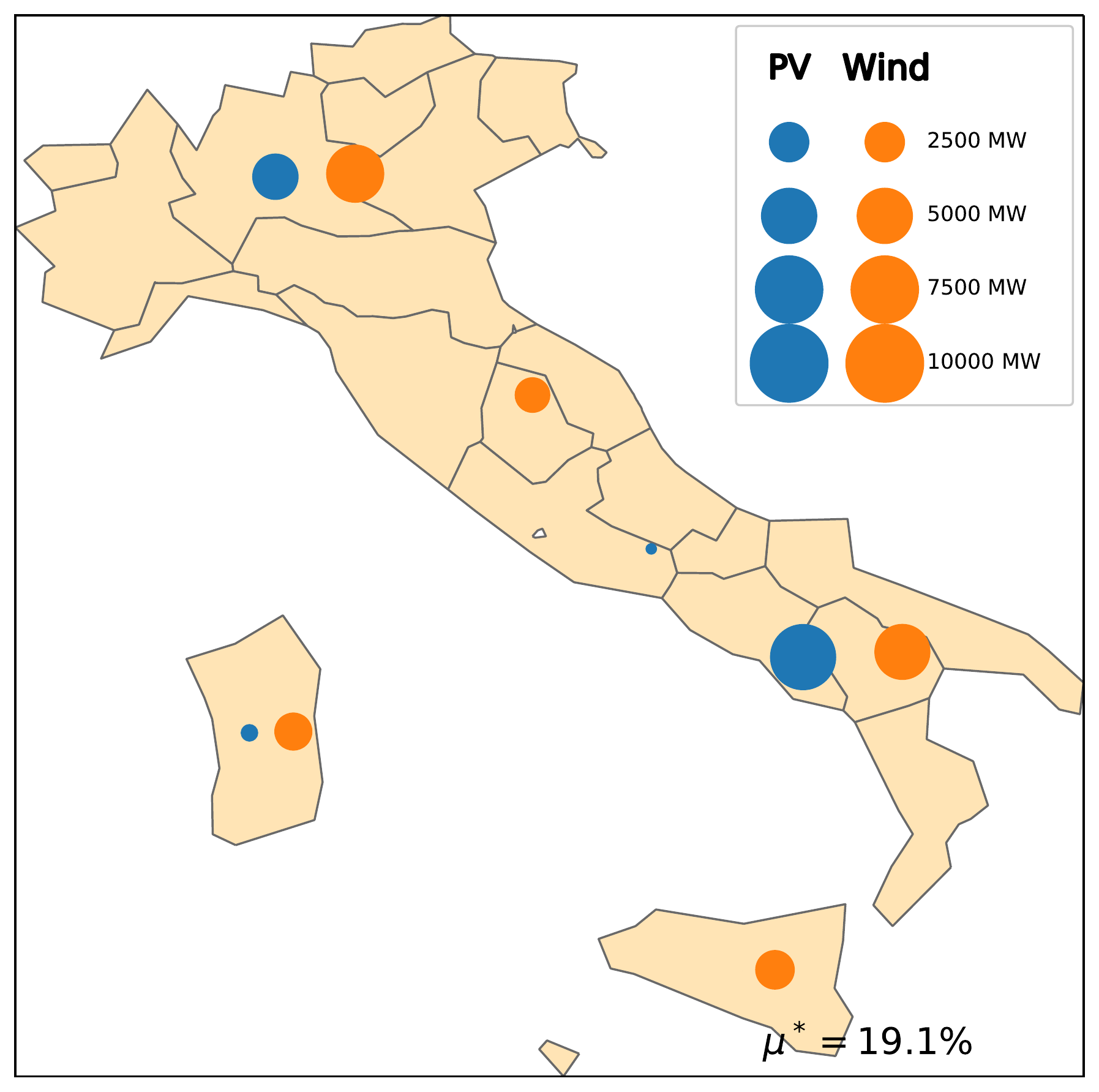}
    \caption{}\label{fig:mapGlobalMaxRatio}
  \end{subfigure}
  \begin{subfigure}{0.32\linewidth}
    \includegraphics[width=\textwidth]{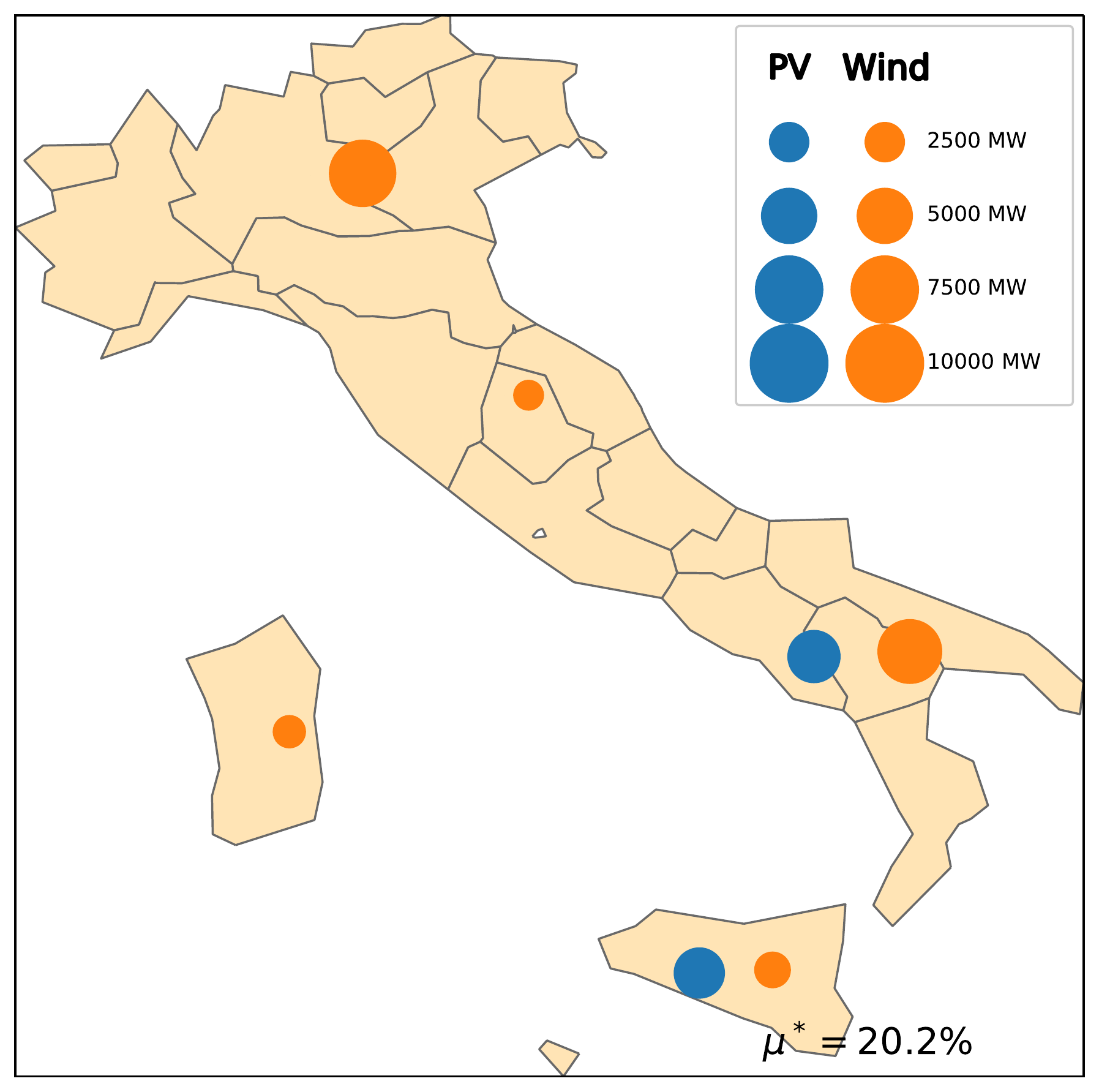}
    \caption{}\label{fig:mapGlobalHighPen}
  \end{subfigure}
  \caption{
    PV-wind capacity distributions obtained for the CORDEX hourly data for the global strategy with the total capacity constraint.
    The top, middle and bottom panels represent the optimal mixes for the minimum-risk, maximum mean-risk-ratio and high-penetration mixes, respectively (blue dot, black dot and blue diamond in Figure~\ref{fig:meanRiskGlobal}).
  }\label{fig:capacityMapOpt}
\end{figure}

Because wind capacity factors are higher than PV ones, the PV ratio is a decreasing function of the mean penetration.
The shortage and the saturation curves (in green) have distinct global minima due to the increase of the probability of occurrence of extremes with the risk.
The vertical lines in Figure~\ref{fig:ratioGlobal} represent the level of mean penetration for the minimum-risk, maximum-mean-risk-ratio and high-penetration mixes.
The minimum-risk, the maximum-mean-risk-ratio and the high-penetration mixes
respectively include $51\%$, $39\%$ and $31\%$ of PV capacity in the mix.
Saturation situations occur less often for the minimum-risk mix, while the maximum mean-risk ratio and the high-penetration mixes are close to the shortage-occurrence minimum.
Representing the corresponding PV-wind capacities in Figure~\ref{fig:capacityMapOpt} allows one to further compare these mixes.

To summarize, optimal mixes and their properties strongly depend on the level of risk that is tolerated.
As a consequence, capacities are not necessarily distributed where they would be expected based on the averaged resource potential only.

\subsection{Comparison with the 2015 Italian mix}\label{sec:comparisonActual}

The 2015 (actual) Italian mix, represented in Figure~\ref{fig:capacityMapGSE}, is composed of 68\% PV and 32\% wind energy capacity.
For historic and economic reasons, the largest fraction of installed PV capacity is in the North of Italy, whereas most of wind capacity is located in the South.

To compare the optimal mixes discussed so far with the actual 2015 Italian mix, it is possible to directly provide the latter to the \emph{e4clim} post-processing step (Fig.~\ref{fig:flow_chart}).
This mix is represented by the gray point in Figure~\ref{fig:meanRiskGlobal}.
It lies to the right of the optimal frontiers.
The actual Italian mix thus appears to be sub-optimal.
For the global problem, this mix reaches a level of mean penetration comparable to that of the minimum-risk mix, but its mean-risk ratio is about $19\%$ smaller than that of the latter and its PV ratio about $72\%$ larger.

\subsection{Choice of the climate data and climate variability}\label{sec:impactClimate}

By estimating the energy production and demand from climate data with \emph{e4clim}, we can discuss the impact of climate variability on mixes.
On the other hand, climate-data biases may also impact the quality of the results, thus calling for multi-model approaches.
Both points are now discussed.

\subsubsection{Dependence on the climate data}\label{sec:choice_climate}

The robustness of the mean-variance analysis presented in Section~\ref{sec:method} depends on the quality of the energy estimates from the climate data.
The latter is in turn impacted by climate-model biases.
With the \emph{e4clim} software, it is possible to use different climate-data sources to test the sensitivity of the results to biases stemming from the climate data.

To illustrate this point, we compare the results of Section~\ref{sec:application} obtained with the daily CORDEX data with intraday parameterizations (Appendix~\ref{sec:data_model_description}) with results obtained from hourly simulations from the MERRA-2 reanalysis (Sect.~\ref{sec:MERRA-2}), over the same period (1989--2012), for which intraday parameterizations are not needed.
Divergence in the results may thus stem both from differences in the climate data and from these parameterizations.

\begin{figure}
  \centering
  \begin{subfigure}{0.49\linewidth}
    \includegraphics[width=\textwidth]{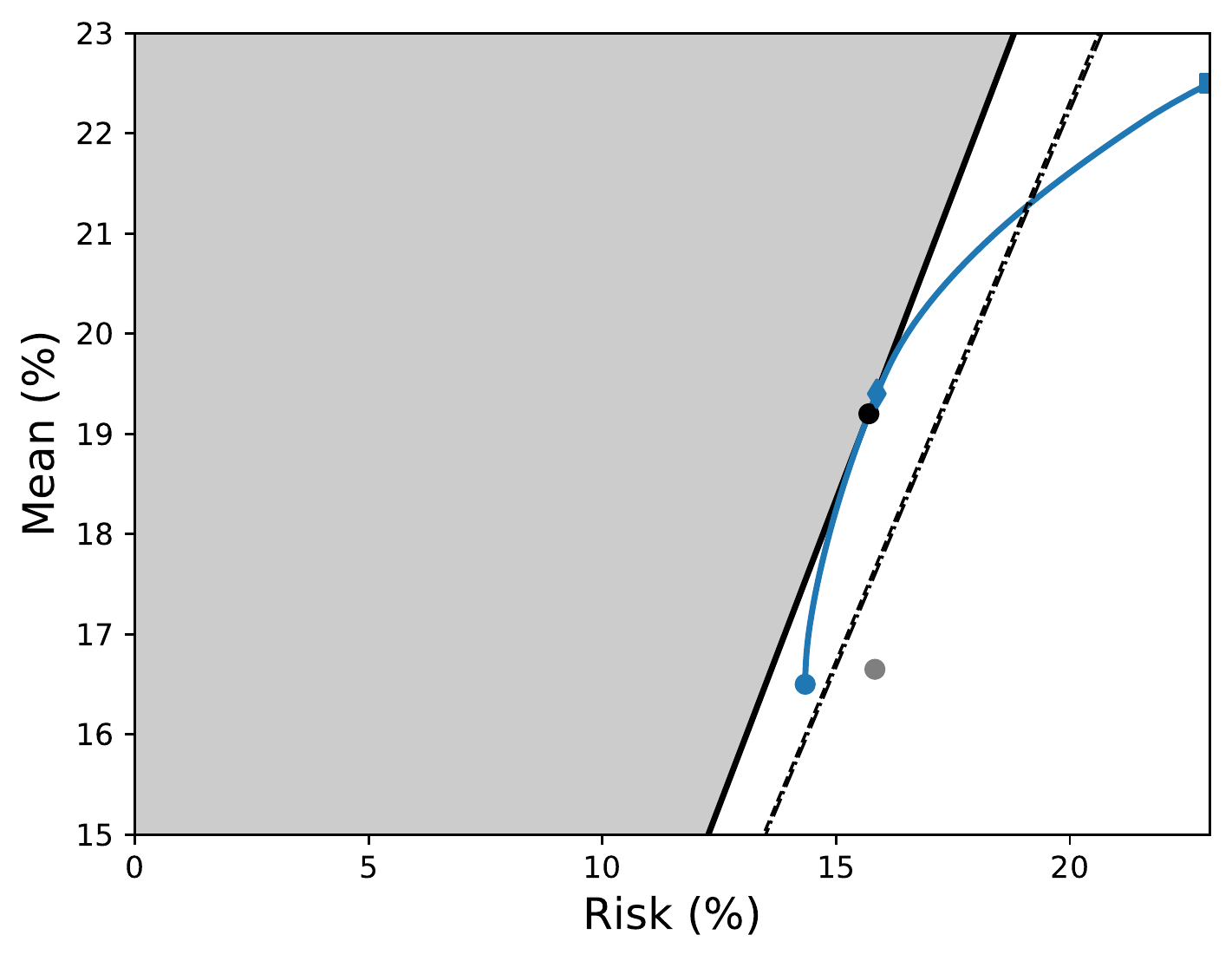}
    \caption{}
  \end{subfigure}
  \begin{subfigure}{0.49\linewidth}
    \includegraphics[width=\textwidth]{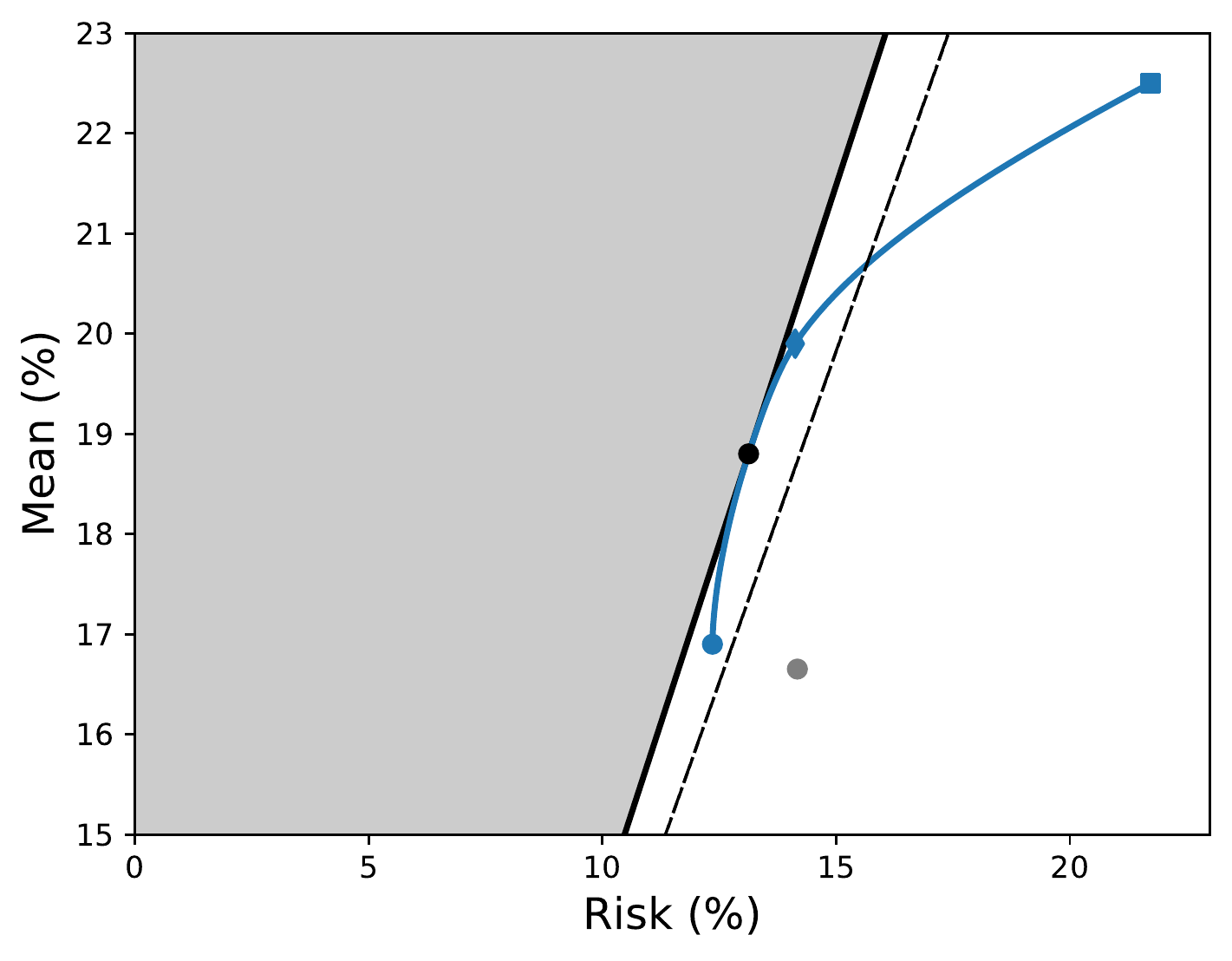}
    \caption{}
  \end{subfigure}\\
  \begin{subfigure}{0.49\linewidth}
    \includegraphics[width=\textwidth]{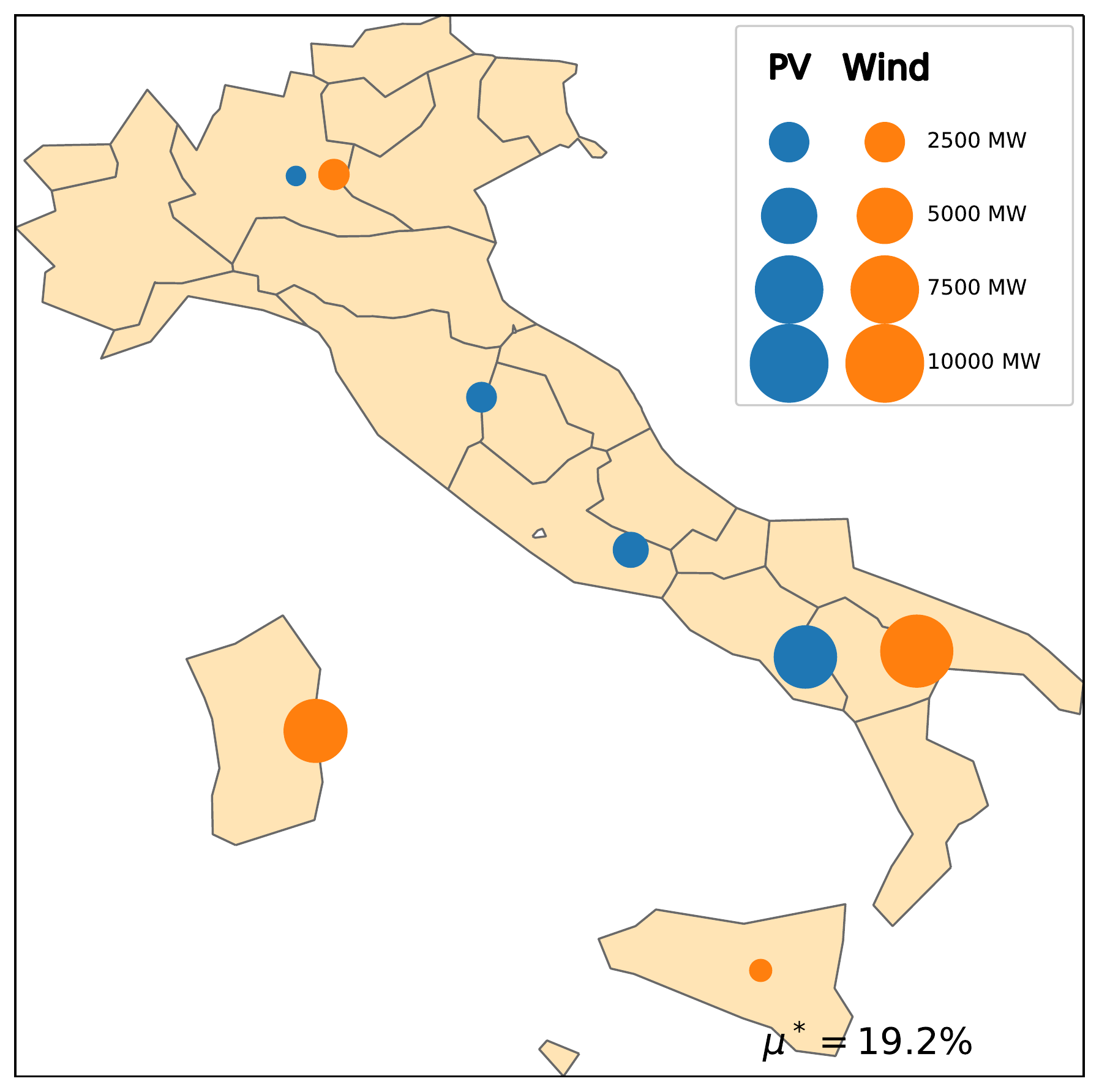}
    \caption{}
  \end{subfigure}
  \begin{subfigure}{0.49\linewidth}
    \includegraphics[width=\textwidth]{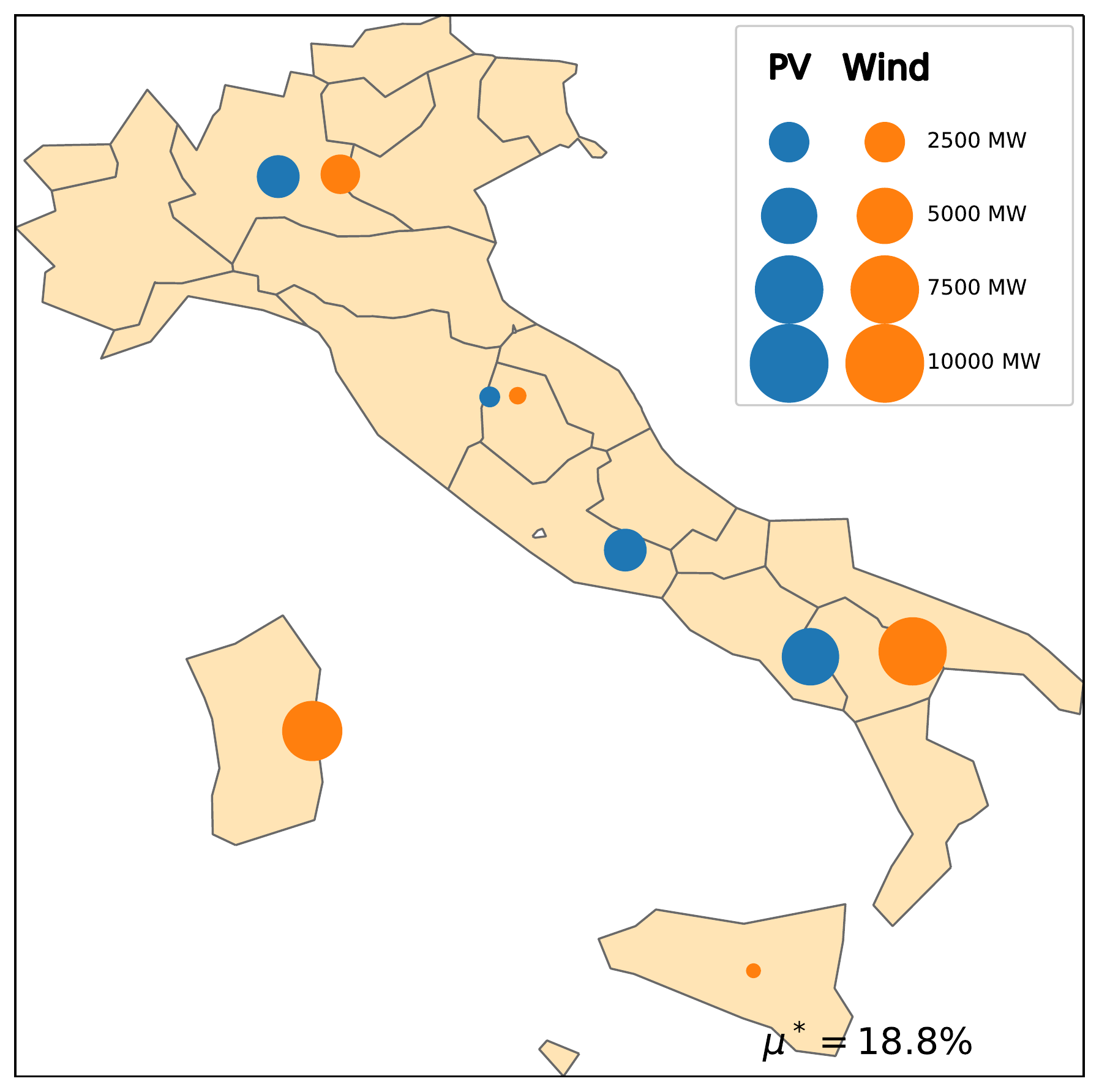}
    \caption{}
  \end{subfigure}
  \caption{
    To be compared with Figure~\ref{fig:mix}.
    Optimal frontiers approximations (top) and PV-wind capacity distributions (bottom) for the global strategy computed using hourly MERRA-2 data with 10m-winds (left) and with 50m-winds (right).
    To be compared with Figure~\ref{fig:mix} and~\ref{fig:mapGlobalMaxRatio}.
  }\label{fig:mixReanalysis}
\end{figure}

Figure~\ref{fig:mixReanalysis} shows the approximated optimal frontiers (top) and the corresponding capacities for the mix maximizing the mean-risk ratio of the global strategy (bottom) obtained by applying the hourly demand and capacity factor models to the MERRA-2 data using 10m-winds (left) and 50m-winds (right).
Overall, the qualitative picture of the frontiers remains unchanged, but important quantitative differences exist.
First, the mean-risk ratio is smaller for the MERRA-2 data with 10m-winds than for the two other cases by about 15\%, a difference which is in fact larger than that of 5\% found in Section~\ref{sec:application} between the global and the technology strategies.
Large differences between the capacity distributions also exist between all three cases.

\subsubsection{Interannual to decadal variability}
\begin{figure}
  \centering
  \begin{subfigure}{0.49\linewidth}
    \includegraphics[width=\textwidth]{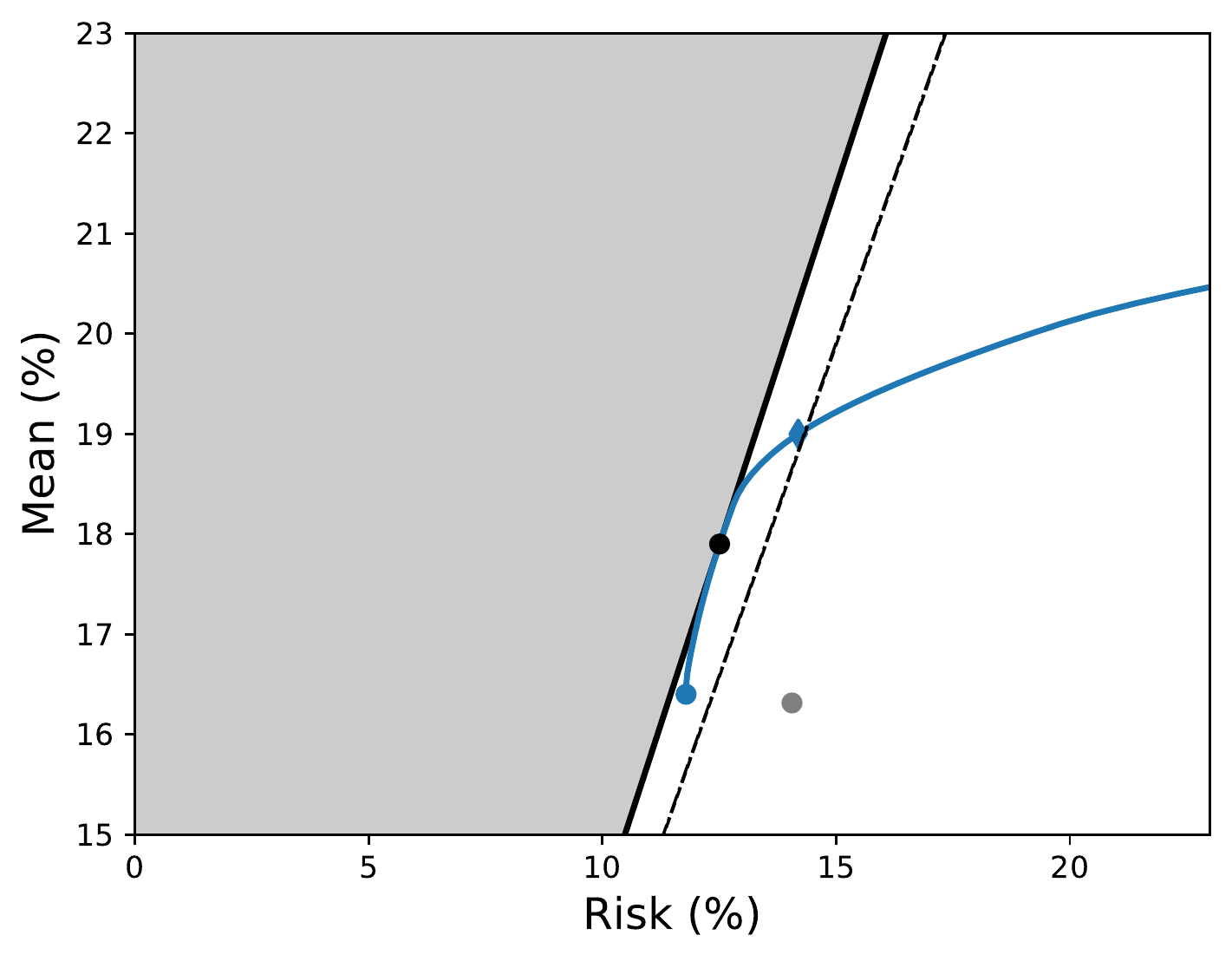}
  \end{subfigure}
  \begin{subfigure}{0.49\linewidth}
    \includegraphics[width=\textwidth]{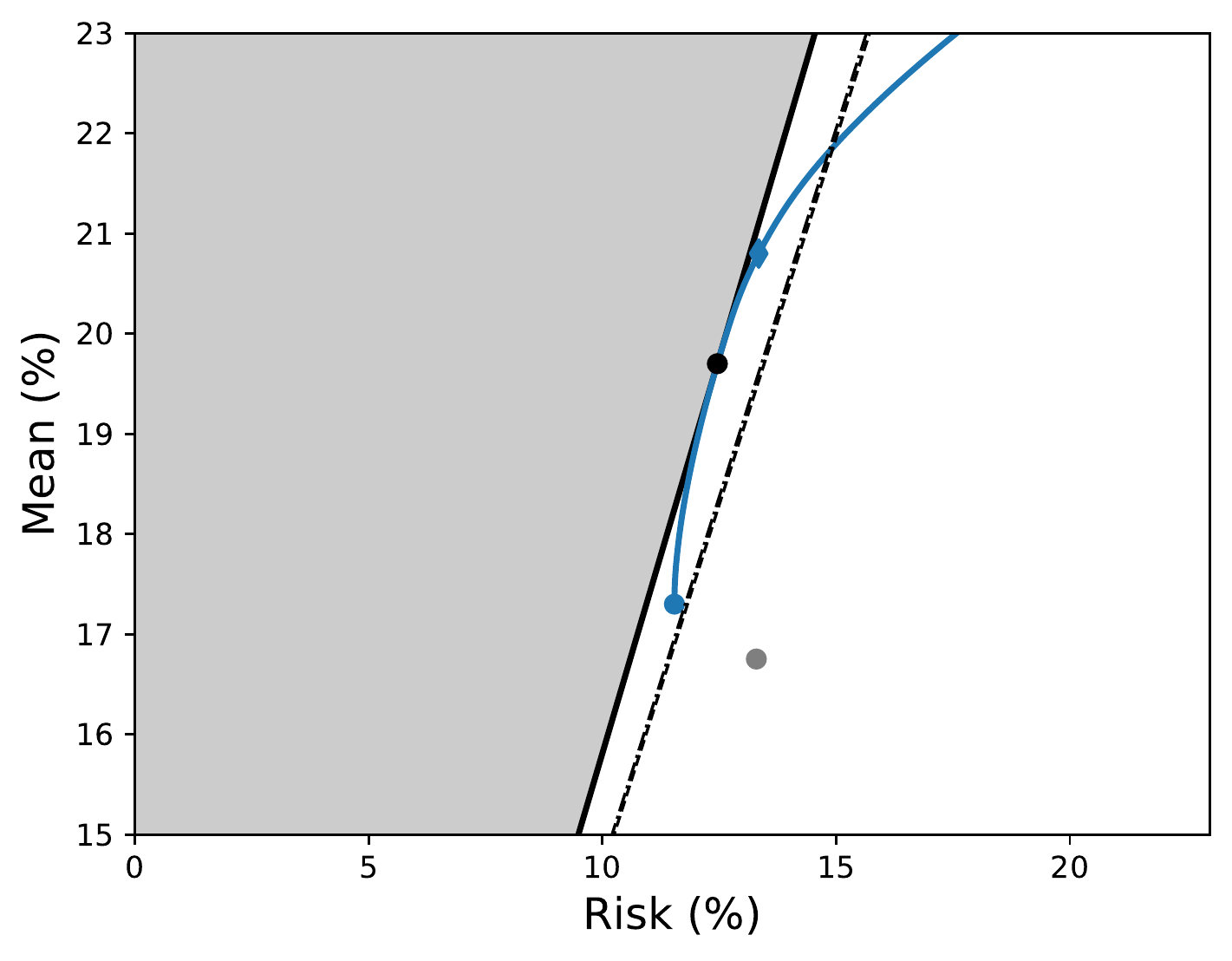}
  \end{subfigure}\\
  \begin{subfigure}{0.49\linewidth}
    \includegraphics[width=\textwidth]{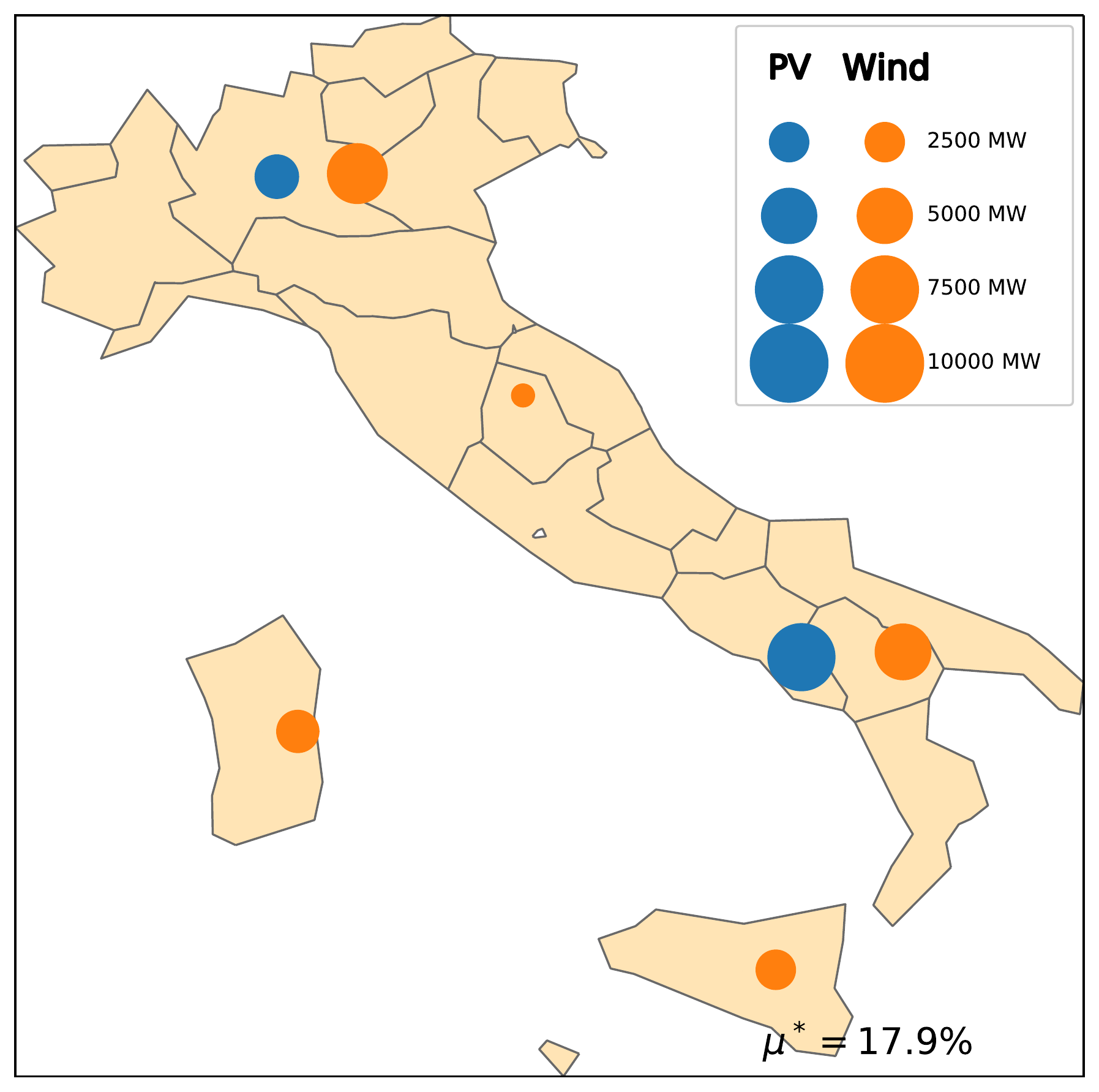}
    \caption*{1989}
  \end{subfigure}
  \begin{subfigure}{0.49\linewidth}
    \includegraphics[width=\textwidth]{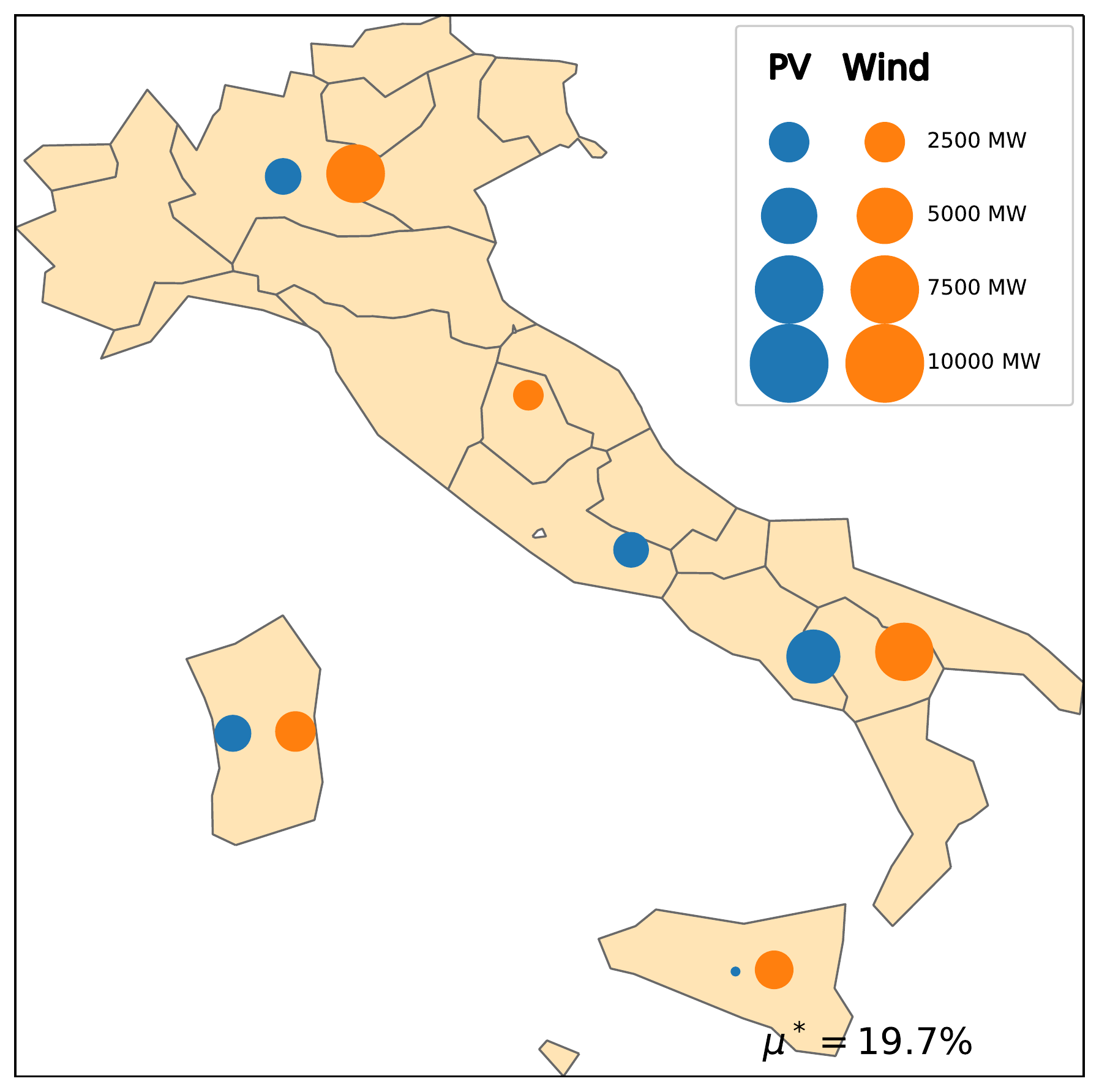}
    \caption*{1996}
  \end{subfigure}
  \caption{
    Optimal frontiers approximations (top) and PV-wind capacity distributions (bottom) for the global optimization problem obtained from the CORDEX hourly data over the years 1989 (left) and 1996 (right).
    To be compared with Figure~\ref{fig:mix} and~\ref{fig:mapGlobalMaxRatio}.
  }\label{fig:compareYearlyMix}.
\end{figure}

To assess the impact of interannual climate variability (as found in the CORDEX data) on energy mixes, we repeat the mean-variance analysis successively using data blocks of one year, from 1989 to 2012, rather than the full 1989--2012 block.
In other words, each of the 23 optimal frontiers are optimized for the climatic conditions of a given year and low-frequency climate variability results in different optimal mixes.
The mean-risk ratio for the unconstrained global frontier can be used as observable of these changes.
It is found to average to $1.46$ with $95\%$ of its realizations belonging to the centered interval $[1.43, 1.49]$.
Thus, even though the average of the yearly mean-risk ratio is close
to the one of 1.43 obtained in Section~\ref{sec:optimization} using the full record, interannual climate variability in the CORDEX data is responsible for year-to-year variations of the mean-risk ratio of up to $4\%$.
As an example, we represent in Figure~\ref{fig:compareYearlyMix}
the geographical and technological distribution of the mixes
for the year 1989, with a particularly low mean-risk ratio of $1.43$, and for the year 1996, with a particularly high mean-risk ratio of $1.49$.

\subsubsection{Intraday variability}
\begin{figure}[!ht]
  \centering
  \begin{subfigure}{0.49\linewidth}
    \includegraphics[width=\linewidth]{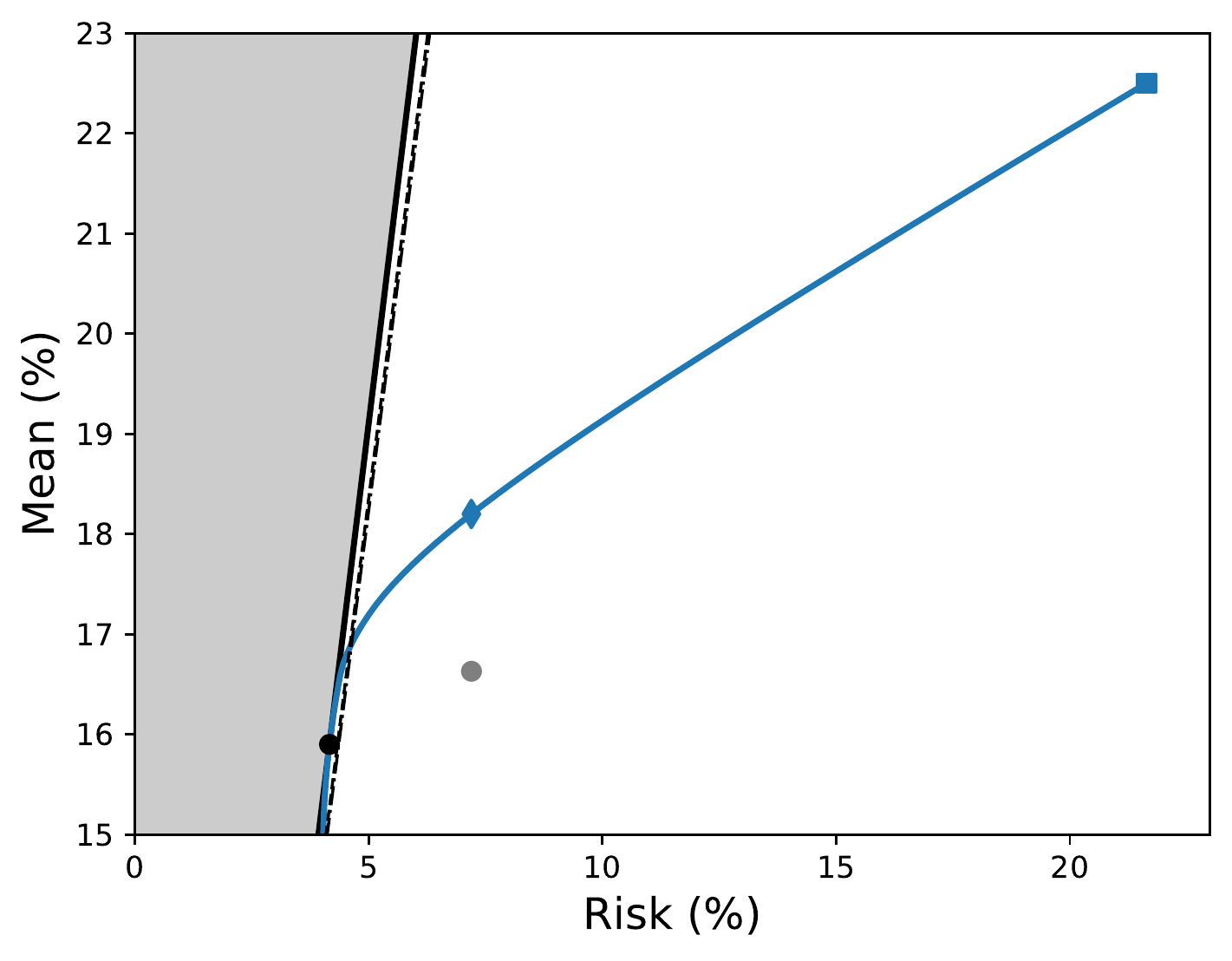}
    \caption{}
  \end{subfigure}
  \begin{subfigure}{0.49\linewidth}
    \includegraphics[width=\linewidth]{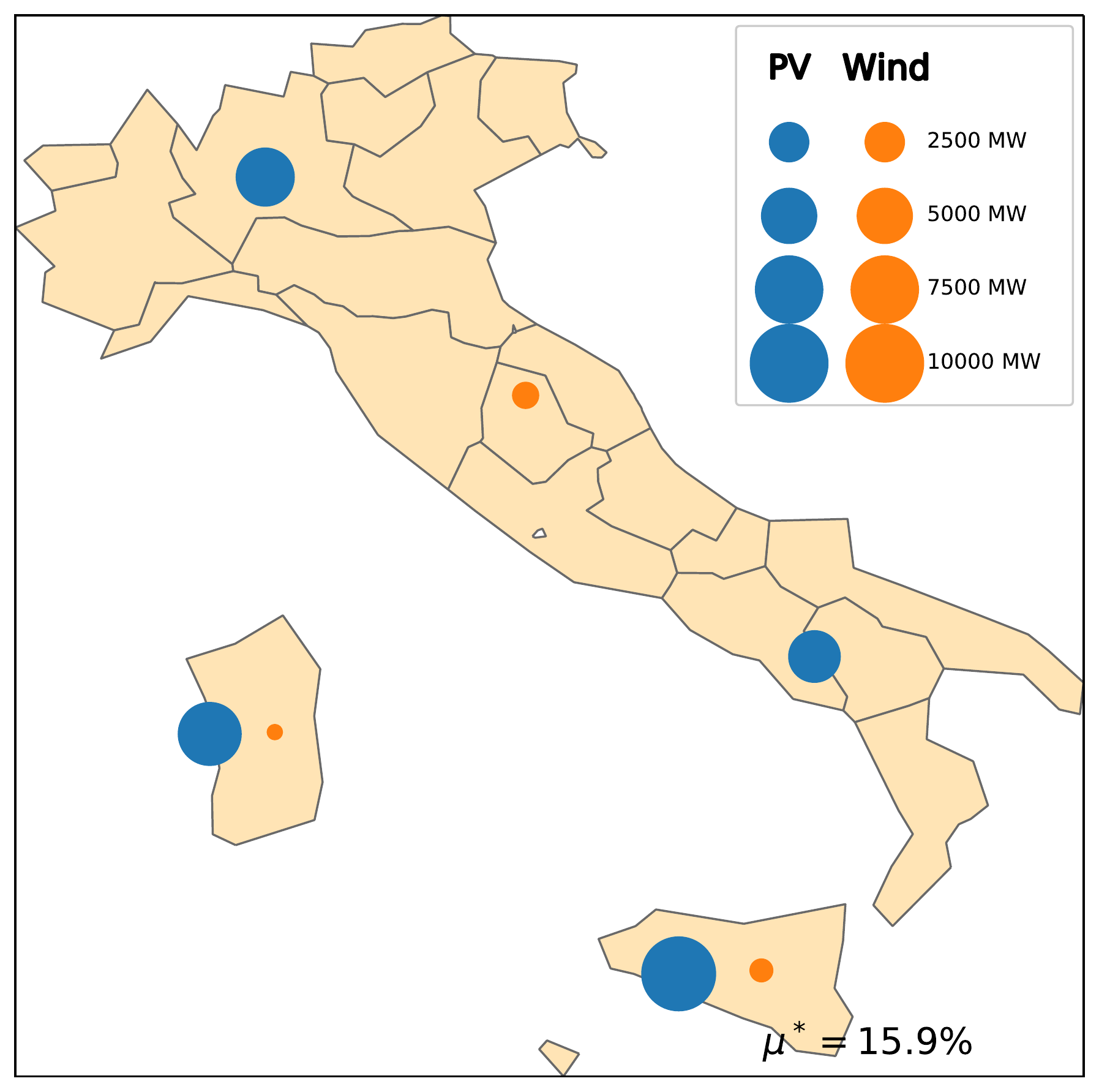}
    \caption{}
  \end{subfigure}
  \caption{
    Optimal frontiers approximations (left) and PV-wind distribution for the maximum-mean-risk-ratio mix (right) computed using daily CORDEX data \emph{without} intraday parameterizations.
    To be compared with Figure~\ref{fig:mix} and~\ref{fig:mapGlobalMaxRatio}.
  }\label{fig:mixDay}
\end{figure}

A large fraction of the PV, demand and wind variance is contained in the intraday range.
Yet, climate data is not always available at an hourly sampling.
This is for instance the case of the CORDEX data used here, for which intraday parameterizations are added to the energy models (Sect.~\ref{sec:model_description}).
To test the impact of ignoring such fluctuations on the optimal mixes, we represent in Figure~\ref{fig:mixDay} optimal frontiers (left) and the PV-wind distribution of the maximum-mean-risk-ratio mix (right) obtained directly from the daily CORDEX data without intraday parameterizations.
It is clear that the risk is underestimated by a factor two or more compared to the risk obtained using hourly data (c.f.~Fig.~\ref{fig:mix} and~\ref{fig:capacityMapOpt}).
This can be understood from the fact that, while the mean capacity factors remain unchanged, the variance in the modeled daily PV and wind capacity factors is dramatically underestimated (see Appendix~\ref{sec:evaluation}).
Because the PV production is more variable during the day than the wind production, ignoring intraday fluctuations results in distributing more PV capacities.

\section{Conclusion}\label{sec:conclusion}

This work is aimed at developing an integrated modelling framework dedicated to the assessment and elaboration of optimal energy mixes taking into account flexibility needs associated with high shares of VRE.
It relies on a methodology extending mean-variance analysis to resolve low-frequency climate variability and allow new technologies to be integrated.
This methodology is implemented as an extensible open-source Python software, \emph{e4clim}.
Its potential is demonstrated with an application to a recommissioning of the 2015 Italian PV-wind mix.
This application is, however, but one example of possible implementations in the \emph{e4clim} system.

The software's flow is divided in three steps: (i) energy time series are first estimated from climate data and fitted to observations; (ii) a VRE mix is then prescribed or optimized by, e.g., mean-variance analysis; (iii) the mix properties are finally analysed.
The first step relies on climate data to take production and demand-variability on a broad range of time scales into account and to allow for the integration of new technologies for which no or little observations are available.
The current version of \emph{e4clim} is also adapted to consider optimal strategies in a warming climate using 21st century regional projections provided by the CORDEX program (not shown here).
Using climate data from multiple independent sources is then recommended to estimate errors stemming from these sources.

Different optimal scenarios are derived in the second step, ranging from maximizing the total renewable energy penetration to minimizing the total risk, and so, flexibility requirements to meet the demand.
Different strategies can quickly be tested, allowing one, for instance, to evaluate benefits from leveraging correlations between zones.

As opposed to cost-minimizing problems for full mixes, the mean-variance analysis allows us to focus on VRE capacities alone, thus limiting the complexity of the algorithm and making it a fast and flexible tool for sensitivity analyses.
This leaves the estimation of associated economic costs and GHG emissions to the third (post-processing) step.
For that purpose, the hydro and the conventional production would have to be modeled with reserve, network and dispatch constraints.
In addition, the generalization of such integrated modeling tool at Euro-Mediterranean scale is a priority.

\section*{Acknowledgement}
AT is grateful to Mathilde Mougeot and J\'er\^ome Collet for insightful discussions on modeling the electricity demand.
This work was conducted in the framework of the TREND-X research program on energy transition at Ecole Polytechnique, which benefited from the support of the Ecole polytechnique fund raising – TREND-X Initiative. This research was also supported by the ANR project FOREWER (ANR-14-414 CE05- 0028) and the HyMeX project (HYdrological cycle in The Mediterranean EXperiment) through the working group Renewable Energy, funded by INSU-MISTRALS program.
The \emph{e4clim} open-source software is available at \url{https://gitlab.in2p3.fr/alexis.tantet/e4clim} and its documentation at \url{https://alexis.tantet.pages.in2p3.fr/e4clim/}.
It includes a reference guide and an example corresponding to the application presented here.

\appendix

Here, we give more information on data sources, models and validation results for the application of the \emph{e4clim} software to the Italian PV-wind mix recommissioning (see~Sect.~\ref{sec:application}).

\section{Data and model description}\label{sec:data_model_description}

An \emph{e4clim} project relies on models to predict energy time series (demand, capacity factors, etc.) from climate data.
These models depend on energy data to be fitted.
The energy data, climate data and demand, PV and wind models are described here.

\subsection{Energy data: GME and GSE databases}\label{sec:GMEGSE}

Time series of the hourly Italian regional electricity
demand and of the yearly regional renewable capacity factors
are used to design the demand and generation models.
See the ``demand'' and ``generation data'' blocks at the top of Figure~\ref{fig:flow_chart}.
These variables are extracted from two publicly available databases
provided respectively by the market operator GME\footnote{
  Gestore del Mercato Elettrico: https://www.gse.it/dati-e-scenari/statistiche}
and the energy operator GSE\footnote{Gestore dei Servizi Energetici:
  https://www.gse.it/dati-e-scenari/statistiche}.
For this reason, we first briefly comment on the structure
of the Italian electricity market and next describe the databases we use.

The Italian power market consists of 7 foreign virtual zones,
6 regional sub-markets, or bidding zones, and 5 poles of limited production.
The 20 administrative regions composing the Italian territory
are aggregated in the 6 bidding zones (Fig.~\ref{ElecRegions}):
Northern Italy (NORD), Central-Northern Italy (CNOR),
Central-Southern Italy (CSUD), Southern Italy (SUD),
Sardinia (SARD) and Sicily (SICI).
Each zone has its own generation mix determined by historical and
geographic reasons and characterized by a given level of efficiency.
For instance, the Northern regions have larger hydroelectric production
due to the proximity to the Alps. Inter-zonal transmission capacities
are not equally distributed either.

The Italian power exchange, which is managed by the GME\footnote{
  The GME manage as well the OTC Registration Platform for
  forward electricity contracts that have been concluded off the bidding system.}
is composed of a spot market,
a forward market and a platform for the physical delivery of contracts
concluded on the financial derivatives segment of the Italian Stock Exchange.
The  spot market is composed of three sub-markets:
the day-ahead, the intraday and the ancillary services markets.
We focus on the day-ahead submarket.
The liquidity of the day-ahead market,
calculated as the ratio of volumes traded on the day-ahead market
to the total volumes (including bilateral contracts) of the Italian power system,
has increased between 2010 to 2015, passing from 62.6\% with 198 operators in 2010
to 67.8\% with 259 operators in 2015.
The peak liquidity has been reached in 2013 with a 71.6\% liquidity
and 214 operators (GME, 2017). 


The GME database encompasses hourly bids and offers in the wholesale
electricity market from 2004 to 2017; the offers are identified by supplier's
technology. The hourly electricity demand is appraised from this source.
GSE annual reports~\citep[e.g.][]{gse_rapporto_2016}
contain information about the yearly electrical production
and the associated installed capacity\footnote{In the GSE reports,
  the capacity for a particular year is the installed capacity at the
  end of this year.}
detailed by region and sources from the beginning of 2008 to the end of 2016.
At the beginning of this period,
the installation of renewable energy capacity has shown a very rapid increase.

The regional time-mean capacity factors for PV and wind
are calculated from this source.
Since 2013, the PV and wind capacity factors are relatively stable.
The demand and the capacity factors for PV and wind for the 2013--2017 period are presented in Table~\ref{tab:demandFactors}.
Figure~\ref{fig:capacityMapGSE} of the article summarizes the current installed capacity at the end of 2015.


\begin{table*}
\begin{tabular}{*{3}{c}}
  Zone & Electrical demand (GWh/day | \%)
  & Capacity Factor (PV | Wind)  \\
    \midrule
     NORD  & 312.2 | 56.5 & {\color{blue} 12.1} | {\color{green} 20.4}   \\
     CNOR  & 50.4 | 9.1 & {\color{blue} 13.3} | {\color{green} 19.2}  \\
     CSUD  & 82.0 | 14.9  &  {\color{blue} 14.1} | {\color{green} 18.8} \\
     SUD  & 44.2 | 8.0  &  {\color{blue} 15.6} | {\color{green} 20.9}   \\
     SARD & 26.7 | 4.8 & {\color{blue} 14.5} | {\color{green} 19.6} \\
     SICI  & 36.8 | 6.7 & {\color{blue} 15.9} | {\color{green} 18.7} \\
\end{tabular}
\caption{
  Regional electrical demand (from GME) and capacity factors for PV (blue) and wind energy (green, both from GSE) averaged over the 2013--2017 period.
}\label{tab:demandFactors}
\end{table*}

\subsection{Climate data}\label{sec:climate_data}

The mean-variance optimization problem (Appendix~\ref{sec:meanVariance}) relies on electricity demand and PV and wind capacity-factor time-series.
Observed time series are only a few years long, too short to resolve low-frequency climate variability.
The models described in Section~\ref{sec:model_description} are thus used to predict these energy time-series from climate data.
See the ``climate data'' block at the top of Figure~\ref{fig:flow_chart}.
In this study, one particular CORDEX regional simulation is mainly used, that we refer to as the CORDEX data.
This choice is motivated by the fact that, contrary to reanalysis products, CORDEX projections for the 21st century are also available, which could be used to apply the \emph{e4clim} software to assess the impact of climate change on energy mixes.
Another climate dataset, the MERRA-2 reanalysis, is also used (i) to parametrize intraday wind-fluctuations not resolved by the CORDEX data (Sect.~\ref{sec:wind_model}) and (ii) to test the robustness of the Italian application results to the choice of the climate dataset (Appendix~\ref{sec:evaluation}).

\subsubsection{CORDEX regional simulations}\label{sec:WRF}

A third variable employed in our study is the multi-year series of  production. The deployment of RES capacity being relatively recent (starting around 2008 in Italy), available time series of observed RES production are not sufficiently long to estimate statistics taking into account low-frequency climate variability.
To take into account climate variability, RES production is instead computed
using regional climate simulations covering the historical 1989 to 2012 period.

\begin{figure}
  \centering
  \begin{subfigure}{0.48\linewidth}
    \includegraphics[width=\linewidth]{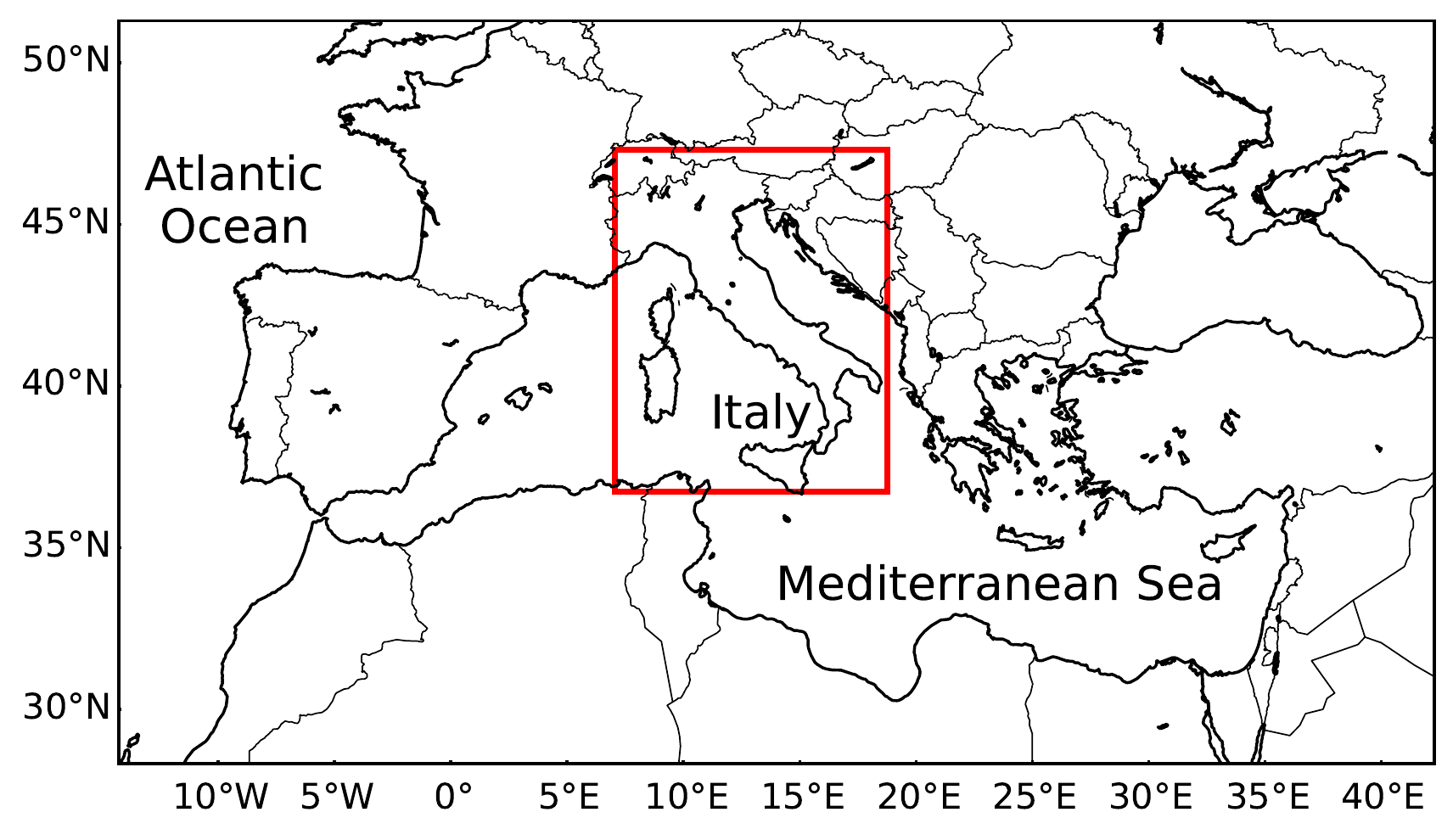}
    \caption{Domain of the HyMeX/MED-CORDEX simulation covering Europe and the Mediterranean region.
      The rectangle indicates the domain of investigation of this study.
    }\label{CordexMap}
  \end{subfigure}
\end{figure}

We use the version 3.1.1 of the Weather Research and Forecasting Model (WRF).
WRF is a limited area model, non-hydrostatic,
with terrain following eta-coordinate mesoscale modeling system designed
to serve both operational forecasting and atmospheric research needs~\citep{skamarock_description_2005}.
The WRF simulation has been performed in the framework of HyMeX~\citep{drobinski_hymex_2014} and MED-CORDEX~\citep{ruti_med-cordex_2016}
programs with a 20~km horizontal resolution over the domain
shown in Fig.~\ref{CordexMap} between 1989 and 2012
with initial and boundary conditions provided by the ERA-interim reanalysis
and updated every 6~hr~\citep{dee_era-interim_2011}.
The WRF simulation has been relaxed towards the ERA-I large scale fields (wind, temperature and humidity) with a nudging time of 6~hr~\citep{salameh_effect_2010, omrani_optimal_2013, omrani_using_2015}. A detailed description of the simulation configuration can be found in e.g.~\citet{flaounas_precipitation_2013}.

The simulation has been evaluated against ECA\&D gridded precipitation
and precipitation at the Mediterranean basin scale~\citep{flaounas_precipitation_2013},
and have been used to study heatwaves~\citep{stefanon_soil_2014, chiriaco_european_2014},
heavy precipitation~\citep{lebeaupin_brossier_ocean_2013, lebeaupin_brossier_regional_2015, berthou_prior_2014, berthou_sensitivity_2015, berthou_influence_2016}
and offshore wind energy potential assessment~\citep{omrani_spatial_2017}
in a configuration coupled or not with a regional ocean model
for the Mediterranean Sea~\citep{drobinski_model_2012}.
The simulation, that we refer to as the \emph{CORDEX} data, is available on the HyMex/MED-CORDEX database\footnote{
  \url{ftp://www.medcordex.eu/MED-18/IPSL/ECMWF-ERAINT/evaluation/r1i1p1/IPSL-WRF311/v1/day/}}.

\subsubsection{MERRA-2 reanalysis}\label{sec:MERRA-2}

The MERRA-2 dataset, used in Appendix~\ref{sec:evaluation}, is a state-of-the-art reanalysis providing, among other products, hourly time series of atmospheric variables from 1980 to present day.
As a reanalysis it combines observation data (from NASA's GMAO) with the NASA's GEOS modeling and analysis system.
See~\citet{gelaro_modern-era_2017} for a full description, and~\citet{fujiwara_introduction_2017}, for a comparison of various reanalyses.
The MERRA-2 product presents the advantage over the CORDEX data of being provided at an hourly sampling, of containing 50m (in addition to 10m) wind data and of overlapping with both the GSE and GME data.
However, we have chosen to use the CORDEX data for the application to Italy, in order to be able to extend this study to climate change scenarios using CORDEX projections in future work.

\subsection{Model description}\label{sec:model_description}

We describe here the wind production, PV production and demand models that are fitted to the energy data and applied to the daily CORDEX data to produce the energy time series taken as input to the optimization problem.
See the ``Demand'', ``PV'' and ``wind prediction'' blocks at the top of Figure~\ref{fig:flow_chart}.

\subsubsection{Wind model}\label{sec:wind_model}

To compute wind energy capacity factors from daily-mean CORDEX data (Sect.~\ref{sec:WRF}), horizontal wind-speeds are first interpolated at hub height (101~m) using an empirical power-law with exponent 1/7~\citep{justus_height_1976}.
A transfer function based on the power curve of a particular wind turbine, the relatively representative Siemens SWT-2.3~MW-101m, is applied to the wind speed to compute the electrical production at each climate-data gridpoint~\citep{omrani_spatial_2017}\footnote{
  Note that, due to the bias correction (see~Sect.~\ref{sec:aggregationBias}),
  only the variability of the wind production may be sensitive
  to this choice of power curve.}
Before applying the transfer function,
the wind speed at hub height is multiplied by a factor
${(\rho / \rho_0)}^{(1/3)}$
accounting for deviations of the daily-mean air density $\rho$ from the standard density $\rho_0$ for which the power curve has been obtained.
The air density $\rho$ is computed from the air temperature, pressure,
and specific humidity at the surface using the ideal gas law for moist air\footnote{This correction is applied
  to the wind speed rather than directly to the wind production in order
  to shift the power curve horizontally rather than scale it vertically
  and hence preserve the cut-in and cut-out behavior of the turbine.}.

In addition, it is essential for the mean-variance analysis (Appendix~\ref{sec:meanVariance}) to take intraday fluctuations of the wind production into account.
To parameterize intraday wind fluctuations at all grid points, we assume that these fluctuations follow a multivariate Weibull distribution with a mean vector given by the daily-mean wind-speed at hub height.
The scale and shape parameters per grid point as well as the correlation matrix of this distribution must thus be estimated.
For each day and for a given vector of shape parameters and correlation matrix, the scale parameter is estimated so as for the mean of the distribution to coincide with the daily-mean wind-speed.
This procedure allows for the daily variance to adapt to changes in the daily-mean wind-speed.
The vector of shape parameters and the correlation matrix are assumed to be constant and are estimated from the MERRA-2 10m-wind data (Sect.~\ref{sec:MERRA-2}).
The validity of our parameterization thus relies on the following assumptions: (i) at each grid point, intraday fluctuations are identically Weibull-distributed and independent; (ii) the shape parameters and the correlation matrix are independent of time; (iii) these parameters are the same for the MERRA-2 and the CORDEX datasets, respectively.

\begin{figure}
  \centering
  \begin{subfigure}{0.48\linewidth}
    \includegraphics[width=\linewidth]{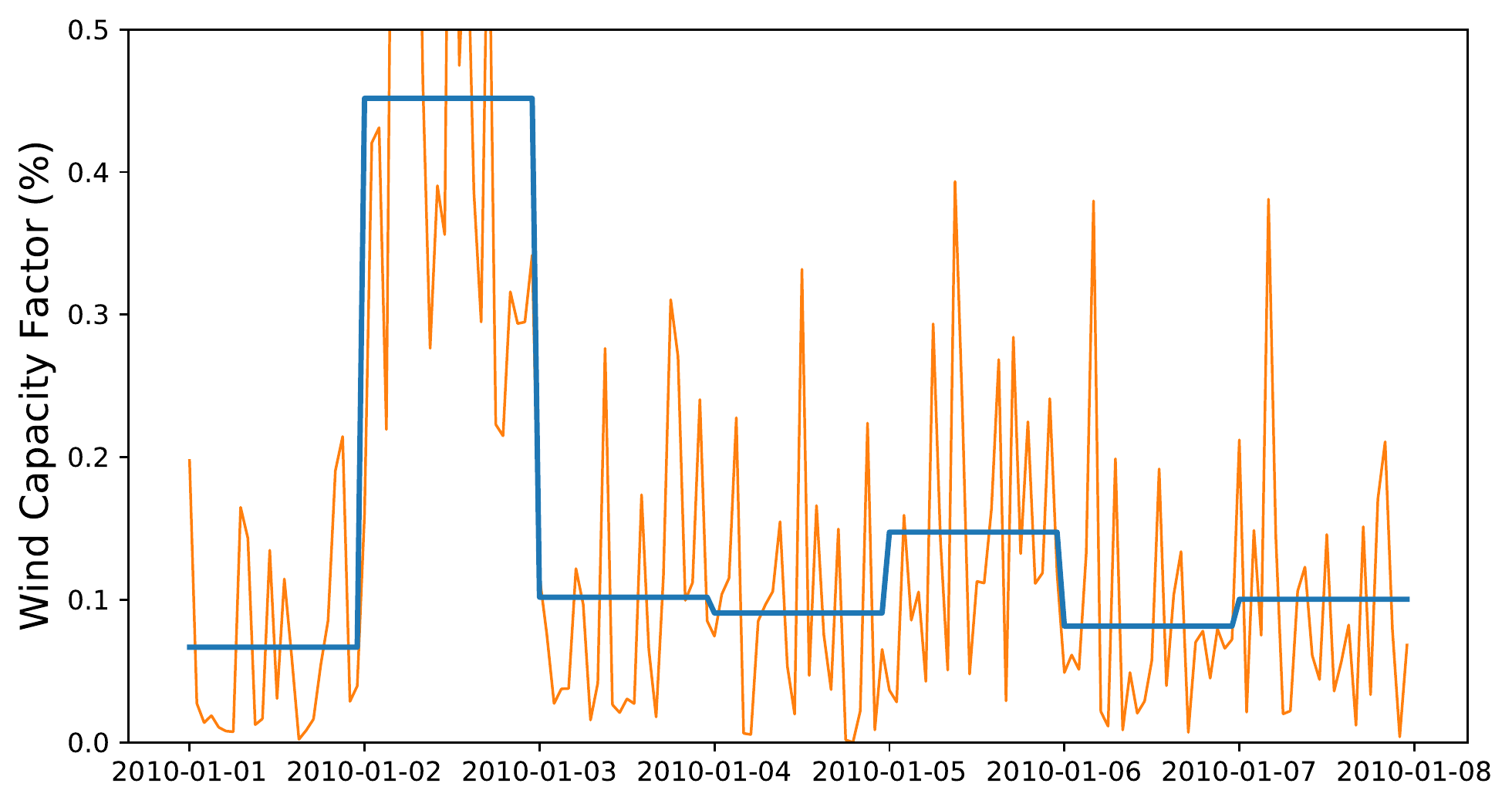}
    \caption{}\label{fig:intradayWindWinter}
  \end{subfigure}
  \begin{subfigure}{0.48\linewidth}
    \includegraphics[width=\linewidth]{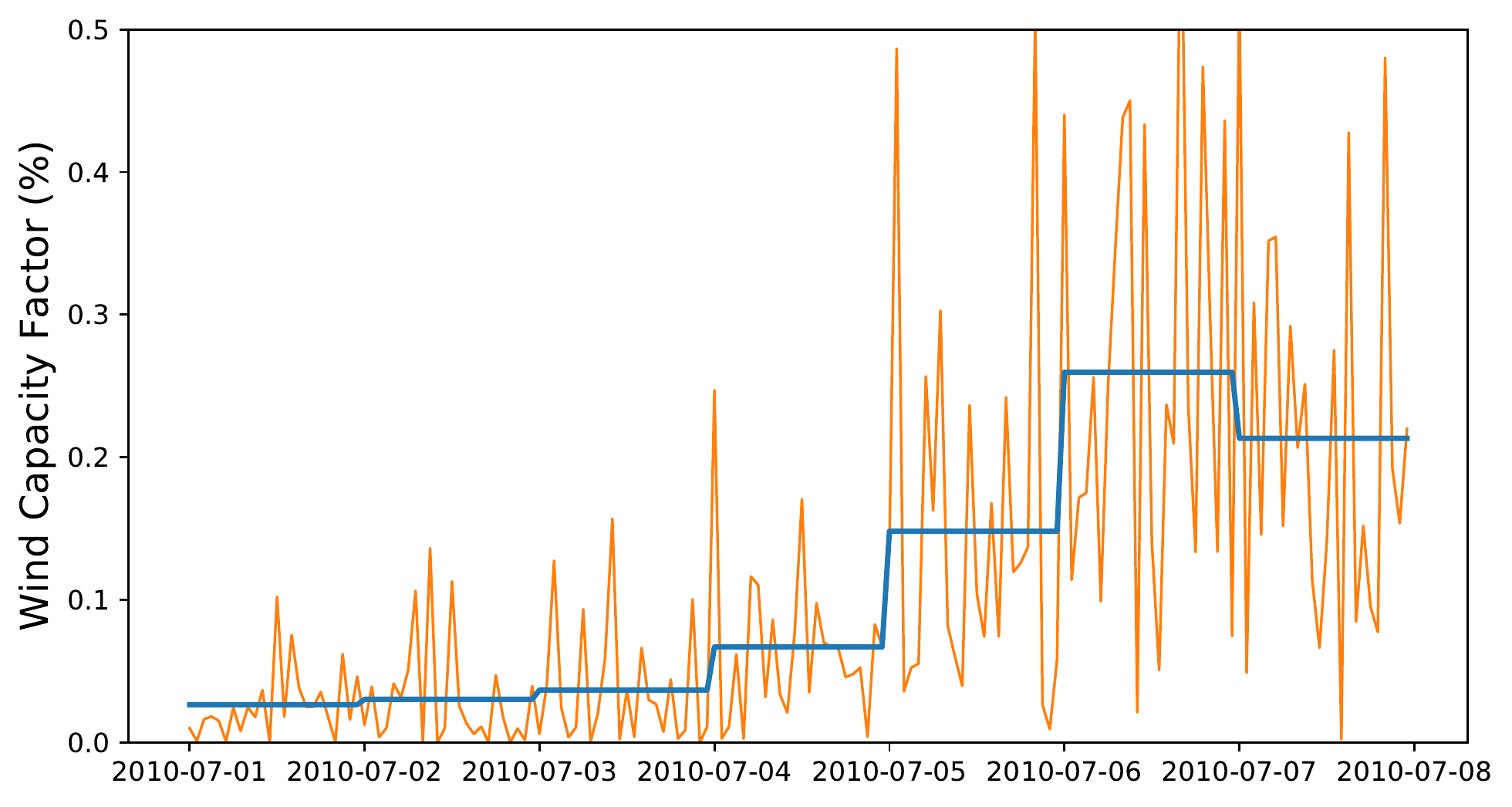}
    \caption{}\label{fig:intradayWindSummer}
  \end{subfigure}\\
  \begin{subfigure}{0.48\linewidth}
    \includegraphics[width=\linewidth]{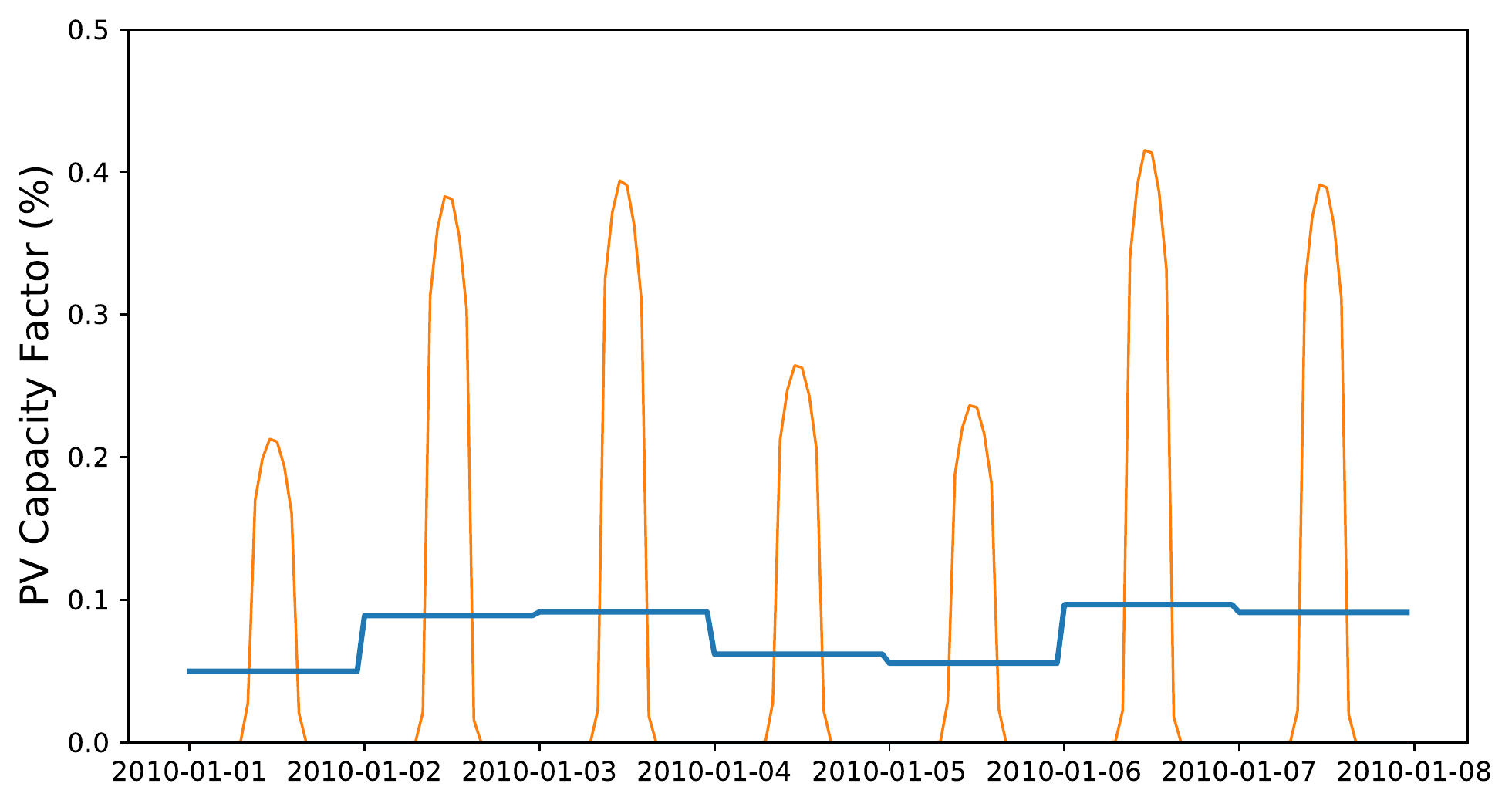}
    \caption{}\label{fig:intradaySolarWinter}
  \end{subfigure}
  \begin{subfigure}{0.48\linewidth}
    \includegraphics[width=\linewidth]{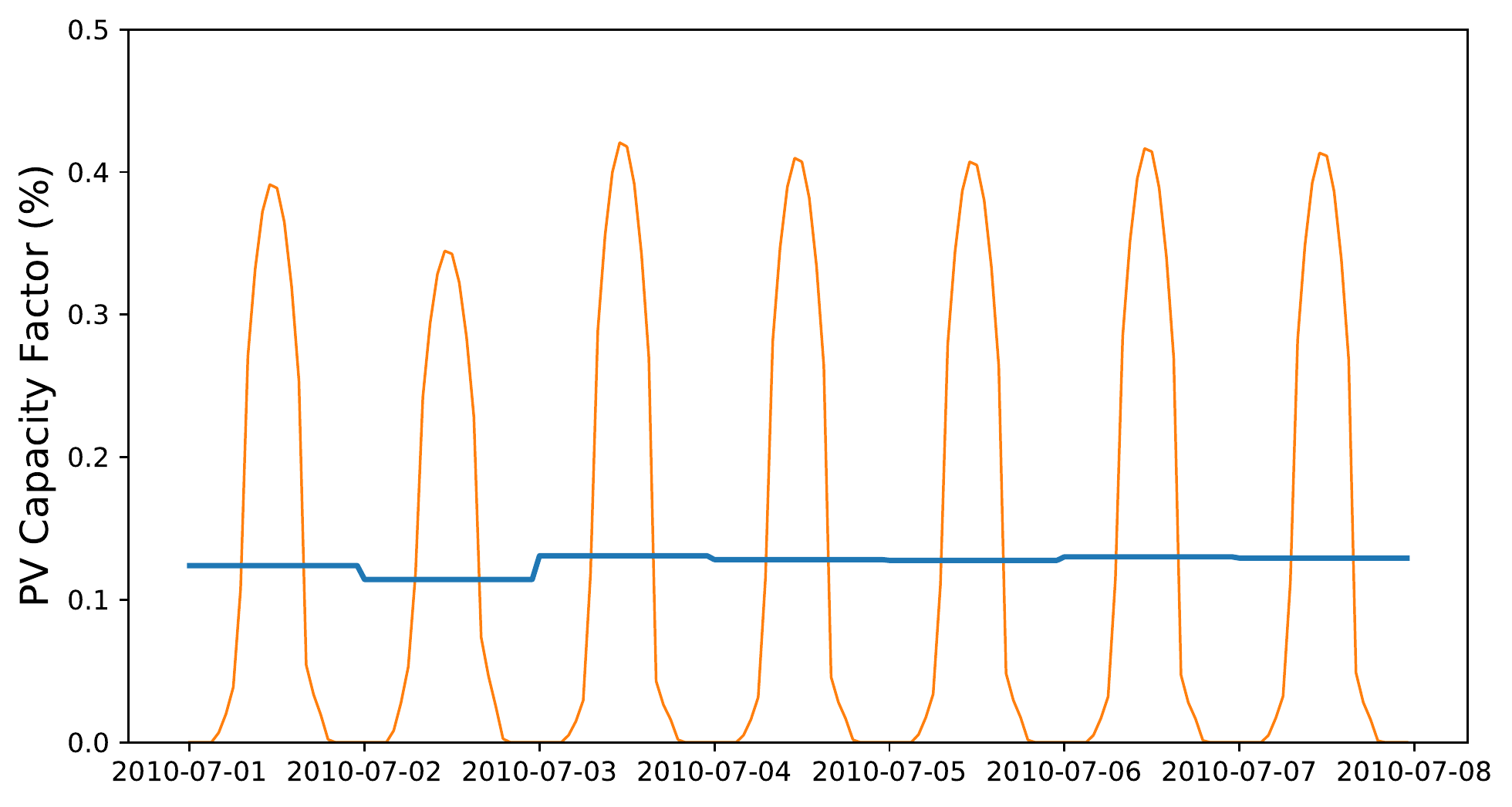}
    \caption{}\label{fig:intradaySolarSummer}
  \end{subfigure}\\
  \begin{subfigure}{0.48\linewidth}
    \includegraphics[width=\linewidth]{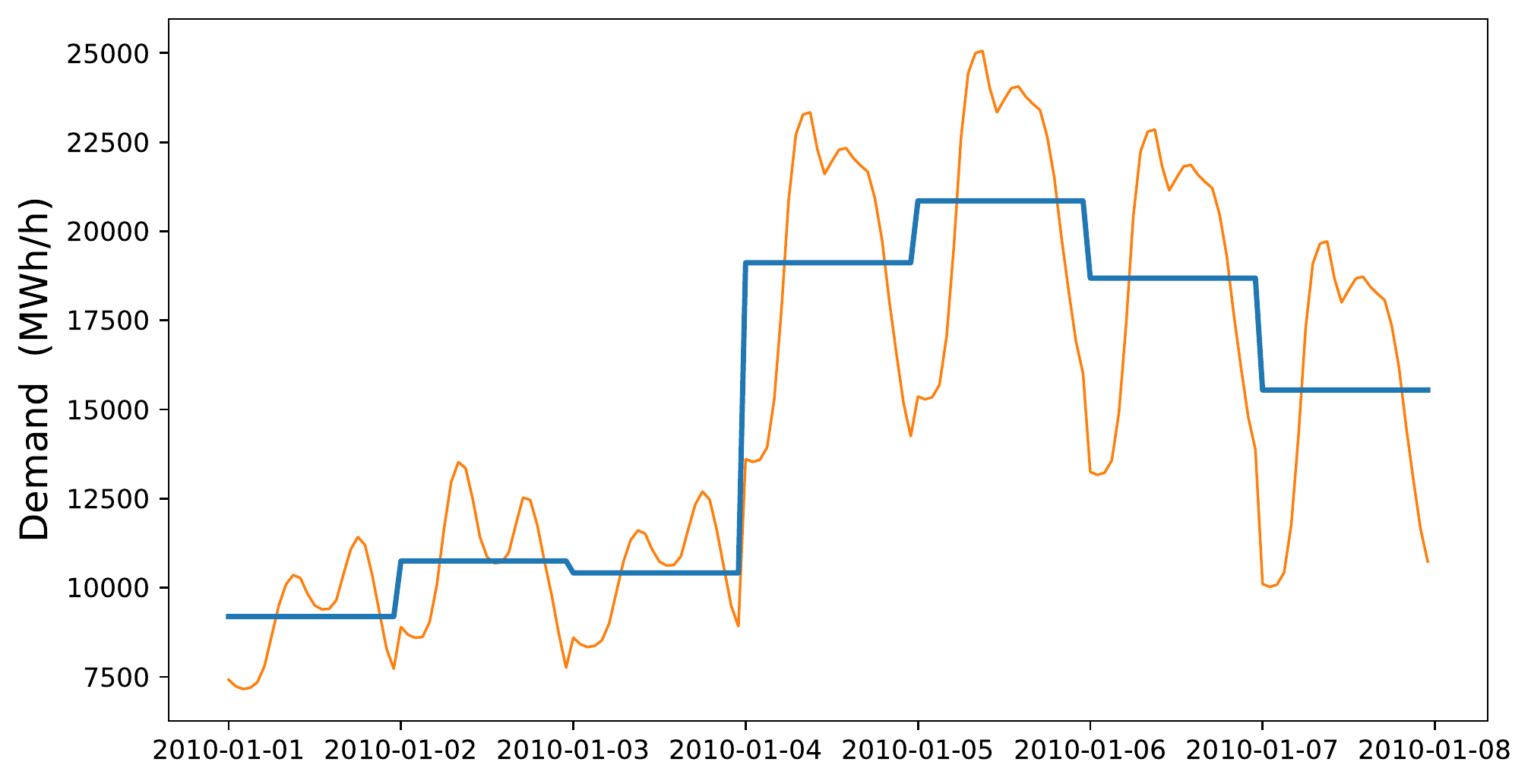}
    \caption{}\label{fig:intradayDemandWinter}
  \end{subfigure}
  \begin{subfigure}{0.48\linewidth}
    \includegraphics[width=\linewidth]{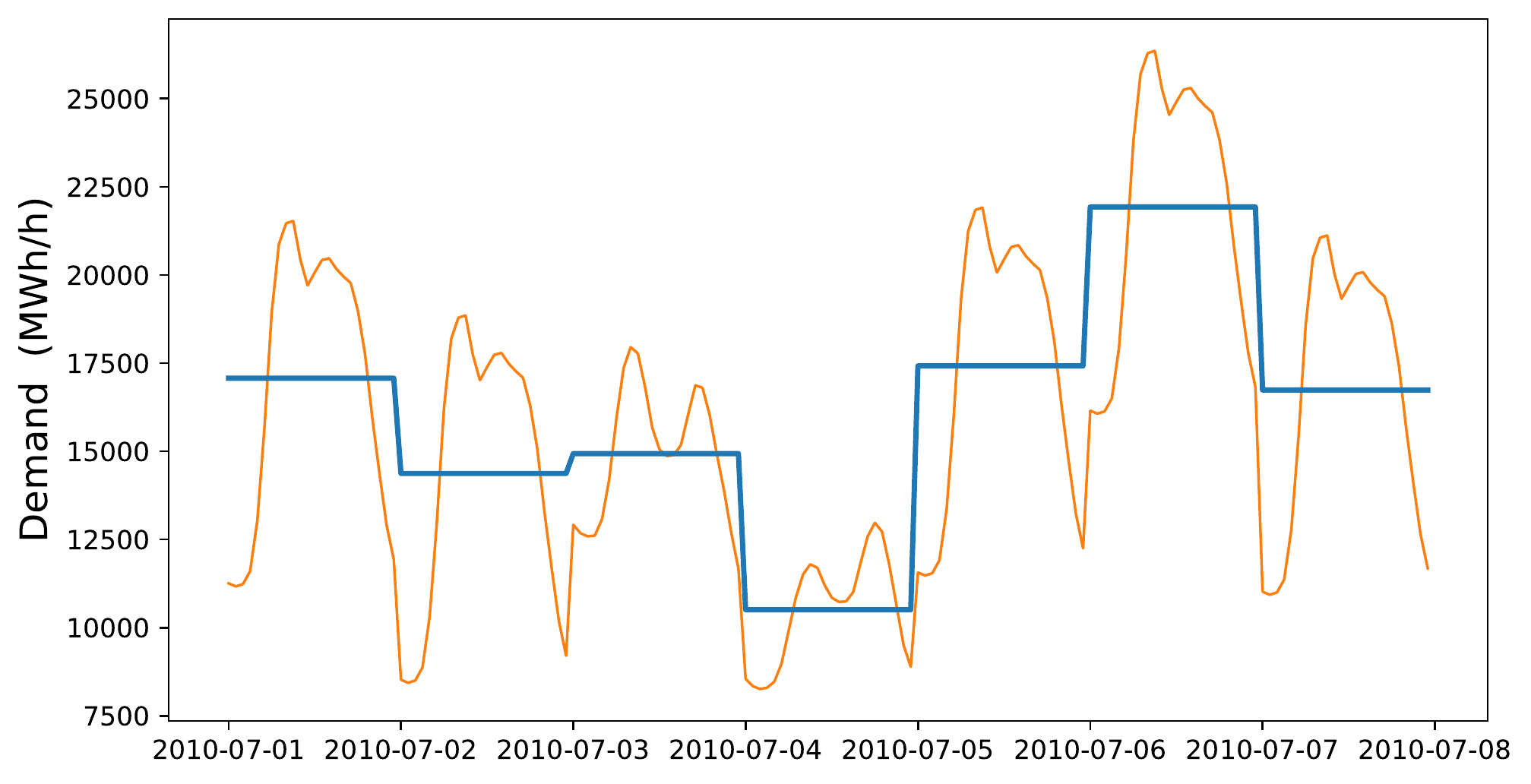}
    \caption{}\label{fig:intradayDemandSummer}
  \end{subfigure}
  \caption{
    Hourly (orange) and daily-mean (blue) wind capacity-factor (top), PV capacity-factor (middle) and electricity demand (bottom) for the north zone, the first week of January (left) and of July (right) 2010.
    The data is predicted from the daily-mean CORDEX data with the models and intra-day parameterizations described in Section~\ref{sec:model_description}.
  }\label{fig:intradayGeneration}
\end{figure}

Hourly realizations to be fed to the transfer function are then obtained by randomly drawing samples from the multivariate Weibull distribution.
In order to estimate the parameters and to draw samples, we use a change of variable from a Weibull to a normal distribution, as described in~\citet{villanueva_multivariate_2013}.
The effect of this parameterization of intraday wind fluctuations on the wind capacity factor of the north region is shown in Figure~\ref{fig:intradayWindWinter} and~\ref{fig:intradayWindSummer}, for a sample week in winter and another in summer 2010, respectively.


\subsubsection{PV model}\label{sec:pv_model}

We simulate the PV production for arrays at each gridpoint composed
of multi-crystalline silicon PV cells.
The crystalline silicon PV cell occupies about 90\% of the PV market,
among which multi-crystalline PV cells have the highest share at 53\%
and mono-crystalline PV cells have a 33\% share~\citep{hosenuzzaman_global_2015}.
Each module has a nominal power of \SI{250}{W}
for an area of \SI{1.675}{m^2},
resulting in a reference efficiency of about 15\%\footnote{The nominal power itself
  is not important for this methodology, as only capacity factors are used.}.
The real efficiency of the cell is, however, dependent on its temperature,
which is itself dependent on the air temperature and the wind from the CORDEX data and on the global tilted irradiance (see below).
This dependence is modeled using the thermal model described in \citet[][Chap.~23]{duffie_solar_2013}\footnote{
  The thermal model is configured for common parameter values for cristalline cells, i.e., for a temperature coefficient of \SI{0.004}{K^{-1}}, a reference temperature of \SI{25}{\celsius} and a cell temperature at nominal operating cell temperature of \SI{46}{\celsius}~\citep{skoplaki_temperature_2009}.
  The efficiency of the overall electrical installation behind the modules is assumed to be of $86\%$.
  Note, however, that constant multiplicative factors such as the electrical efficiency do not play a role in this study, due to the bias correction of the capacity factors presented in Section~\ref{sec:aggregationBias}.}.

Solar radiation from CORDEX is partitioned into direct, diffuse and reflected components~\citep[][Chap.~2.16]{duffie_solar_2013} at every gridpoint.
This partitioning depends on the clearness index $\bar{K}_T$ and elevation angle of the sun at the gridpoint.
The quantity $\overline{K}_T(d)$, for some day $d$, is defined as the ratio of the horizontal radiation at ground level, $\overline{I}(d)$, to the corresponding radiation available at the top of the atmosphere, i.e.~the extraterrestrial radiation $\overline{I}_0(d)$.

Only daily-mean extraterrestrial and surface irradiances are available in the CORDEX data.
Yet, the effect of the diurnal cycle on the tilted irradiance accounts for most of the variance of the PV production (see~Appendix~\ref{sec:evaluation}).
In order to take the diurnal cycle into account, the hourly extraterrestrial solar radiation, $I_0(d, h)$, is instead computed for every hour $h$ from the calendar data.
The hourly horizontal radiation at the surface, $I(d, h)$, for the hour $h$ of the day $d$ is then computed by multiplying the hourly extraterrestrial radiation $I_0(d, h)$ by the clearness index $\overline{K}_T(d)$, assumed constant throughout the day.
In other words,

\begin{align}
  \label{eq:hourlyIrradiance}
  I(d, h) &= \overline{K}_T (d)~I_0 (d, h)\\
  \mathrm{with} \quad \overline{K}_T (d) &= \frac{\overline{I}(d)}{\overline{I}_0(d)}.
\end{align}
Fluctuations associated with intraday variations of the clearness index, e.g.~associated with changes in the cloud cover, are, however, still ignored.

PV arrays are assumed to be tilted by an angle equal to the latitude of the array and to face due South.
To separate the diffuse component from the direct component of the global horizontal irradiance, the model from \citet{reindl_diffuse_1990} is used.
For solar elevations below \ang{10} and when the sun is behind the array, the direct horizontal irradiance is set to zero.
The diffuse component of the tilted irradiance is computed following the model of \citet{reindl_evaluation_1990}.
The reflected component of the tilted irradiance depends on the zenith angle and follows the usual formula given by \citet[][Chap.~2.16]{duffie_solar_2013} with an albedo of 0.2\footnote{The global tilted irradiance tends, however,
  to be dominated by its direct and diffuse components.}.

The effect of the diurnal cycle on the PV capacity factor of the north region
is shown in Figure~\ref{fig:intradaySolarWinter} and~\ref{fig:intradaySolarSummer},
for a sample week in winter and another in summer 2010, respectively.

\subsubsection{Aggregation and bias correction}\label{sec:aggregationBias}

The regional wind and PV capacity factors are obtained by dividing the computed production at each grid point of the climate data by its nominal value and then summing it over the zone on an hourly basis.
In so doing, a strong bias (up to 100\%) is found between the yearly-averages of the computed capacity factors and the region's capacity factors computed from the GSE data (Tab.~\ref{tab:demandFactors}).

To correct this bias, we use the fact that the second moment of the capacity factors roughly scales with their mean (not shown here) and re-scale the computed capacity factors so as their average over the climate-data period (1989--2012) to coincide with the GSE averages over the relatively slow-growth 2013--2017 period (see Sect.~\ref{sec:GMEGSE}).
\begin{figure}
  \centering
  \begin{subfigure}{0.48\linewidth}
    \includegraphics[width=\linewidth]{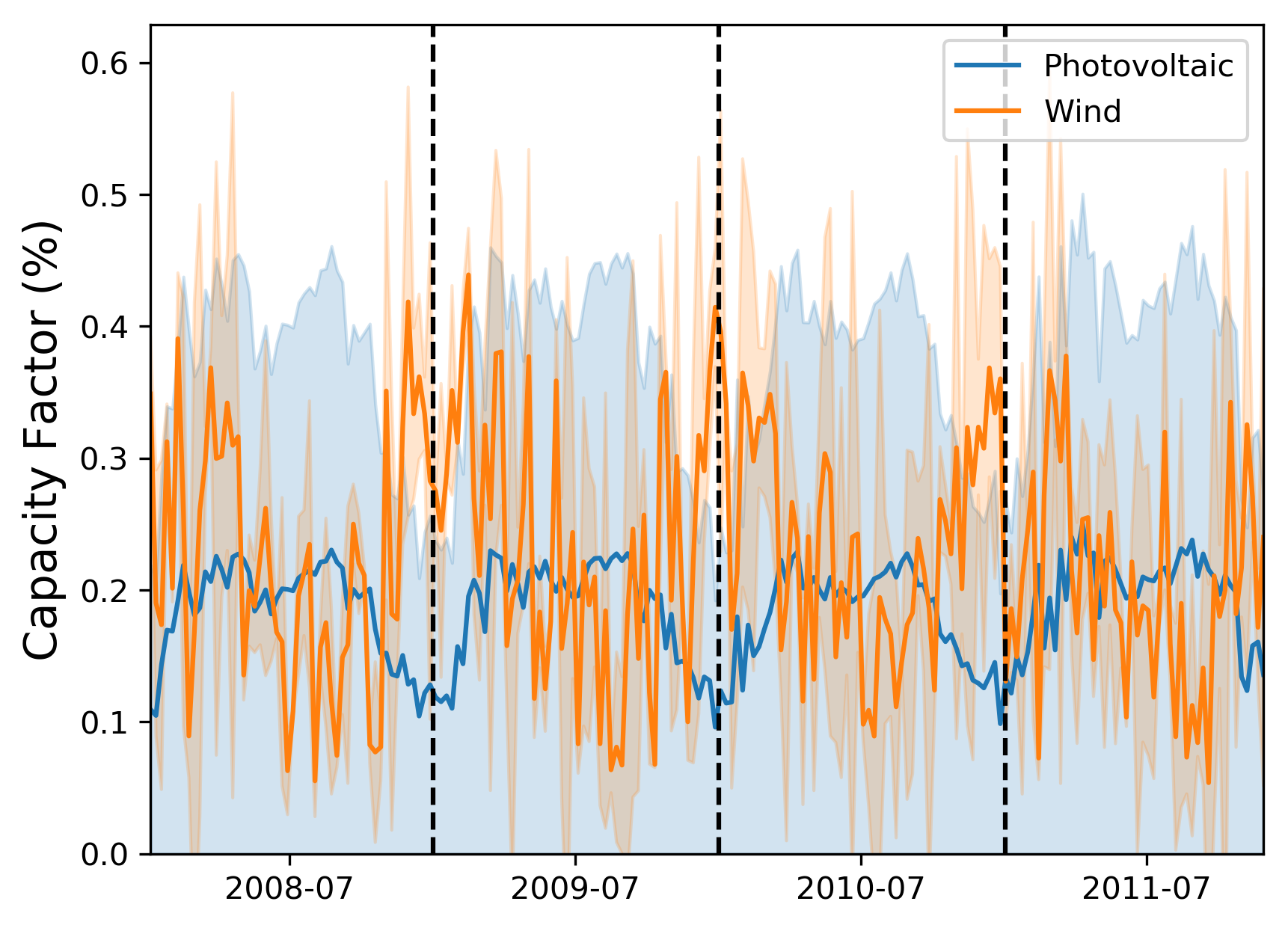}
    \caption{}\label{fig:simCF}
  \end{subfigure}
  \begin{subfigure}{0.48\linewidth}
    \includegraphics[width=\linewidth]{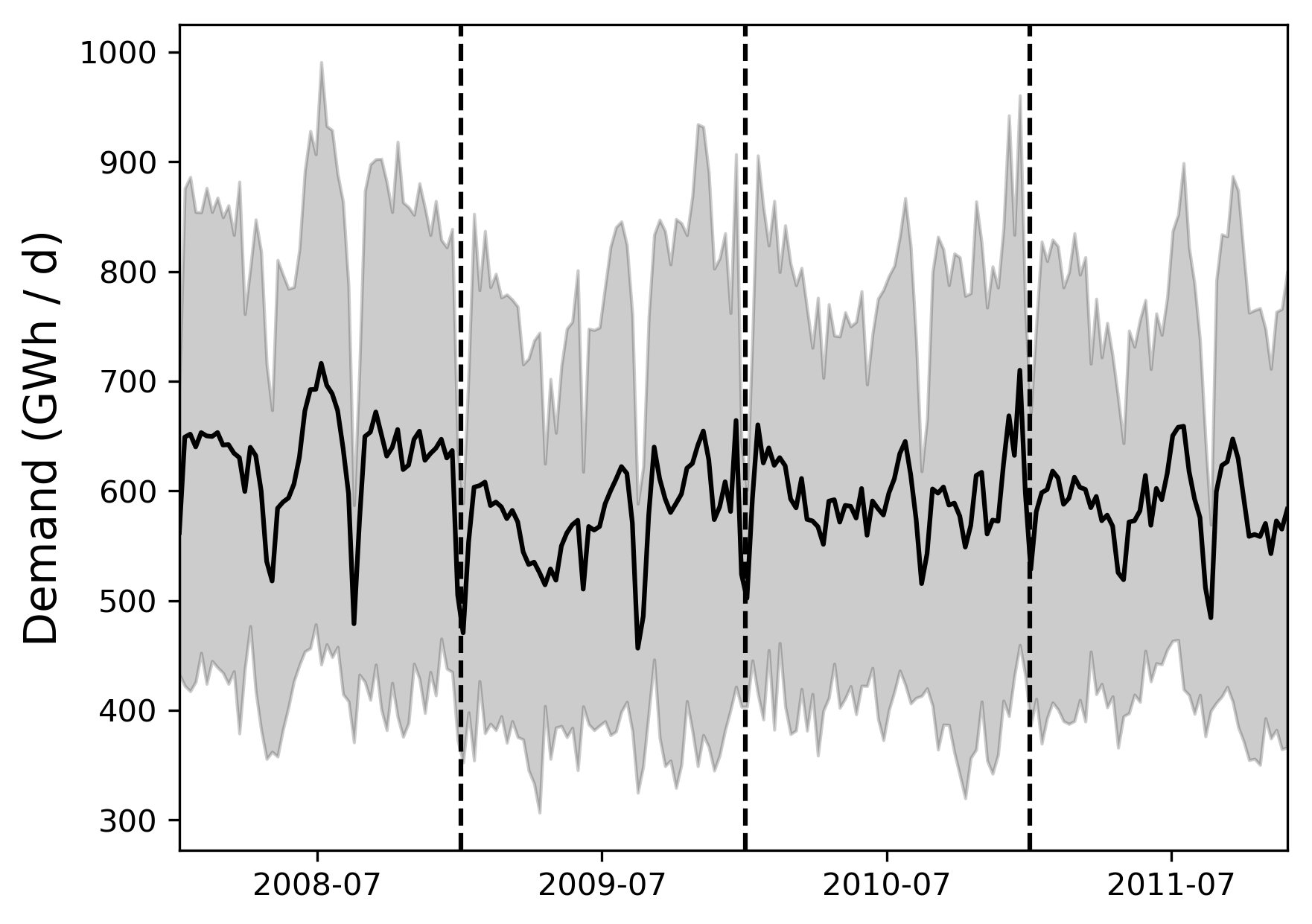}
    \caption{}\label{fig:simDem}
  \end{subfigure}
  \caption{Time evolution of weekly-averaged PV (orange)
    and wind (blue) capacity factors (a)
    and demand (b). The shadings represent
    the standard deviation of the time series.}\label{fig:simProdCons}
\end{figure}

Figure~\ref{fig:simCF} represents the weekly and regionally-averaged PV (orange)
and wind (blue) capacity factors for a few sample years.
Seasonal cycles of PV and wind energy production are phase shifted, with wind energy production (PV) peak yield in winter (summer).
Wind energy production is also characterized by a stronger sub-seasonal variability when compared to PV at large-scale.
Both the mean and the variance of the capacity factors for the wind production tend to be larger than those of the photovoltaic production.
Albeit complementary over a typical year the wind and PV production requires additional energy inputs from other sources to counterbalance some recurrent short term deficit between the demand and the wind and PV production.

Let us insist that the present bias correction only corrects
for differences in the first moment of the capacity factors with the observed values.
For the analysis of Section~\ref{sec:optimization},
higher moments, in particular the variance and covariance, are also important.
Although one expects the variance of the capacity factors to scale with their mean,
discrepancies may persist.
To estimate biases in the variance, our computations are tested against observations
in Appendix~\ref{sec:evaluation}.


Note finally that the conventional production, which includes here thermal
as well as hydropower plants, is not explicitly modeled in this study, thus preventing the translation of the computed electricity mixes to be translated into greenhouse gas or economic costs.
This is left for future work.

\subsubsection{Electricity-demand model}\label{sec:demandModel}



Since we are primarily interested in the impact of climate variability and change
on the demand, the objective of the model is to predict
the part of the daily regional demand depending on climate,
in particular on the surface air temperature\footnote{
  Other variables, such as the specific humidity, the wind, or the irradiance
  where not found to significantly affect the demand, in this case.},
while preserving the statistics associated with other factors.

We follow a statistical learning approach~\citep{hastie_elements_2009}
whereby the model is trained against the regionally-averaged temperature CORDEX data as input and the regionally-averaged demand data from GME
as output, from the beginning of 2005 to the end of 2011
(i.e.~the intersection of the climate with the demand record).


Let the electricity demand $D_{in}$ for the zone $i$ at the time step $n$
and for a particular type of day (see below) be given by

\begin{align}
  D_{in}(T_i) = f_i(T_{in}) + \epsilon_{in}.
  \label{eq:demandModel}
\end{align}
where $f_i, 1 \le i \le N$ is some real-valued function of the daily-mean temperature $T_{in}$ in the zone $i$ at the time step $n$ ($T_{in}$ is constant within a day), and the residual $\epsilon$ accounts for other factors impacting the demand, such as changes in the population, the economy, tourism, individuals decisions, etc.

It is known that the demand has a nonlinear dependence on the temperature.
Electric heating is switched on only for lower temperatures,
while air conditioning is switched on only for higher temperatures.
This can be seen in Figure~\ref{fig:simDem} for Italy.
The demand has two main peaks per year, one during winter and one during summer,
and lows in spring and fall and during holidays.
In winter, the consumption peak is due to heating,
especially in the northern part of Italy (see below).
In summer, the consumption peak is due to tourism and air conditioning~\citep{terna_sustainability_2016}.
Figure~\ref{fig:simProdCons} shows that, except for the summer period,
wind energy production is well correlated with the demand,
whereas PV production is negatively correlated with the latter.

Once an individual appliance is switched on, its electricity
consumption is to a first approximation linear in the temperature.
Assuming, that all consumers behave in the same way and that
a consumer switches the heater (air conditioning) for a
constant temperature threshold $T_H$ ($T_C$),
we define the functions $f_i$ as a piecewise-linear function
of the temperature.
In addition, the behaviour of the consumers differs
significantly for the week days, Saturdays, and Sundays and holidays
(respectively marked \emph{work}, \emph{sat} and \emph{off}, in the following)
and the demand is known to strongly depend on the
hour of the day (see Appendix~\ref{sec:evaluation}).
We thus choose to modulate the daily demand by a composite cycle which only depends on the hour of the day and the day type.
This cycle is computed from the GME data by averaging all 24h daily-cycles over the years for each day type.
The resulting model is given by,

\begin{align}
  f_i(T_{in}) &=
  \begin{cases}
    f^{\mathrm{work}}_i(T_{in})
    \quad &\mbox{if the day at } n \mbox{ is a working day} \\
    f^{\mathrm{sat}}_i(T_{in})
    \quad &\mbox{if the day at } n \mbox{ is a Saturday} \\
    f^{\mathrm{off}}_i(T_{in})
    \quad &\mbox{if the day at } n \mbox{ is a holiday},
  \end{cases} \label{eq:demandFunction} \\
  \mathrm{with} \quad
  f^{\mathrm{work}|\mathrm{sat}|\mathrm{off}}_i(T_{in}) &=
  a_H^{\mathrm{work}|\mathrm{sat}|\mathrm{off}}
  \Theta(T_H - T_{in}) (T_H - T_{in})
  \enskip g^{\mathrm{work}|\mathrm{sat}|\mathrm{off}}_n \nonumber \\
  &+ a_C^{\mathrm{work}|\mathrm{sat}|\mathrm{off}}
  \Theta(T_{in} - T_C) (T_{in} - T_C)
  \enskip g^{\mathrm{work}|\mathrm{sat}|\mathrm{off}}_n \nonumber \\
  &+ a_0^{\mathrm{work}|\mathrm{sat}|\mathrm{off}}
  \enskip g^{\mathrm{work}|\mathrm{sat}|\mathrm{off}}_n \nonumber 
\end{align}
where $\Theta$ is the Heaviside step function\footnote{
  The relationship between the demand and the ambient temperature
  in European countries is smoother than a piecewise-linear
  function~\citep{bessec_non-linear_2008}, in part due to the non-homogeneous 
  behaviour of the consumers. Here, however, we prefer to keep the
  model as simple as possible using the above linear basis.}
and the coefficients $g^{\mathrm{work}|\mathrm{sat}|\mathrm{off}}_n$
are given by the average --- over all days of the same day type
and all hours of the same hour of the day ---
of the observed demand on which the model is trained.
The model~\eqref{eq:demandFunction} has a total of $9$ parameters
$a_H^{\mathrm{work}|\mathrm{sat}|\mathrm{off}}$,
$a_C^{\mathrm{work}|\mathrm{sat}|\mathrm{off}}$
and $a_0^{\mathrm{work}|\mathrm{sat}|\mathrm{off}}$
to be adjusted, for each zone.

The resulting linear model is fitted assuming that the thresholds
$T_H$ and $T_C$ are constant over all zones and all day types.
These thresholds constitute two hyper-parameters
that we select via a grid-search with a cross-validation~\citep{hastie_elements_2009}
over seven blocks of one year.

The linear model is fitted using the Bayesian ridge regression method
~\citep{mackay_bayesian_1992} both to avoid over-fitting
and to take into account the variance arising from factors
that are not fully resolved by the deterministic part of the model\footnote{
  The implementation from \emph{scikit-learn}~\citep{buitinck_api_2013}
  of the Bayesian ridge regression is used, whereby the residual and the weights
  are given zero-mean isotropic Gaussian priors.
  The variances of the latter are given as priors gamma distributions.}.
A time series of the hourly regional demand over 1989--2012
is predicted from the full length of the temperature record by randomly
drawing samples from the posterior distribution of the model at each time step.

\begin{figure}[!ht]
  \centering
  \begin{subfigure}{0.32\textwidth}
    \includegraphics[width=\textwidth]{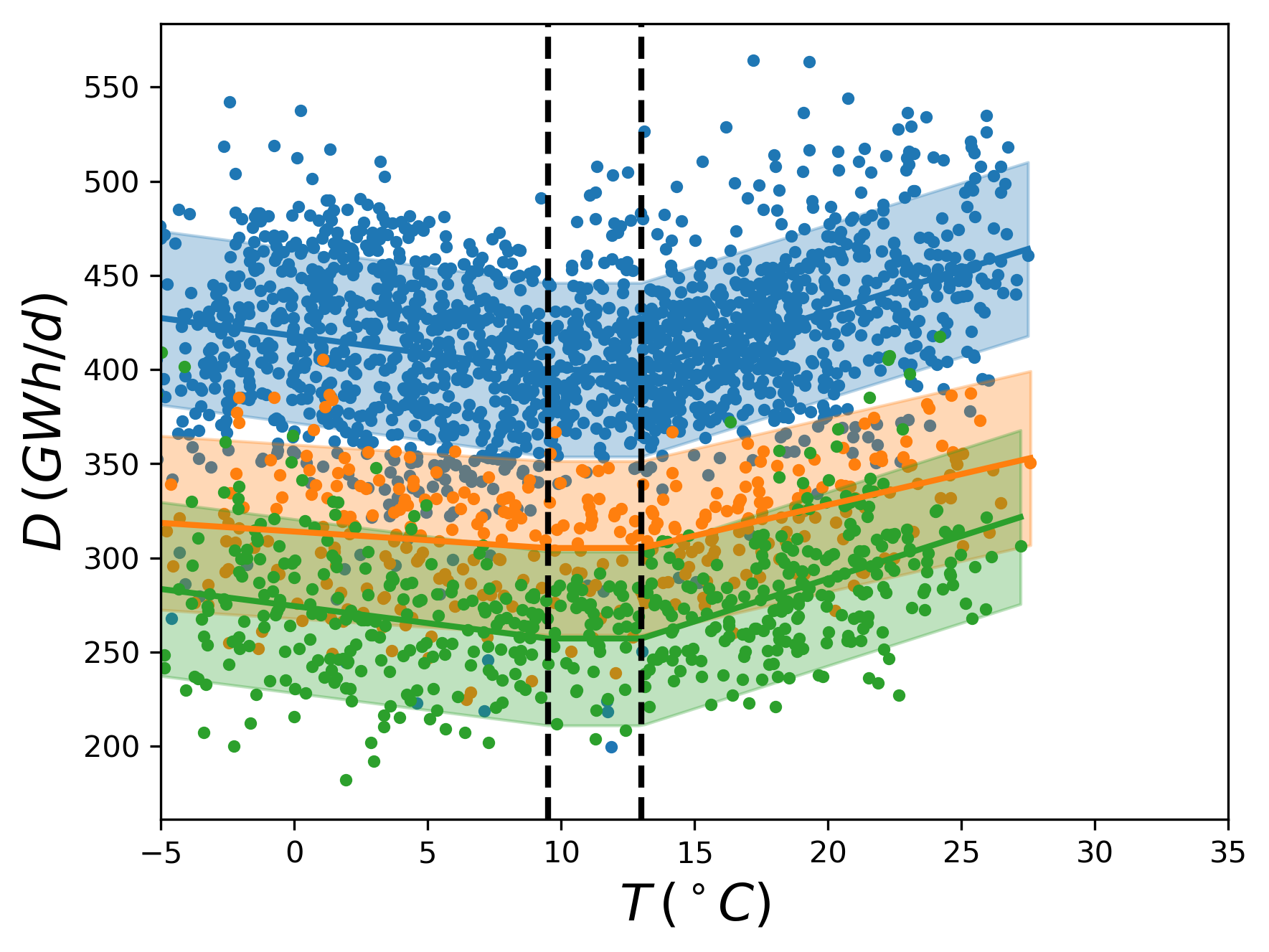}
    \subcaption{NORD}
  \end{subfigure}
  \begin{subfigure}{0.32\textwidth}
    \includegraphics[width=\textwidth]{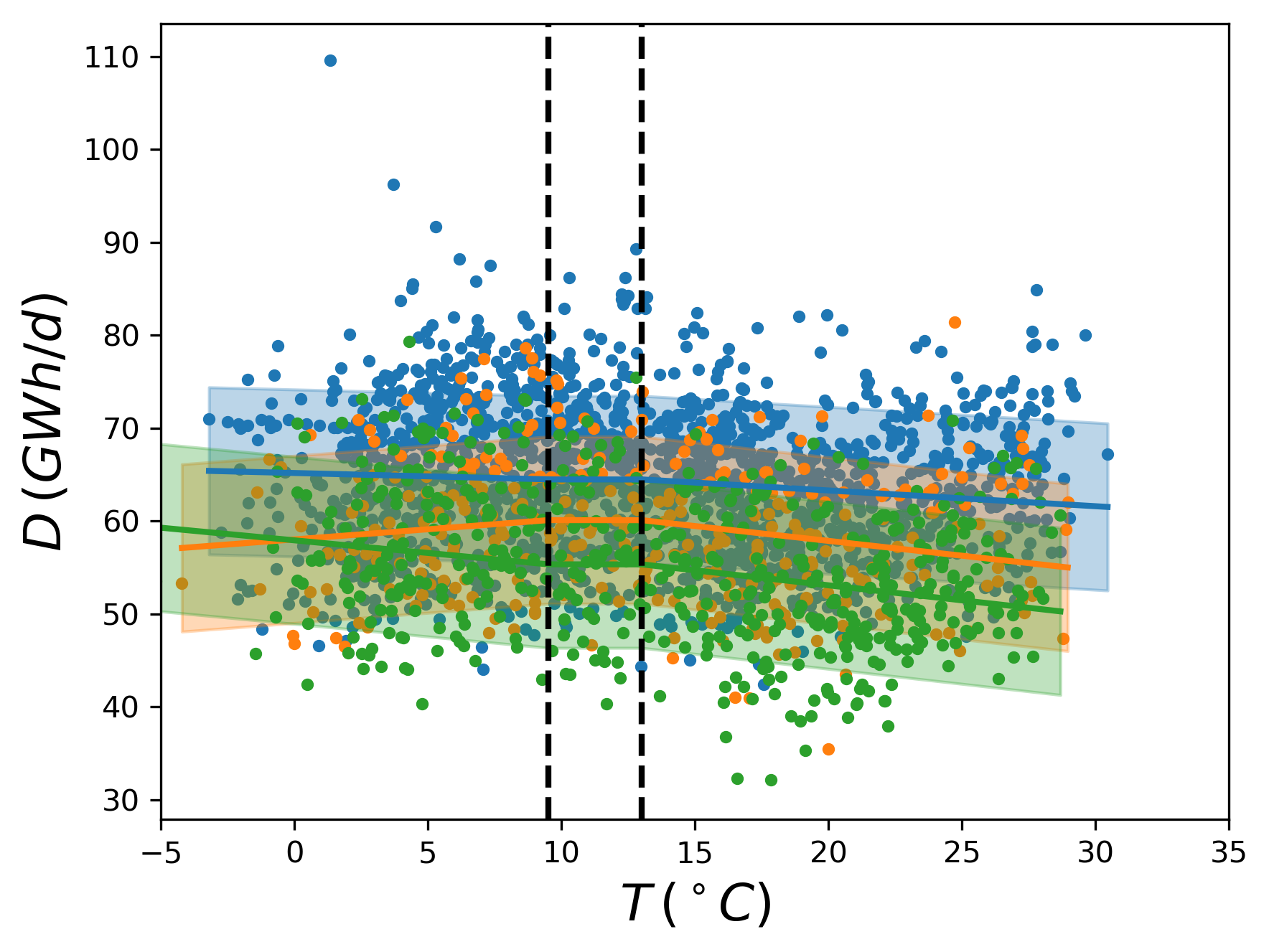}
    \subcaption{CNOR}
  \end{subfigure}
  \begin{subfigure}{0.32\textwidth}
    \includegraphics[width=\textwidth]{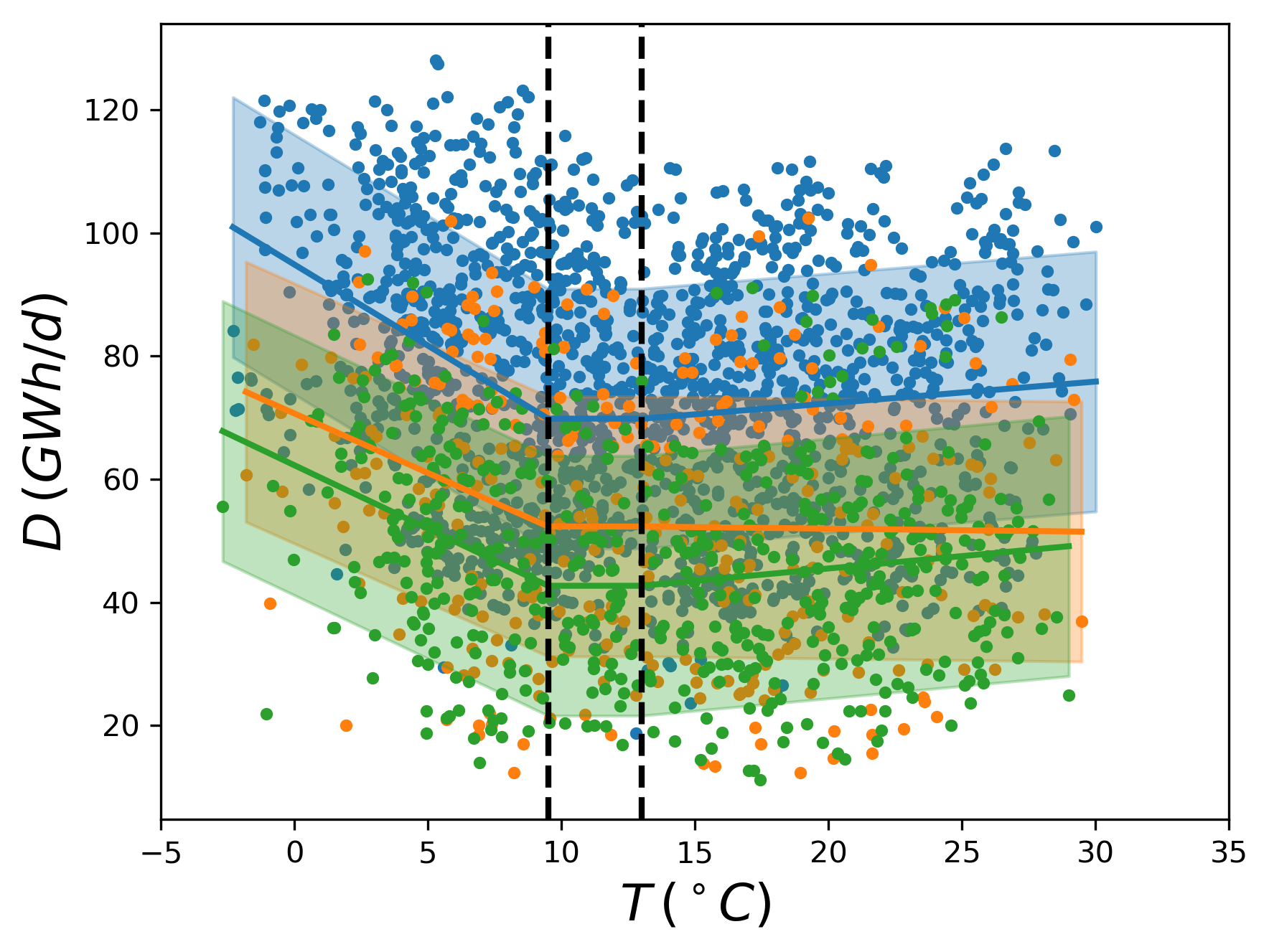}
    \subcaption{CSUD}
  \end{subfigure}\\
  \begin{subfigure}{0.32\textwidth}
    \includegraphics[width=\textwidth]{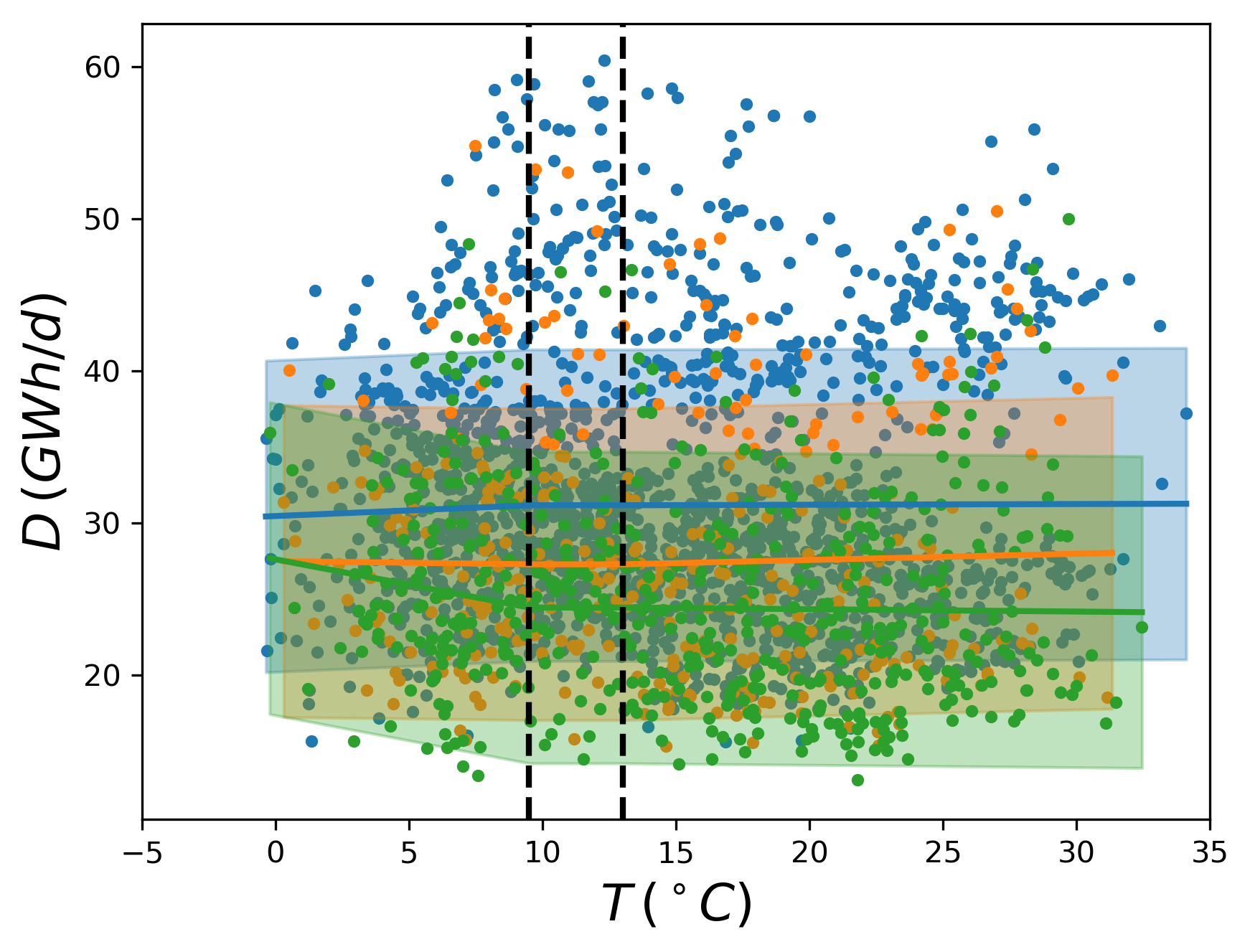}
    \subcaption{SUD}
  \end{subfigure}
  \begin{subfigure}{0.32\textwidth}
    \includegraphics[width=\textwidth]{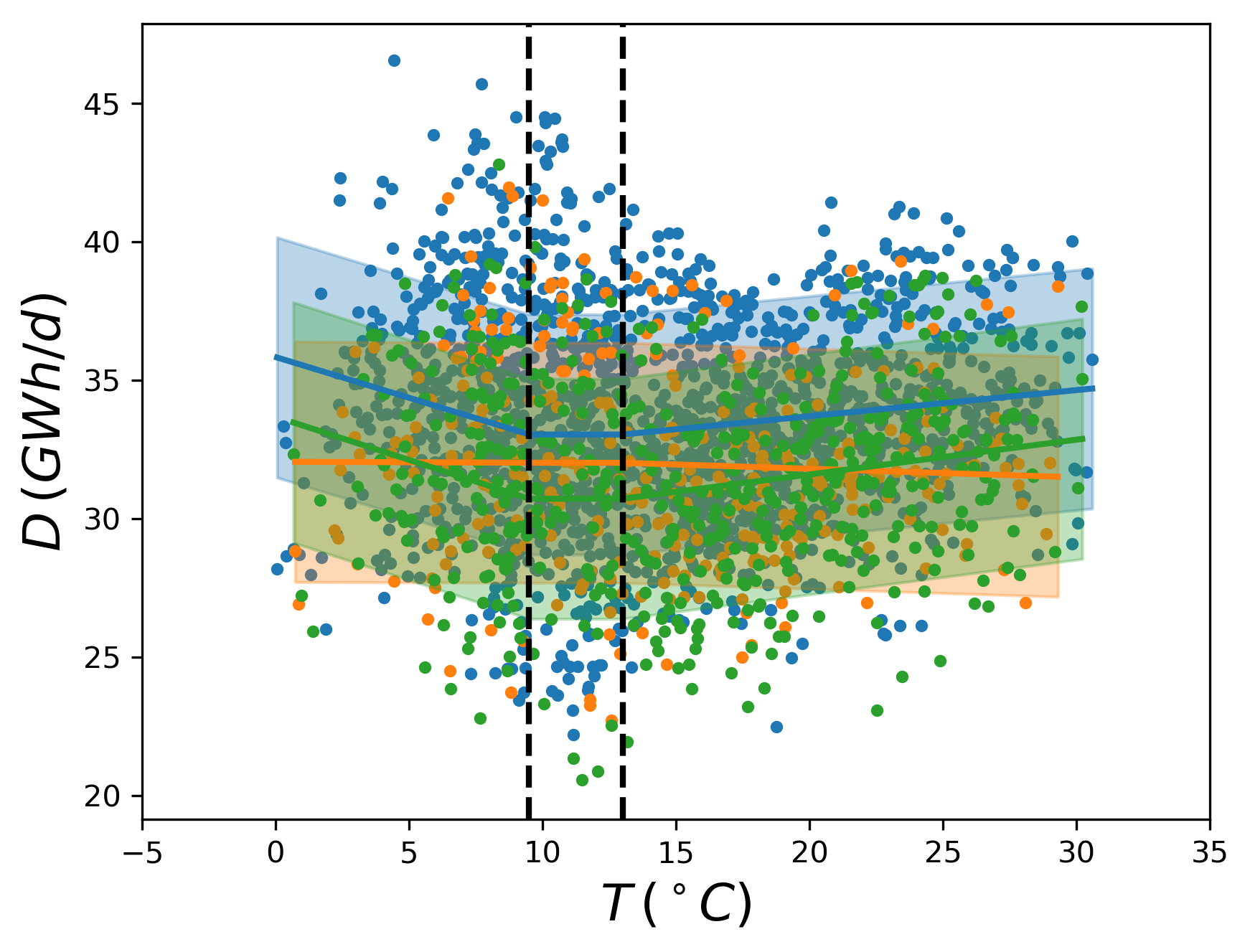}
    \subcaption{SARD}
  \end{subfigure}
  \begin{subfigure}{0.32\textwidth}
    \includegraphics[width=\textwidth]{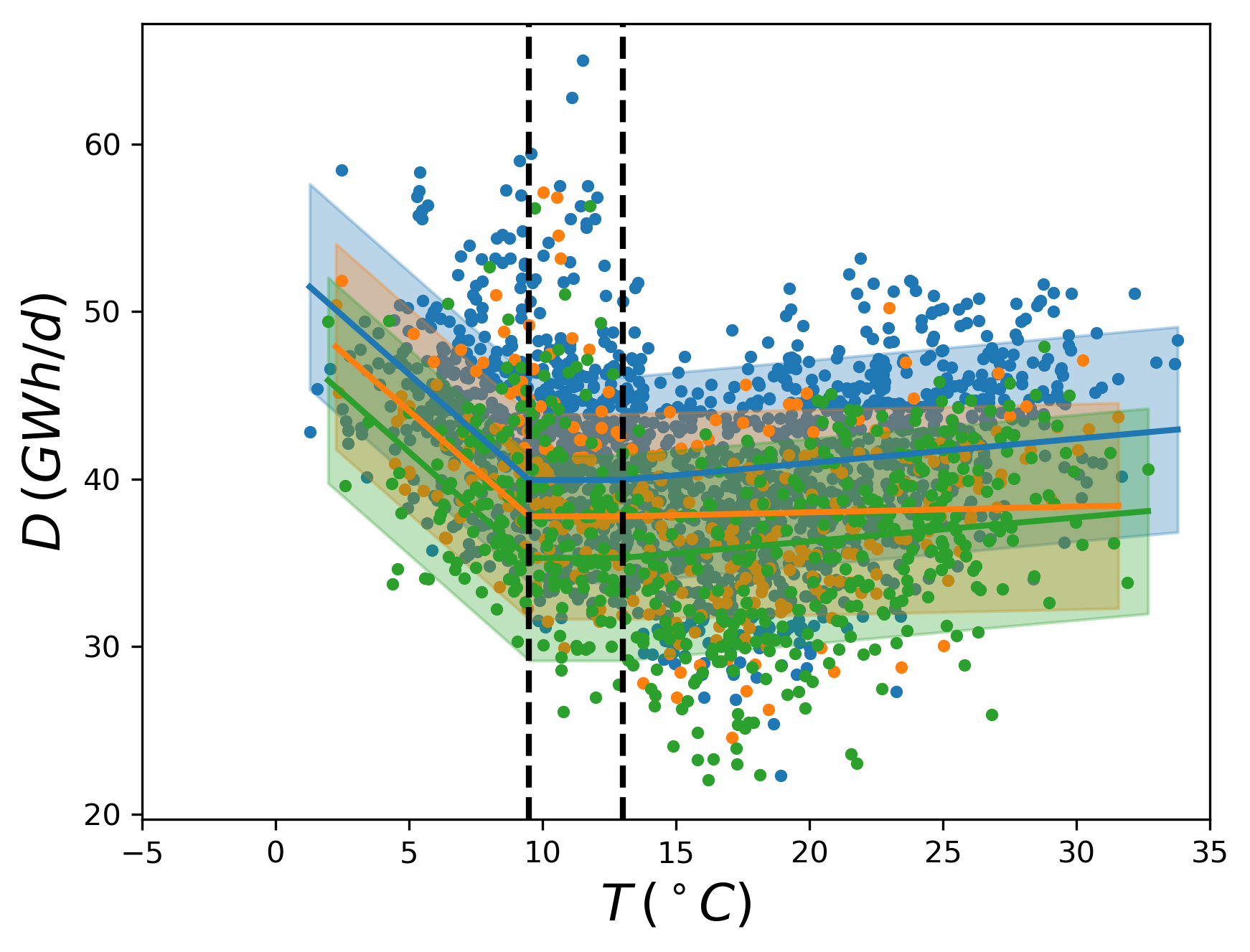}
    \subcaption{SICI}
  \end{subfigure}
  \caption{Daily-mean electricity demand for each zone versus the surface temperature.
    Each point is an observed realization of temperature and demand.
    The lines represent the functions $f_i^{\mathrm{work}|\mathrm{sat}|\mathrm{off}}$ of the demand model, while the associated shaded regions represent the variance of the prediction.
    Blue, orange and green data points and functions correspond to working days, Saturdays, and Sundays and holidays, respectively.
    The two vertical dashed lines represent the temperature thresholds $T_H = 9.5$ and $T_C = 13.0$.
  }\label{fig:demandPrediction}
\end{figure}

Snippets of this prediction for the north zone the first week of January and of July 2010 are represented as a time series in Figure~\ref{fig:intradayGeneration}.
We also represent the daily-means of the demand prediction for each zone in Figure~\ref{fig:demandPrediction} versus the input temperature from the CORDEX data.
The overall coefficient of determination is 0.73.
One can see that the temperature and type of day dependence of the demand is most clear for the economically most dynamic north zone.
This is also true, yet to a lesser extent, for the central south zone.
The shaded regions show that the part of the demand that is not explained by the temperature model is compensated by the Bayesian perturbations, although in a random fashion.

\section{Mean-variance analysis}\label{sec:meanVariance}

Geographic and technological diversification of renewable power plants is
based on a mean-variance analysis that is inspired by
Markowitz's modern portfolio theory\footnote{
  See, for example, \citet{mencarelli_complex_2018} for a survey
  on mathematical programming approaches for the portfolio selection problem.}.
We now give further details on the optimization problem associated with the mean-variance analysis implemented in \emph{e4clim} and applied in Section~\ref{sec:application}.
See the ``optimize mean-variance'' block at the top of Figure~\ref{fig:flow_chart}.

\subsection{Mean-variance optimization problem}\label{sec:mean_variance_optimization_problem}

In our context, the mean-variance analysis refers to the process of finding optimal spatial and technological distributions of renewable energy production achieving a trade-off between the mean penetration rate and some measure of the variance in the renewable energy production.
The variance is a proxy for the risk: minimizing the variance corresponds to maximizing the diversification of the renewable configuration, which in turn lowers the variability of renewable energy penetration and improves the flexibility of the system and its resistance to shocks.
In particular, a lower variance in the renewable energy mix is less demanding in services from conventional production (for which start up and shutting down services have a cost) or demand management.

The mean VRE penetration $\mu$ is given by~\eqref{eq:meanPenetration}, while the squared risk for the global strategy is given by~\eqref{eq:globalRisk}.
In the technology strategy, covariances between regions are ignored.
The squared risk thus becomes:

\begin{align}
  \sigma_\mathrm{technology}^2(\mathbf{w})
  &:= \sum_i \mathbb{V}\left[\frac{w_{(i, \mathrm{PV})} \eta_{(i, \mathrm{PV})}(\cdot)
    + w_{(i, \mathrm{wind})} \eta_{(i, \mathrm{wind})}(\cdot)}{\sum_i D_i(\cdot)}\right],\label{eq:regionalRisk}
\end{align}
For the base strategy, all covariances are ignored, to wit:
      
\begin{align}
  \sigma_\mathrm{base}^2(\mathbf{w})
  &:= \sum_i \mathbb{V}\left[\frac{
    w_{(i, \mathrm{PV})} \eta_{(i, \mathrm{PV})}(\cdot)}{\sum_i D_i(\cdot)
    }\right]
    + \mathbb{V}\left[\frac{
    w_{(i, \mathrm{wind})} \eta_{(i, \mathrm{wind})}(\cdot)}{\sum_i D_i(\cdot)
    }\right].\label{eq:baseRisk}
\end{align}

The classical method used in the following Section~\ref{sec:application} to approximate the optimal frontier numerically is explained in~\ref{sec:biobjective}.
In this study, we refer to the optimal frontier as the curve
$\left(\sigma_{\mathrm{global} | \mathrm{technology} | \mathrm{base}}(\hat{\mathbf{w}}),
  \mu(\hat{\mathbf{w}})\right)$,
where the $\hat{\mathbf{w}}$ are the optimal solutions, although, strictly speaking, it is the risk squared that is minimized rather than the risk itself.
The numerical results of the following section suggest that the so-defined frontier of the bi-objective problem~\eqref{eq:optimization} without the total capacity constraint~\eqref{eq-biobj:constr_sum_w} is a half line with a positive slope that we refer to as the \emph{mean-risk ratio} $\alpha_{\mathrm{global} | \mathrm{technology} | \mathrm{base}}$.
In other words, the optimal mixes for this problem are such that

\begin{align}
  \label{eq:meanRiskRatio}
  \mu(\hat{\mathbf{w}})
  = \alpha_{\mathrm{global} | \mathrm{technology} | \mathrm{base}} \enskip
  \sigma_{\mathrm{global} | \mathrm{technology} | \mathrm{base}}(\hat{\mathbf{w}}),
\end{align}
In the following, we assume that this is indeed the case and the mean-risk ratio $\alpha_{\mathrm{global} | \mathrm{technology} | \mathrm{base}}$ is used to diagnose the variants of the optimization problem.
The proof of a rigorous mathematical result is left for future work.

\subsection{Method to find an approximation of the optimal frontier}\label{sec:biobjective}
The results of the following sections are valid for all three strategies.
First of all, let us define two single objective subproblems which represent a restriction of the bi-objective problem we aim at solving.
We follow the well-known method called $\epsilon$-constraint, see, for example,~\citet[][Chap.~II.3]{miettinen_nonlinear_1999}.

The first subproblem ${(P)}^{\mbox{min}}$ is defined by
\begin{eqnarray}
  \min_\mathbf{w}
  & \sigma^2(\mathbf{w}) \label{eq-spmin:of1}\\
  \mbox{subject to} & \sum_\mathbf{k} w_\mathbf{k}
                      = w_{\mbox{total}}\label{eq-spmin:constr_sum_w}\\
         & w_\mathbf{k} \geq 0 \quad \forall~\mathbf{k} \label{eq-spmin:positive_w}\\
         & \sum_\mathbf{k} w_\mathbf{k} \mathbb{E}[\eta_\mathbf{k}]
           \geq \mu^* \mathbb{E} \left[ \sum_i D_i \right] \label{eq-spmin:constr_of2}
\end{eqnarray}
where $\mu^*$ ranges from $l^{of2}/(\sum_i D_i)$ to $u^{of2}/(\sum_i D_i)$ where $l^{of2}$ and $u^{of2}$ are the lower and the upper bound on the value of the second objective function~\eqref{eq-spmax:of2} (below), respectively.

The second subproblem ${(P)}^{\mbox{max}}$ is defined by
\begin{eqnarray}
  \max_\mathbf{w} & \sum_\mathbf{k} w_\mathbf{k}
           \mathbb{E}[\eta_\mathbf{k}]\label{eq-spmax:of2}\\
  \mbox{subject to} & \sum_\mathbf{k} w_\mathbf{k}
                      = w_{\mbox{total}}\label{eq-spmax:constr_sum_w}\\
         & w_\mathbf{k} \geq 0 \quad \forall~\mathbf{k} \label{eq-spmax:positive_w}\\
         & \sigma^2(\mathbf{w})
           \leq  {(\sigma^*)}^2 \label{eq-spmax:constr_of1}
\end{eqnarray}
where ${(\sigma^*)}^2$ is defined in $[l^{of1},u^{of1}]$ where $l^{of1}$ and $u^{of1}$ are the lower and the upper bound on the value of the first objective function~\eqref{eq-spmin:of1}, respectively.

\subsubsection{The algorithm}\label{sec:algorithm}
The idea is to find the best value of~\eqref{eq-spmin:of1} by solving ${(P)}^{\mbox{min}}$ for each value of $\mu^*$. As $\mu^*$ is continuously defined, it is, of course, impossible to solve it for each possible value of it.
Thus, we discretize the possible values of $\mu^*$ with a step of $0.1\%$ and find just a subset of the optimal frontier.

Note that solving ${(P)}^{\mbox{min}}$ for different values of $\mu^*$ is not enough to guarantee that the solutions found are not dominated by any other solution.
For this reason, it is necessary to alternate between solving ${(P)}^{\mbox{min}}$ for a given value of $\mu^*$, then solving ${(P)}^{\mbox{max}}$ by setting ${(\sigma^*)}^2$ equal to the objective function value found by solving ${(P)}^{\mbox{min}}$, then update $\mu^*$ accordingly and solving ${(P)}^{\mbox{max}}$ again, and so on, until the values $\mu^*$ and ${(\sigma^*)}^2$ cannot be updated anymore. 

If we do not perform this alternating update and solve of the two subproblems it might happen that solving only ${(P)}^{\mbox{min}}$ or ${(P)}^{\mbox{max}}$ might produce a dominated solution. Take the case of Fig.5(a), for example: for a value of risk equal to $15$ there are two points in the light blue curve, i.e., mean equal to $15$ and equal to $22$. If we solve ${(P)}^{\mbox{min}}$ with constraint~\eqref{eq-spmin:constr_of2} corresponding to mean at least equal to $15$, we will find a solution of objective functino value $15$, but it might correspond either to a mean of value $15$ (and in this case the solution would be dominated) or to a mean of value $22$.

\subsubsection{How to find the bound on the RHS of~\eqref{eq-spmin:constr_of2} and~\eqref{eq-spmax:constr_of1}}
We aim at finding the lower and upper bound
of the objective functions~\eqref{eq-spmin:of1} and~\eqref{eq-spmax:of2}
so as to be able to define the interval over which we can vary
the right-hand-side of constraints~\eqref{eq-spmin:constr_of2}
or~\eqref{eq-spmax:constr_of1}.

For the lower bound of~\eqref{eq-spmin:of1}
(resp.~the upper bound of~\eqref{eq-spmax:of2}) it is simple:
we drop~\eqref{eq-spmax:of2} (resp.~\eqref{eq-spmin:of1})
and solve the corresponding single objective problem,
which is a relaxation of original bi-objective problem.
To find the upper bound of~\eqref{eq-spmin:of1} (resp.~the lower bound of~\eqref{eq-spmax:of2}) it is sufficient to we drop~\eqref{eq-spmax:of2} (resp.~\eqref{eq-spmin:of1}),
invert the direction of~\eqref{eq-spmin:of1} (resp.~\eqref{eq-spmax:of2})
and solve the corresponding single objective problem.

\section{Evaluation against observations}\label{sec:evaluation}

The robustness of the results presented in Section~\ref{sec:optimization} is tested here.
We evaluate the capacity of the demand and capacity factor models to reproduce
the spectral characteristics of observations.
For that purpose, we compare in Table~\ref{tab:varFreqSolar} the percentage of the variance of the Italian average of the electricity demand (top), PV capacity factor (middle) and wind capacity factor (bottom) explained by periods greater than a year,
less or equal than a year and greater than a day and less or equal than a day.
These estimates are (i) directly from observed time series; (ii) from the models applied to the daily CORDEX data; (iii) and from the models applied to the MERRA-2 data with 10m and 50m wind\footnote{
  The variance explained by each frequency band is calculated
  from the variance of the respectively low-pass, band-pass
  and high-pass filtered time series using running averages as filters.}.
Because the demand observations from GME, the CORDEX and the MERRA-2 data overlap over the 2005--2012 period, it is possible to select a common period over which to compare the demand variance.
However, this is not the case of capacity factor observations, as the GSE and the CORDEX data do not overlap (the GSE is selected over the 2013--2017 period, see Sect.\ref{sec:GMEGSE}).
Then, the common period of 2013--2018 is selected for the GSE and the MERRA-2 capacity factors, but the 1989--2012 period is used for the CORDEX capacity factors.

We can see that, regardless of the climate data, the demand model does not resolve the little interannual variability of the observed demand.
This is to be expected if this variability is due to socio-economial factors rather than interannual climate variability.
In fact, the variance from year-to-year unresolved factors is modeled by the Bayesian model as intraday random perturbations, so that part of the interannual variance is transferred to the intraday variance.
Moreover, when using the CORDEX (MERRA-2) data, the intraday variance is under-estimated (over-estimated).
Note, however, that these results are sensitive to the realizations of the noise in the Bayesian model (not shown here).

We now analyse the variance of the capacity factors (Tab.~\ref{tab:varFreqSolar}).
The seasonal variability of the PV capacity factors appears to be underestimated, regardless of the climate data used, although less so with the MERRA-2 data.
In addition, the variability of the wind capacity factor is relatively well estimated when using the MERRA-2 data with 10m data.
The intraday variability of the wind capacity factor is underestimated when using the MERRA-2 data with 50m wind, suggesting that 50m winds are much less turbulent than 10m winds and than observations.
On the other hand, it is overestimated when using the CORDEX data.
This suggests that the parameterization of the intraday variability of the wind capacity factors is not entirely satisfactory.
However, because the CORDEX and the GSE/MERRA-2 data do not overlap, it is not possible to tell whether differences in the capacity factors are due to modelling errors or to interannual variability.
\begin{table}
  \centering
  \begin{tabular}{r c c c}
    & Interannual & Seasonal & Intraday \\
    \midrule
    Electricity Demand (2005--2012) & & & \\
    \cmidrule{1-1}
    GME & 3.3 & 44.9 & 51.8 \\ 
    CORDEX & 0.0 & 54.1 & 45.9 \\
    MERRA-2 & 0.0 & 33.8 & 66.2 \\
    \midrule
    PV Capacity Factor & & & \\
    \cmidrule{1-1}
    ENTSO-E \& GSE (2013--2018) & 1.1 & 11.8 & 87.1 \\
    CORDEX (1989--2012) & 0.0 & 4.3 & 95.7 \\
    MERRA-2 (2013--2018) & 0.0 & 5.6 & 94.3 \\
    \midrule
    Wind Capacity Factor & & & \\
    \cmidrule{1-1}
    ENTSO-E \& GSE (2013--2018) & 1.4 &  64.4 & 34.2 \\
    CORDEX (1989--2012) & 0.4 & 46.9 & 52.8 \\
    MERRA-2 10m wind (2013--2018) & 0.2 & 65.9 & 33.8 \\
    MERRA-2 50m wind (2013--2018) & 0.3 & 74.0 & 25.6 \\
  \end{tabular}
  \caption{
    Comparison, for the North zone, of the percentage of the variance  of the electricity demand (top), PV capacity factor (middle) and wind capacity factor (bottom) explained by periods greater than a year, less or equal than a year and greater than a day and less or equal than a day, from the GME/ENTSO-E \& GSE, CORDEX and MERRA-2 datasets.
  }\label{tab:varFreqSolar}
\end{table}

\begin{figure}
  \centering
  \begin{subfigure}{0.32\linewidth}
    \includegraphics[width=\textwidth]{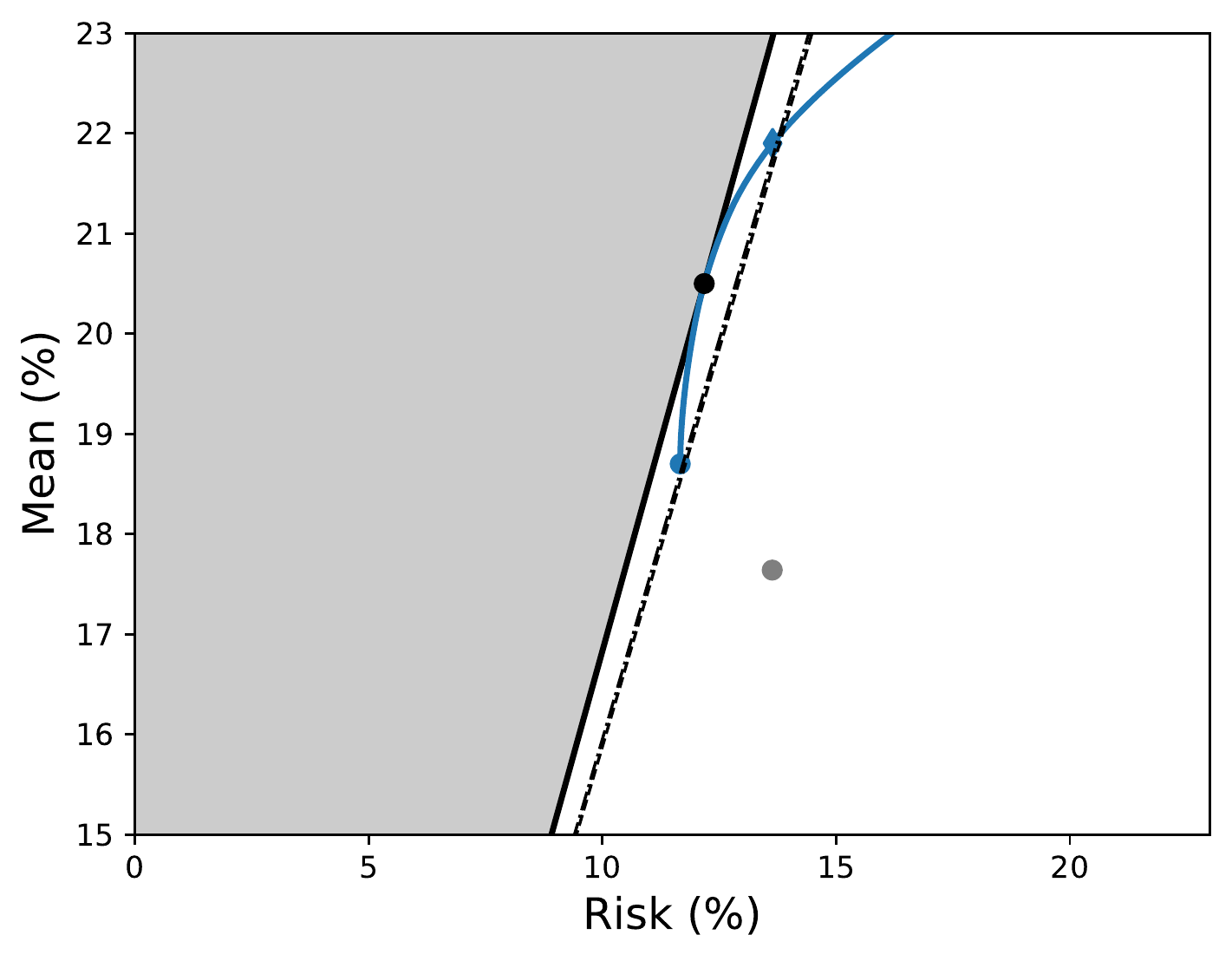}
    \caption{}\label{fig:meanRiskGlobalObs}
  \end{subfigure}
  \begin{subfigure}{0.32\linewidth}
    \includegraphics[width=\textwidth]{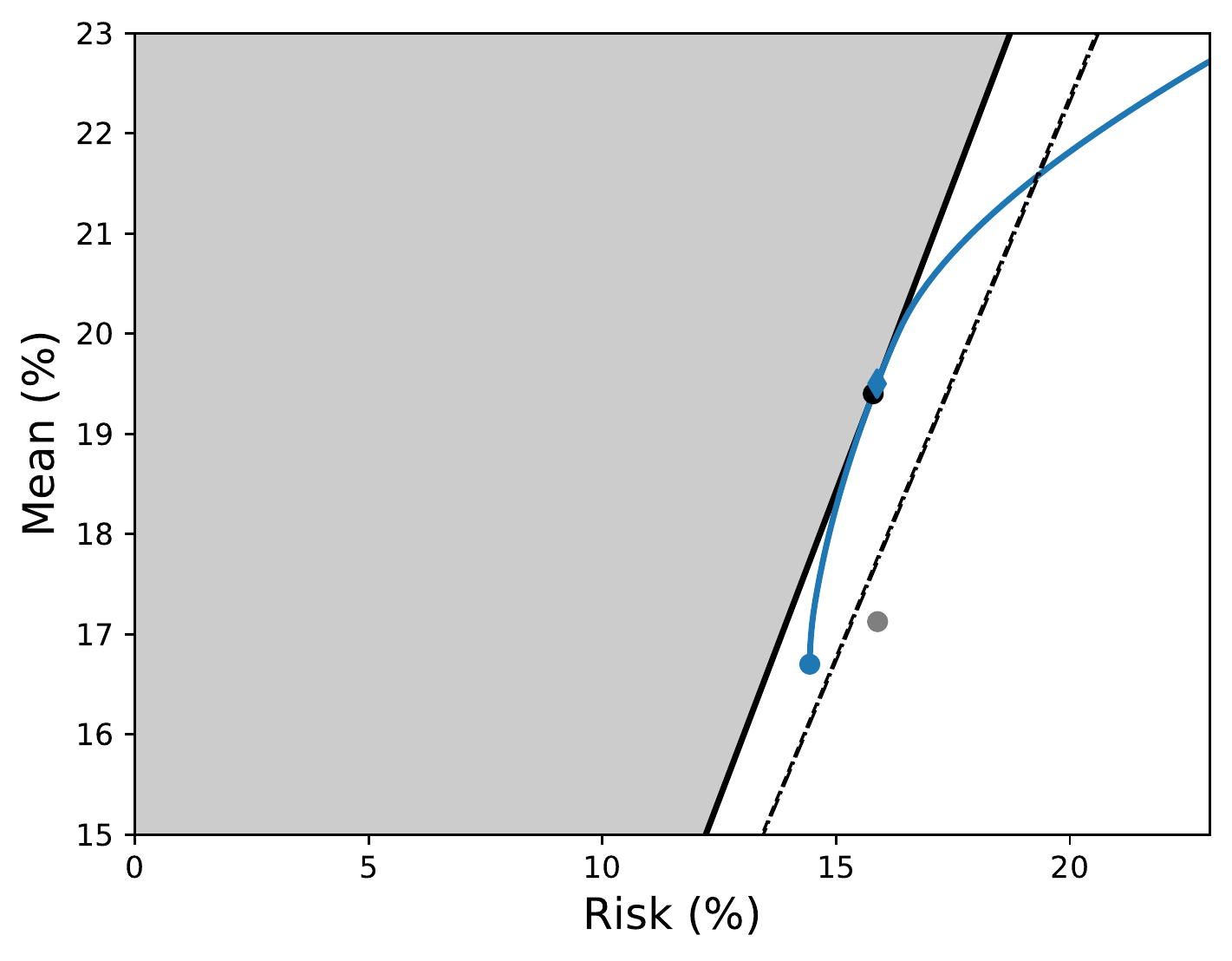}
    \caption{}
  \end{subfigure}
  \begin{subfigure}{0.32\linewidth}
    \includegraphics[width=\textwidth]{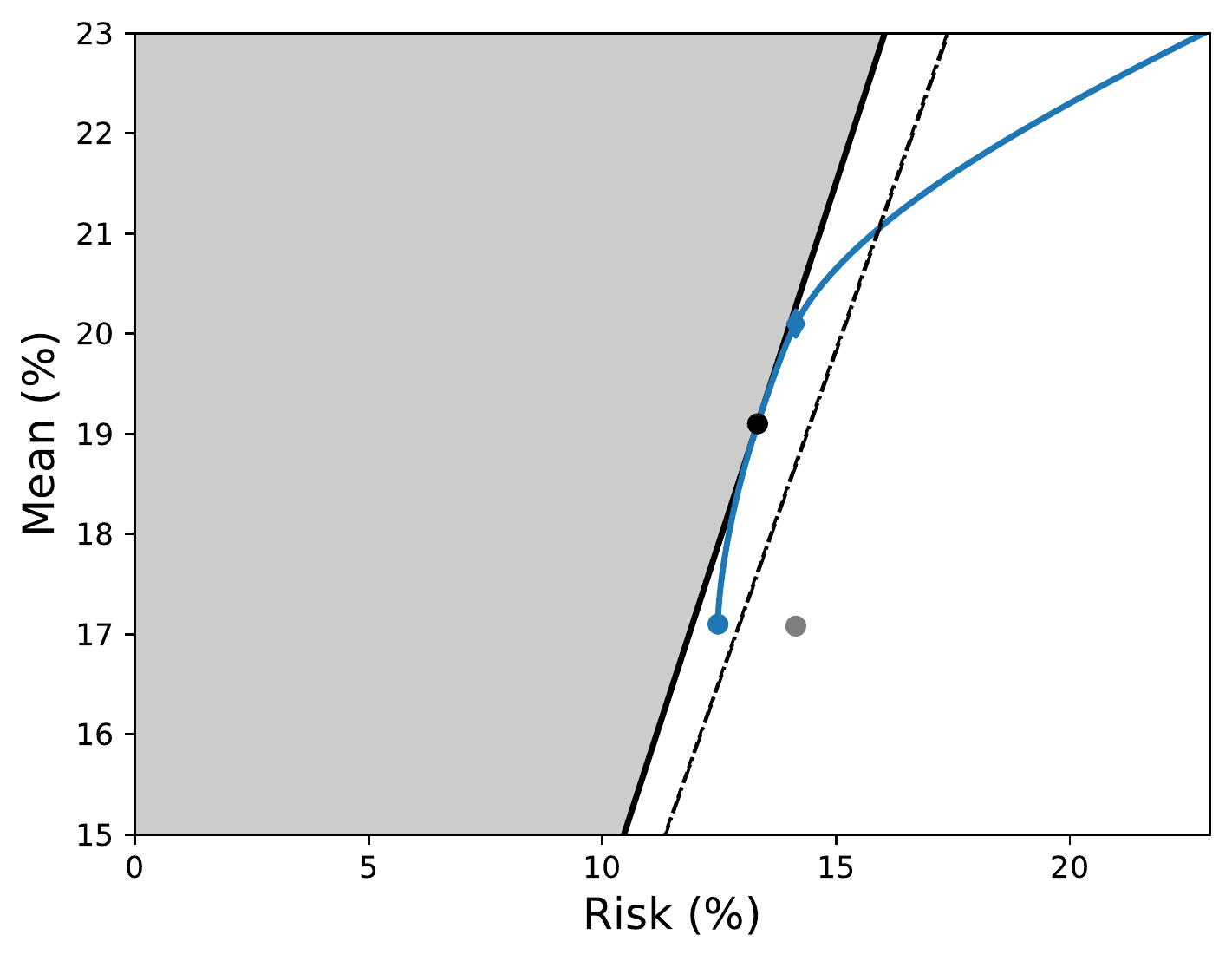}
    \caption{}
  \end{subfigure}
  \caption{
    Approximated optimal frontiers for the global strategy directly solved from observations (left), MERRA-2 with 10m wind (center) and MERRA-2 with 50m wind (right).
    To be compared with the approximated optimal frontiers and mix characteristics obtained from the CORDEX data in Figure~\ref{fig:mix}.
  }\label{fig:mixObs}
\end{figure}

\begin{figure}
  \centering
  \begin{subfigure}{0.32\linewidth}
    \includegraphics[width=\textwidth]{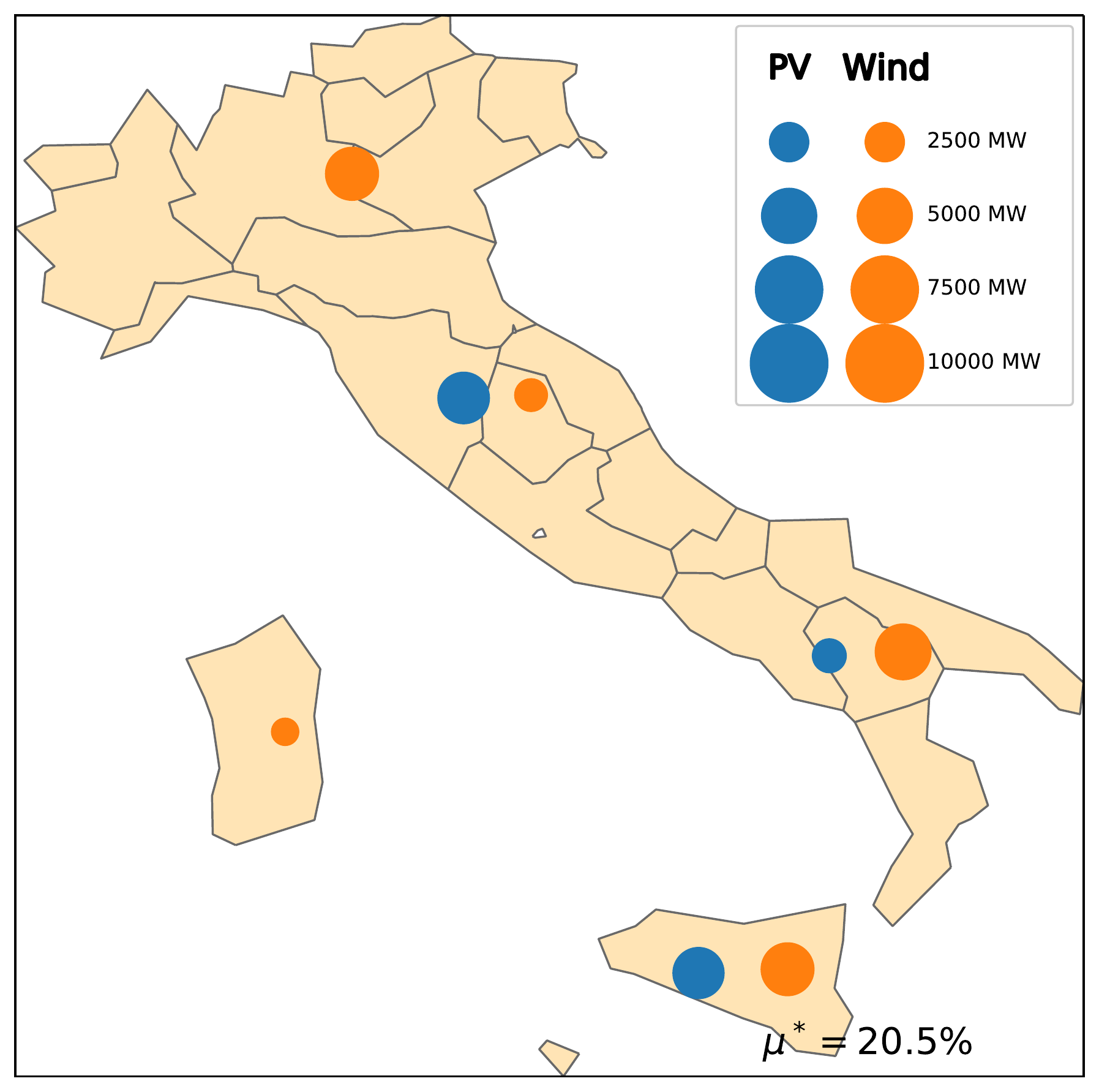}
  \end{subfigure}
  \begin{subfigure}{0.32\linewidth}
    \includegraphics[width=\textwidth]{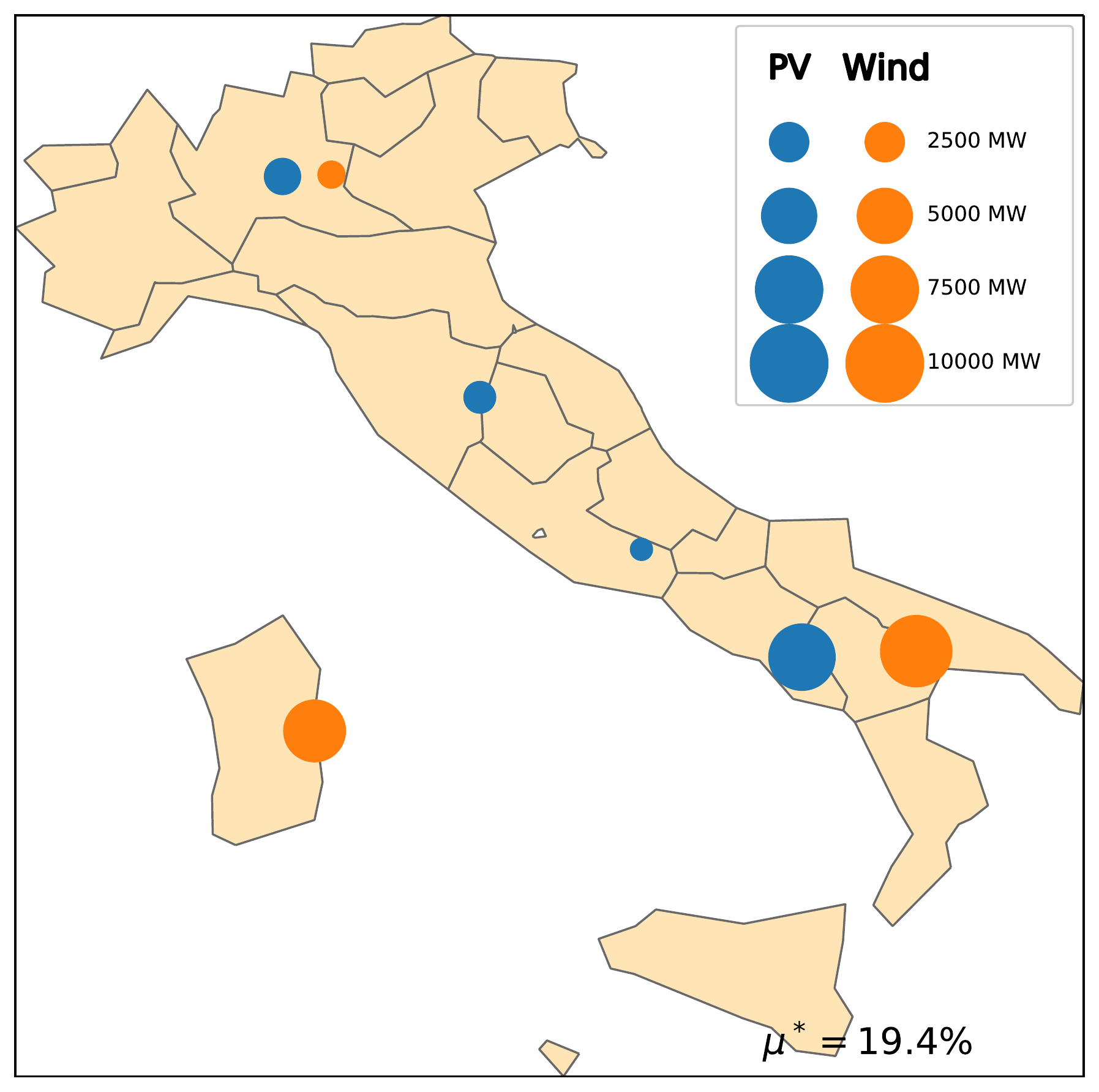}
  \end{subfigure}
  \begin{subfigure}{0.32\linewidth}
    \includegraphics[width=\textwidth]{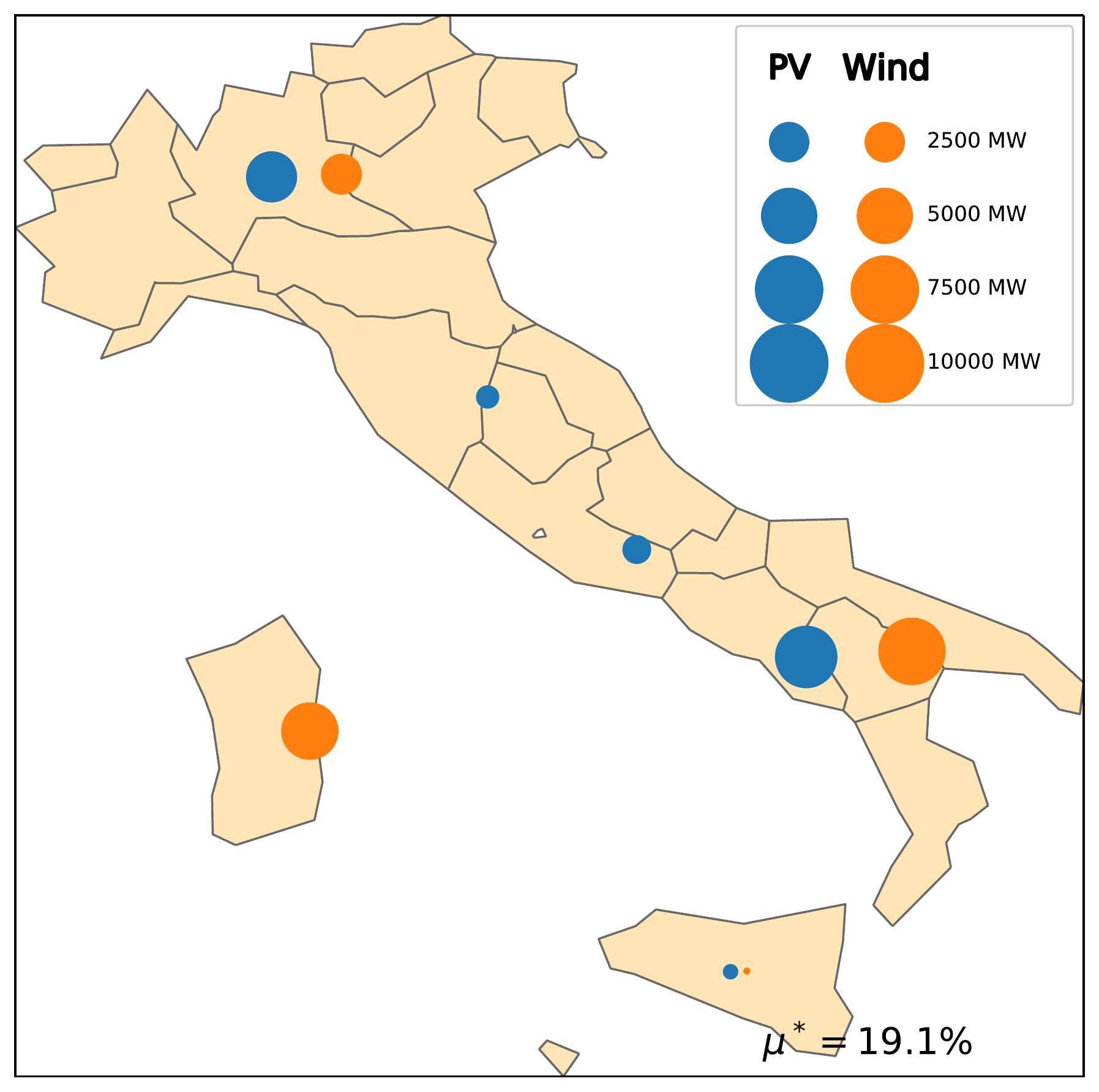}
  \end{subfigure}
  \caption{
    Geographical and technological distribution of the PV and wind capacity maximizing the mean-risk ratio for the global strategy obtained from observations (left), MERRA-2 with 10m wind (center) and MERRA-2 with 50m wind (right).
    To be compared with the RES distribution obtained from the CORDEX data in  Figure~\ref{fig:mapGlobalMaxRatio}.
  }\label{fig:capacityMapObs}
\end{figure}

To compare the optimal mixes obtained from direct observations of the capacity factors with those obtained from the MERRA-2 data over the same 2013--2018 period, the approximated optimal frontiers and the maps of the mix minimizing the mean-risk ratio obtained from the observed capacity factors and demand data are represented in Figures~\ref{fig:mixObs} and~\ref{fig:capacityMapObs}, respectively.
The mean-risk ratio obtained from observations, MERRA-2 with 10m wind and MERRA-2 with 50m wind is of 1.68, 1.23 and 1.43, respectively.
We can thus see that large differences exist between the observations and the MERRA-2 frontiers, which translate into qualitatively different capacity distributions.

These results suggest that, while using climate data to perform the mean-risk analysis allows one to study the impact of climate variability on mixes, care should be taken when interpreting the optimal capacity distributions in absolute terms.
The fact that biases exist between the results obtained from the different climate datasets shows that, while some of the differences with observations may be due to the energy models, some of them originate from the climate data.
As suggested in Section~\ref{sec:impactClimate}, we recommend analysing results using multiple, relatively independent, climate data sources to estimate biases stemming from the climate data.

\section*{References}
\bibliographystyle{plainnat}
\bibliography{201807_EnergyMix_Italy}

\end{document}